\documentclass[12pt,english]{article}
\usepackage[T1]{fontenc}
\usepackage[latin9]{inputenc}
\usepackage{geometry}
\usepackage{float}
\usepackage{url}
\usepackage{amsmath}
\usepackage{amssymb}
\usepackage{graphicx}
\usepackage{rotfloat}
\usepackage{setspace}
\usepackage{esint}
\makeatletter

\providecommand{\tabularnewline}{\\}
\floatstyle{ruled}
\newfloat{algorithm}{tbp}{loa}
\providecommand{\algorithmname}{Algorithm}
\floatname{algorithm}{\protect\algorithmname}

\newtheorem{theorem}{Theorem}

\newtheorem{condition}{Assumption}

\newtheorem{example}{Example}

\newtheorem{lemma}{Lemma}

\newtheorem{proposition}{Proposition}

\newenvironment{proof}[1][Proof]{\textbf{#1.} }{\ \rule{0.5em}{0.5em}}

\makeatother

\usepackage{caption}
\usepackage[labelfont={bf,sf}]{caption}

\usepackage{babel}
\usepackage{comment}
\usepackage{adjustbox}
\usepackage{booktabs}
\usepackage{rotating}
\usepackage{import}

\usepackage{amsmath}
\usepackage{graphicx}
\usepackage{subfig}

\usepackage{tkz-graph}  
\usetikzlibrary{arrows.meta}

\makeatother

\usepackage{babel}
\begin{document}
\title{Consistent Causal Inference for High-Dimensional Time Series\thanks{We are grateful to the Editor Serena Ng and the Referees for comments
that have led to corrections and improvements in content and presentation.
We are also grateful to Yanqin Fan for having shared the latest version
of Fan et al. (2022) and useful discussions. We thank the participants
at the Model Evaluation and Causal Search workshop at the University
of Pisa, the Lancaster Financial Econometrics Conference in honour
of Stephen Taylor, and the 2023 SoFiE Conference at Sungkyunkwan University.
The first author acknowledges financial support  from MIUR Progetti
di Ricerca di Rilevante Interesse Nazionale (PRIN) Bando 2017. Both
authors acknowledge financial support from the Leverhulme Trust Grant
Award RPG-2021-359. } }
\author{Francesco Cordoni\thanks{Department of Economics, Royal Holloway University of London, Egham
TW20 0EX, UK. Email: francesco.cordoni@rhul.ac.uk} and Alessio Sancetta\thanks{Corresponding Author. Department of Economics, Royal Holloway University
of London, Egham TW20 0EX, UK. Email: asancetta@gmail.com}}
\maketitle
\begin{abstract}
A methodology for high dimensional causal inference in a time series
context is introduced. It is assumed that there is a monotonic transformation
of the data such that the dynamics of the transformed variables are
described by a Gaussian vector autoregressive process. This is tantamount
to assume that the dynamics are captured by a Gaussian copula. No
knowledge or estimation of the marginal distribution of the data is
required. The procedure consistently identifies the parameters that
describe the dynamics of the process and the conditional causal relations
among the possibly high dimensional variables under sparsity conditions.
The methodology allows us to identify such causal relations in the
form of a directed acyclic graph. As illustrative applications we
consider the impact of supply side oil shocks on the economy, and
the causal relations between aggregated variables constructed from
the limit order book on four stock constituents of the S\&P500.


\end{abstract}

\section{Introduction\label{Section_introduction}}

Identifying and estimating causal relations is a problem that has
received much interest in economics. In the last two decades the statistical
and machine learning literature has made a number of advances on the
front of identification and estimation within the framework of causal
graphs (Comon, 1994, Hyv\"{a}rinen and Oja, 2000, Pearl, 2000, Spirtes
et al., 2000, Hyv\"{a}rinen et al., 2001, Shimizu et al., 2006, Meinshausen
and B\"{u}hlmann, 2006, Kalisch and B\"{u}hlmann, 2007, Cai et al.,
2011, B\"{u}hlmann et al., 2014, Peters et al., 2014), where the
data generating process can be characterized as a system of structural
equations. This complex causal relations system might be represented
through the causal graph, which conveys essential topological information
to estimate causal effects.

However, the true data generating process is often a latent object
to researchers, which can only rely on finite sample observations
to infer the causal structure and mechanism of the true system. A
causal model entails a probabilistic model from which a researcher
can learn from observations and outcomes about changes and interventions
of the system variables (Pearl, 2000, Peters et al., 2014). Thus,
causality can be formally defined using the do-notation of Pearl (2000)
in terms of intervention distributions. This definition of causality
is quite different from the well known concept of Granger causality.
However, causal relations in economics and finance require to account
for time series dependence. 

In this paper we develop a methodology to extract the causal relations
of time series data, conditioning on the past in a flexible way. We
assume that there is a monotone transformation of the data that maps
the original variables into a Gaussian vector autoregressive (VAR)
model (see also Fan et al., 2022). There are a number of advantages
to this approach. First, we are able to retain the interpretability
of VAR models building on the rich econometrics literature on structural
VAR models. Second, we do not need any assumptions on the marginal
distribution of the data. This means that the procedure is robust
to fat tails, as we do not make any assumption on the existence of
any moments. For instance, given that the existence of a second moment
for financial data has been a much debated topic in the past (Mandelbrot,
1963, Clarke, 1973, for some of the earliest references) dispensing
all together of this unverifiable condition should be welcomed. Third,
we can model variables that take values in some subset of the real
line, for example variables that only take positive values or are
truncated. This is not possible using a standard VAR model. 

The estimation of the contemporaneous causal structure of a time series
is equivalent to solving the identification problem of a structural
VAR model. The latter can be achieved by finding a unique Choleski
type decomposition of the covariance matrix of the VAR innovations
(Rigobon, 2003, Moneta et al., 2013, Gouri\'{e}roux et al., 2017,
Lanne et al., 2017). Recent advances in the identification problem
under general conditions and linearity exploit the use of internal
and external instruments and the method of local projections (Stock
and Watson, 2018, Plagborg-M\o{}ller and Wolf, 2021). However, the
time series dynamics of economic and financial data may not be captured
well by a linear VAR model when the data is not Gaussian. For example,
some variables may only be positive. The problem of estimation is
exacerbated if the data have fat tails. This may distort the estimates.
Such problems reflect negatively on the estimation of causal relations
for time series data. Furthermore, due to the curse of dimensionality,
SVAR analysis is only feasible in a low-dimensional context. Restricting
the VAR model only to a few variables may lead to unreasonable adverse
effects such as `price-puzzles' in impulse responses (Sims, 1992,
Christiano et al., 1999, Hanson, 2004). Moreover, models of the global
economy tend to be high dimensional. To avoid the curse of dimensionality,
factor augmented VAR (FAVAR) models (Bernanke et al., 2005) and dynamical
factor models (Forni et al., 2000, Forni et al., 2009) are often employed.
However, the interpretation of the causal relations with factor models
is not always straightforward. Along these approaches, we also mention
the GVAR methodology, originally proposed by Pesaran et al. (2004),
where country specific VAR models are stacked together in a way that
maintains ease of interpretation at the cost of some assumptions.
Our methodology does not require the machinery of factor models or
assumptions on how to join lower dimensional models into a higher
dimensional one. However, this is achievable at the cost of certain
restrictions. We envisage that our methodology could work in conjunction
with the the existing ones to shed further light on structural relation
in high dimensional VAR models. We also point out that high dimensional
VAR models may even arise in practice as a result of a large number
of lags. 

This paper builds on a number of previous contributions and develops
a methodology to address the aforementioned problems. Our approach
is tantamount to the assumption that the cross-sectional and transition
distribution of the variables can be represented using a Gaussian
copula. The procedure builds on the work of Liu et al. (2012) and
does not require us to estimate any transformation of the variables
or the marginal distribution of the data, as commonly done when estimating
a copula. In fact, our procedure bypasses the estimation of the innovations
of the model altogether. Our methodology is built for high dimensional
time series, as commonly found in some economics and financial applications.
What we require is some form of sparsity in the partial dependence
of the data. This is different from assuming that the covariance matrix
of innovations or the matrix of autoregressive coefficients are sparse.
Such two restrictions can be restrictive. We shall make this clear
in the text when we discuss our assumptions. Finally, even when not
all causal relations are identified, we are able to identify the largest
number of causal relations. This statement is formalized by the concept
of complete partially acyclic graph using the PC algorithm (Spirtes
et al., 2000, Kalisch and B\"{u}hlmann, 2007). These concepts are
reviewed in the main body of the paper (Section \ref{Section_identification}). 

We conclude this introduction with a few remarks whose aim is to put
the goals of this paper into a wider perspective. The process of scientific
discovery is usually based on 1. the observation of reality, 2. the
formulation of a theory, and 3. tests of that theory. The plethora
of data available allows the researcher to observe different aspects
of reality that might have been precluded in the past. High dimensional
estimation methods are particularly suited to explore the present
data-centric reality. However, the next step forward requires formulation
of a theory or hypothesis. Such theory needs to be able to explain
rather than predict in order to enhance our understanding. This very
process requires the identification of a relatively small number of
explanatory causes for the phenomenon that we are trying to understand.
The problem's solution, in a complex and rather random environment,
should then be a simple approximation. This approximation can then
be tested in a variety of situations in order to verify its applicability.
The program of this paper is to follow this process of scientific
discovery. We start from possibly high dimensional dynamic datasets.
We aim to provide a reduced set of possible contemporaneous causes
conditioning on the past.

\subsection{Relation to Other Work}

One of the main empirical econometric tools for the study of policy
intervention effects is the VAR approach (Sims, 1980, Kilian and L\"{u}tkepohl,
2017). In the first step, the so called reduced form model is estimated.
Then, the structural counterpart needs to be recovered. This gives
rise to an identification problem, which is equivalent to finding
the contemporaneous causal relations among the variables. 

Traditionally, the identification of Structural Vector Autoregressive
(SVAR) models was achieved by imposing model restrictions. Such restrictions
can be derived from an underlying economic model, such as short and
long-run restrictions on the shocks impact (Bernanke, 1986, Blanchard
and Quah, 1989, Faust and Leeper, 1997), or imposing sign restrictions
on impulse response functions (Uhlig, 2005, Chari et al. 2008). 

The success of the VAR approach is its reliance on data characteristics,
thus allowing the validation of economic models under reasonably weak
assumptions. However, standard restrictions necessary for the identification
invalidate the data-driven nature of SVAR.\textbf{ }Recently, researchers
have explored alternative methods to achieve identification in SVAR
models by exploiting different statistical features of the data. For
instance, identification can be obtained by relying on heteroskedasticity
(Sentana and Fiorentini, 2001, Rigobon, 2003, L\"{u}tkepohl and Net\v{s}unajev,
2017) or non-Gaussianity of the residuals (Moneta et al., 2013, Gouri\'{e}roux
et al., 2017, Lanne et al., 2017). On the other hand, another popular
method used for identification, which however does not exploit specific
statistical properties of the data distribution as the previously
mentioned, is the instrumental variables approach (Mertens and Ravn,
2013, Stock and Watson, 2018, Plagborg-M\o{}ller and Wolf, 2021). 

Our method is related to approaches that rely on the graphical causal
model literature (Swanson and Granger, 1997, Demiralp and Hoover,
2003, Moneta, 2008), where identification can be achieved by exploiting
the set of conditional and unconditional independence relations in
the data. Our work is also related to the statistical and machine
learning literature for the identification of causal graph structures
in a high dimensional setting (Meinshausen and B\"{u}hlmann, 2006,
Kalisch and B\"{u}hlmann, 2007, Liu et al., 2009, Zhou et al., 2011,
B\"{u}hlmann et al., 2014, Harris and Drton, 2013). In particular
the latter reference combines the use of rank correlations with the
PC algorithm, as we do in the present paper. However, none of these
approaches accounts for contemporaneous causal inference conditioning
on the past, as required for time series problems. 

To account for the time series dependence, we employ a modelling assumption
that can be viewed as a Gaussian copula VAR model, a definition that
will be made clear in the text. We recently discovered that Fan et
al. (2022) have used the same time series assumption for the analysis
of high dimensional Granger causality. The present paper is concerned
with conditional causal relations and identification of the Gaussian
copula VAR. Moreover, some basic assumptions are also different. For
example, Fan et al. (2022) assume that the autoregressive matrix of
the Gaussian copula VAR is sparse. We instead assume that the inverse
of the scaling matrix of the Gaussian copula that leads to a VAR representation
is sparse. This is a very different assumption. Hence, the contributions
are related, but complementary. 

\subsection{Outline of the Paper}

The plan for the paper is as follows. In Section \ref{Section_Model},
we introduce the model and briefly discuss its statistical properties.
In Section \ref{Section_identification} we discuss identification
of the model and the causal relations. In Section \ref{Section_algos}
we describe algorithms to find estimators for the population quantities,
including the complete partially acyclic graph. In Section \ref{Section_asymptotics}
we state conditions and results for the consistency of the quantities
derived from the algorithms. Section \ref{Section_empirics} provides
two empirical illustrations. First,\textbf{ }we investigate the identification
of the effect of supply side shocks on economic activity. Then, we
analyze the causal relations of order book variables in electronic
trading. Section \ref{Section_conclusion} concludes. Additional explanatory
material can be found in the Appendix. There, we provide more details
on the model and its identification under possibly mixed data types.
We also discuss calculation of impulse response functions for our
nonlinear model. All the proofs and other additional details can be
found in the Electronic Supplement to this paper. There we also present
the main conclusions from a simulation study as evidence of the finite
sample properties of our methodology (Section \ref{Section_summarySimulations}
in the Electronic Supplement). 

\paragraph{Software.}

The algorithms presented in this paper are implemented in the R scripting
language. The code is available from the URL \url{https://github.com/asancetta/CausalTimeSeries}.
Most of the code is based on existing R packages and also includes
a cross-validation procedure to choose tuning parameters. 

\section{The Model\label{Section_Model}}

Let $X:=\left(X_{t}\right)_{t\in\mathbb{Z}}$ be a sequence of stationary
random variables taking values in $\mathbb{R}^{K}$ or some subset
of it. For each $k=1,2,...,K$, we suppose that there is a monotone
function $f_{k}$ such that $Z_{t,k}=f_{k}\left(X_{t,k}\right)$ is
a standard Gaussian random variable such that $Z_{t}=\left(Z_{t,1},Z_{t,2},...,Z_{t,K}\right)'$
\begin{equation}
Z_{t}=AZ_{t-1}+\varepsilon_{t}\label{EQ_gaussianVAR}
\end{equation}
where $A$ has singular values in $\left(0,1\right)$ and $\left(\varepsilon_{t}\right)_{t\in\mathbb{Z}}$
is a sequence of independent identically distributed random variables
with values in $\mathbb{R}^{K}$ and covariance matrix $\Sigma_{\varepsilon}$.
Throughout, the prime symbol $'$ denotes transposition. All vectors
in the paper are arranged as column vectors. We do not require knowledge
of the functions $f_{k}$. We also note that there is always a monotone
transformation that maps any univariate random variable into a standard
Gaussian. We provide details about this in Section \ref{Section_remarksGaussianTransform}
in the Appendix. Here, the main assumption is that such transformed
variables satisfy the VAR dynamics in (\ref{EQ_gaussianVAR}). Under
stationarity assumptions, all the information of the model can be
obtained from the covariance matrix of the $2K$-dimensional vector
$\left(Z_{t}',Z_{t-1}'\right)'$, which we denote by $\Sigma$. We
can then partition $\Sigma$ as 
\begin{equation}
\Sigma=\left(\begin{array}{cc}
\Sigma_{11} & \Sigma_{12}\\
\Sigma_{21} & \Sigma_{22}
\end{array}\right)=\left(\begin{array}{cc}
\Gamma & A\Gamma\\
\Gamma A' & \Gamma
\end{array}\right)\label{EQ_copulaScalingMatrixW}
\end{equation}
with obvious notation, once we note that $A$ is as in (\ref{EQ_gaussianVAR})
and $\Gamma:=\mathbb{E}Z_{t}Z_{t}'$. Clearly, $\Sigma_{\varepsilon}:=\Gamma-A\Gamma A'$
(recall $\Sigma_{\varepsilon}:=\mathbb{E}\varepsilon_{t}\varepsilon_{t}'$). 

The above setup can be recast into a formal probabilistic framework
using the copula function to model Markov processes (Darsow et al.,
1992). The copula transition density would be the ratio of two Gaussian
copulae: one with scaling matrix $\Sigma$ and one with scaling matrix
$\Gamma$. Given that we shall not use this in the rest of the paper,
we omit the details. However, given this fact, for short, we refer
to our model as a Gaussian copula VAR. We note that when $X_{t}$
has an invariant distribution with marginals that are continuous,
the functions $f_{k}$ are necessarily equal to the unconditional
distribution of $X_{t,k}$, by Sklar's Theorem (Joe, 1997). 

We consider a high dimensional framework, where $K$ can go to infinity
with the sample size. Formally, this would require us to consider
a family of models (\ref{EQ_gaussianVAR}) indexed by the sample size
$n$ to allow for increasing dimension $K$ (Han and Wu, 2019, for
more details). We do not make explicit this in the notation. Next,
we summarise the main properties of the model under the possibility
that $K\rightarrow\infty$. 

\begin{proposition}\label{Proposition_strongMixing}Define $Z_{t,k}=f_{k}\left(X_{t,k}\right)$
for some increasing monotonic transformation $f_{k}:\mathbb{R}\rightarrow\mathbb{R}$
, $k=1,2,...,K$, such that $\left(Z_{t}\right)_{t\in\mathbb{Z}}$
follows a Gaussian VAR as described in (\ref{EQ_gaussianVAR}). Furthermore,
suppose that the singular values of $A$ are in a compact interval
inside $\left(0,1\right)$ and the eigenvalues of $\Sigma_{\varepsilon}$
are in a compact interval inside $\left(0,\infty\right)$, uniformly
in $K$. Then, $\left(X_{t}\right)_{t\in\mathbb{Z}}$ is a stationary
Markov chain with strong mixing coefficients that decay exponentially
fast, uniformly in $K$ even for $K\rightarrow\infty$.\end{proposition}

Recall that the singular values of a matrix $A$ are the square root
of the eigenvalues of $A'A$. Hence, the condition means that $A$
is full rank with eigenvalues inside the unit circle. We note that
for fixed $K$ the model is not only strong mixing, but also absolutely
regular (beta mixing), with exponentially decaying coefficients (Doukhan,
1995, Theorem 5, p.97). However, when $K$ is allowed to increase,
this is not the case anymore (Han and Wu, 2019, Theorem 3.2). Nevertheless,
allowing for increasing dimension $K$, it is still strong mixing
with exponentially decaying coefficients. 

The extension of (\ref{EQ_gaussianVAR}) to a VAR($p$), for fixed
finite $p$, has been considered by Fan et al. (2022, Appendix B).
The process remains geometrically strong mixing if the singular values
of the autoregressive matrices are all in a compact interval inside
$\left(0,1\right)$. For simplicity, we shall restrict attention to
the VAR(1) case. The methodological implementation for a higher order
VAR is simple, but we will still provide some remarks on this as it
is relevant to the high dimensional framework. 

\section{Identification\label{Section_identification}}

In the next section, we briefly review causal graph terminology. While
these concepts are not widely used in econometrics, they do simplify
some discussion when stating assumptions and contemporaneous relations
(Section \ref{Section_empiricalOilPriceShock} for an empirical illustration
to oil price shocks). In Section \ref{Section_identificationGaussianCopula}
we show how these concepts relate to the more familiar language and
setup of structural vector autoregressive models. 

\subsection{Preliminary Concepts\label{Section_preliminaryConcepts}}

A graph $G=\left(\mathcal{V},\mathcal{E}\right)$ consists of a set
of vertices $\mathcal{V}=\left\{ 1,2,...,p\right\} $, where $p$
is the number of vertices, and edges $\mathcal{E}\subseteq\mathcal{V}\times\mathcal{V}$.
The edges are a set of ordered pairs of distinct vertices. The edges
are directed if the order matters, $\left(k,l\right)\in\mathcal{E}$
but $\left(l,k\right)\notin\mathcal{E}$, otherwise it is undirected.
Arrows are commonly used to define the direction when there is one.
In our context, $\mathcal{V}$ is the set of indices of $W_{t}=\left(X_{t}',X_{t-1}'\right)'$
, i.e. $p=2K$, while $\mathcal{E}$ contains the direction in the
causal relations if any. For example, we know that we cannot have
$X_{t,i}\rightarrow X_{t-1,i}$ while the other way around is possible
if $X_{t-1,i}$ Granger causes $X_{t,i}$. In the language of graphs
we say that $X_{t-1,i}$ is a parent of $X_{t,i}$. In this paper
we focus on the causal relations of $X_{t}$ conditioning on $X_{t-1}$.
This is different from Granger causality. Given that the statistical
relations of the elements in $X_{t}$ conditioning on $X_{t-1}$ are
defined by $\varepsilon_{t}$, we focus on finding the set of parents
of each $\varepsilon_{t,i}$. For example, $\varepsilon_{t,1}$ is
a parent of $\varepsilon_{t,2}$ if $\varepsilon_{t,1}$ causes $\varepsilon_{t,2}$
and not the other way around. We write $\varepsilon_{t,1}\rightarrow\varepsilon_{t,2}$.
When the variables $\varepsilon_{t,k}$ are jointly Gaussian, it is
well known that conditional independence is not enough to identify
the direction of the relation (Moneta et al., 2013, Peters et al.,
2014). 

In the case when all causal relations are identified with no cycles,
the causal graph is a directed acyclic graph (DAG): all edges are
directed and there are no cycles. There are no cycles if no descendant
can be a parent of their ancestor. When the direction cannot be fully
identified, we shall content to obtain some undirected edges. It is
possible that no directed edge can be identified. The graph where
we do not consider the directions is called the skeleton. When we
use observational data, we work with their distribution, possibly
under model assumptions as in (\ref{EQ_gaussianVAR}). We say that
the distribution of the data is faithful to the graph if the set of
all (possibly conditional) independence relations of the distribution
of the data and the graph coincide. The (possibly conditional) independence
relations of the graph are defined as the set of vertices for which
there is no edge between them. Such relations only require to identify
the skeleton. Unfortunately, a given distribution of data can generate
an infinite number of DAG's. In the case of a VAR this is equivalent
to say that the structural VAR cannot be identified. This means that
we cannot draw arrows for all edges. Hence, we may need to content
ourselves with a complete partially directed acyclic graph (CPDAG),
which is a graph where some edges are undirected because they cannot
be identified. In summary, in the more familiar language of econometrics,
identification of the DAG of the $K$-dimensional innovations $\varepsilon_{t}$
means that the system of simultaneous equations for $\varepsilon_{t}$
is recursive. This is equivalent to finding a permutation of the variables
such that the covariance matrix of the permuted innovations is the
product of a lower triangular matrix times its transpose (Lemma \ref{Lemma_SVAR_identification}
in Section \ref{Section_identificationGaussianCopula}). We shall
use a sample based version of the PC algorithm (Kalisch and B\"{u}hlmann,
2007) to identify the CPDAG under the assumption that the underlying
causal structure is recursive. For high dimensional time series data,
we require special tools as devised in the present paper. 

\subsubsection{Remarks on the PC Algorithm}

A full description of the PC algorithm can be found in (Spirtes et
al., 2000). Here, we provide a short overview assuming knowledge of
$\Sigma_{\varepsilon}$. The PC algorithm identifies as many causal
relations as possible and its output is a CPDAG. In the present case,
it exploits the assumption that the system of simultaneous equations
of the innovations is recursive (i.e. the causal graph is a DAG).
It then proceeds into two steps. The first step exploits the set of
all conditional independence relations in the data as follows. It
identifies the so called moral graph, which is the set of all edges
implied by the nonzero entries in $\Theta_{11}:=\Sigma_{\varepsilon}^{-1}$.
Note that the $\left(i,j\right)$ entry in $\Theta_{11}$ is zero
if and only if $\varepsilon_{t,i}$ and $\varepsilon_{t,j}$ are independent
when conditioning on all other variables (Proposition 5.2 in Lauritzen,
1996). Using the zero entries in $\Sigma_{\varepsilon}$, it removes
all those edges in the moral graph that correspond to variables that
are unconditionally independent, i.e. independent when conditioning
on the empty set. This produces the skeleton. It then uses a set of
logical rules to direct as many arrows as possible. 

We give a straightforward example of identification strategy used
by the PC algorithm. Suppose that we only have a set of three variables
$\left\{ \varepsilon_{t,1},\varepsilon_{t,2},\varepsilon_{t,3}\right\} $.
Suppose that any pair of variables from this set is dependent when
conditioning on the third one. According to the aforementioned remarks
on $\Theta_{11}$, we have that this matrix has no zero entries. However,
suppose that when we condition on the empty set, $\varepsilon_{t,1}$
and $\varepsilon_{t,3}$ are independent. This means that these two
variables are unconditionally independent. This is tantamount to saying
that $\left(1,3\right)$ and $\left(3,1\right)$ entries in $\Sigma_{\varepsilon}=\Theta_{11}^{-1}$
are zero. In this case, we must have that $\varepsilon_{t,1}$ and
$\varepsilon_{t,3}$ are related to each other only through a common
effect $\varepsilon_{t,2}$. The PC algorithm would then produce the
following DAG $\varepsilon_{t,1}\rightarrow\varepsilon_{t,2}\leftarrow\varepsilon_{t,3}$.
This conclusion does not assume that the underlying causal structure
be representable by a DAG. Other logical rules used by the PC algorithm
assume that the causal relations between the variables be representable
by a DAG (Algorithm 2 in Kalisch and B\"{u}hlmann, 2007, for the
full list of rules).

In the next section, we relate these concepts to SVAR identification
and existing methods based on instruments. We do so to show how our
methodology adds to the arsenal of already existing methods.

\subsection{Identification of the Gaussian Copula VAR\label{Section_identificationGaussianCopula}}

We conclude with two results that show the identification strategy
in our methodology. We define the precision matrix $\Theta=\Sigma^{-1}$.
As we did for $\Sigma$ in (\ref{EQ_copulaScalingMatrixW}), we partition
it with same dimensions as in (\ref{EQ_copulaScalingMatrixW}): 
\begin{equation}
\Theta=\left(\begin{array}{cc}
\Theta_{11} & \Theta_{12}\\
\Theta_{21} & \Theta_{22}
\end{array}\right).\label{EQ_precisionMatrixPartition}
\end{equation}

The parameters in (\ref{EQ_gaussianVAR}) are identified from the
precision matrix (\ref{EQ_precisionMatrixPartition}). The following,
is a consequence of the classical result on graphical Gaussian models
(Lauritzen, 1996, eq. C3 and C4). 

\begin{lemma}\label{Lemma_identificationVarParms}Suppose that the
conditions of Proposition \ref{Proposition_strongMixing} hold. Then,
$A=$ $-\Theta_{11}^{-1}\Theta_{12}$ and $\Sigma_{\varepsilon}=\Theta_{11}^{-1}$.\end{lemma}

When the DAG is identified, we can identify the SVAR. In the more
common language used in econometrics, this is the same as saying that
the structural equation system of the innovations is recursive, as
it will be formally defined in (\ref{EQ_structuralEquationSystemRecursive}).
To this end, we introduce some notation. Let $\Pi$ be a $K\times K$
matrix that can be transformed into the identity by simple permutation
of its rows. We call $\Pi$ a permutation matrix as it permutes the
rows of the conformable matrix that it premultiplies. We have the
following result for identification of the SVAR. 

\begin{lemma}\label{Lemma_SVAR_identification}Suppose that the conditions
of Proposition \ref{Proposition_strongMixing} hold and that the causal
graph for $\varepsilon_{t}$ in (\ref{EQ_gaussianVAR}) is a DAG.
Then, we can find a permutation matrix $\Pi$ such that 
\begin{equation}
\Pi Z_{t}=D\Pi Z_{t}+\left(I-D\right)\Pi AZ_{t-1}+\xi_{t}\label{EQ_SVAR}
\end{equation}
where $D$ is lower triangular with diagonal elements equal to zero,
and $\xi_{t}$ is a vector of independent Gaussian random variables
such that $\mathbb{E}\xi_{t}\xi_{t}'$ is a diagonal full rank matrix.
In particular, the innovation in (\ref{EQ_SVAR}) satisfies $\Pi\varepsilon_{t}=H\xi_{t}$
where $H:=\left(I-D\right)^{-1}$ is a full rank lower triangular
matrix with diagonal elements equal to one. Furthermore, the process
admits the infinite moving average representation 
\begin{equation}
Z_{t}=\sum_{s=0}^{\infty}\Upsilon_{s}\xi_{t-s},\text{ where }\ensuremath{\Upsilon_{s}}=A^{s}\Pi'H.\label{EQ_maInfinitySVAR}
\end{equation}

\end{lemma}

From the causal DAG we can derive the permutation matrix $\Pi$, where
each row describes the recursive order of the nonzero entry in such
row. The ordering is often nonunique. In what follows, we shall always
refer to the $\Pi$ matrix as the one that is obtained from the least
number of row permutations of the identity matrix. In this case $\Pi$
is unique. Hence, estimation of the DAG is equivalent to estimation
of the permutation matrix $\Pi$. From Lemma \ref{Lemma_SVAR_identification}
we deduce that 
\begin{equation}
\Pi\varepsilon_{t}=D\Pi\varepsilon_{t}+\xi_{t}\label{EQ_structuralEquationSystemRecursive}
\end{equation}
where the above is a structural equation system for the innovations
$\varepsilon_{t}$. The $\varepsilon_{t}$ variables on the right
hand side are the cause of the left hand side variables. 

From the structural model in (\ref{EQ_SVAR}) it is clear that the
shock specific to $Z_{t,l}$ is the $l^{th}$ entry in $\Pi'\xi_{t}$,
using the fact that $\Pi'=\Pi^{-1}$. By this remark and (\ref{EQ_maInfinitySVAR}),
the impact on $Z_{t+s,k}$ of intervening on $Z_{t,l}$ (via the $l^{th}$
entry in $\Pi'\xi_{t}$) is computed as $\Upsilon_{s}\Pi e_{l}$ where
$e_{l}$ is the $K\times1$ vector of zeros, but for the $l^{th}$
entry, which is one. Given that the structural shock $\xi_{t}$ has
diagonal matrix with possibly different diagonal elements, we may
use $\Upsilon_{s}\Sigma_{\xi}^{1/2}\Pi e_{l}$ in place of $\Upsilon_{s}\Pi e_{l}$,
where $\Sigma_{\xi}:=\mathbb{E}\xi_{t}\xi_{t}'$. It is clear that
the representation in (\ref{EQ_maInfinitySVAR}) in terms of the shocks
$\xi_{t-s}$ is not sufficient to carrying out causal inference in
the sense of the structural equation system (\ref{EQ_structuralEquationSystemRecursive}).
Knowledge of the permutation matrix $\Pi$ is necessary. Working with
observational data, we start from a reduced form model (\ref{EQ_gaussianVAR})
and obtain (\ref{EQ_SVAR}) when identification is possible. In turn,
identification is only possible if $\Pi$ can be identified.

When interest lies on the impulse response functions, we need to account
for nonlinearity. The model in (\ref{EQ_gaussianVAR}) is linear only
after applying a transformation to each variable. Koop et al. (1996)
address such problem focusing on generalized impulse response functions
for reduced form models (Kilian and L\"{u}tkepohl, 2017, Ch.18 for
a discussion on structural models). An explicit discussion on the
calculation within our framework can be found in Section \ref{Section_impulseResponseFunctions}
of the Appendix. However, by linearization, the impulse response function
is approximately equal to a constant multiple of $\Upsilon_{s}\Pi$
(Lemma \ref{Lemma_impulseResponse} in the Appendix, and discussion
therein). 

\subsubsection{Identification Using External Instruments\label{Section_identificationWithExternalInstruments}}

The identification strategy based on the PC algorithm (Section \ref{Section_preliminaryConcepts}),
is one additional method to be added to the arsenal of existing strategies
based on internal and external instruments, possibly using local projections
(Stock and Watson, 2018, Plagborg-M\o{}ller and Wolf, 2021). This
follows from the fact that the latent VAR in (\ref{EQ_gaussianVAR})
is Gaussian. Hence, expectations and projections are just functions
of $\Theta$ in (\ref{EQ_precisionMatrixPartition}). The latter is
one of the quantities of interest in this paper.

We note that the methodology based on external instruments can have
nontrivial implications for a recursive system, when projections and
conditional expectations coincide, as in the Gaussian case. Suppose
an augmented VAR so that an external instrument is included in the
VAR as first variable $Z_{t,1}$ to identify the effect of a shock
of $Z_{t,l}$ on $Z_{t,k}$. Being an instrument, $Z_{t,1}$ satisfies
the usual instrumental variable exclusion assumption for a SVAR (Assumption
LP-IV in Stock and Watson, 2018, Assumption 4 in Plagborg-M\o{}ller
and Wolf, 2021). Adapting Assumption 4 in Plagborg-M\o{}ller and
Wolf (2021) to our notation and using the Markov assumption implied
by (\ref{EQ_gaussianVAR}), this means that $Z_{t,1}$ conditional
on $\left\{ Z_{t-s}:s\geq1\right\} $ takes the form $\varepsilon_{t,1}=\alpha e_{l}'\Pi'\xi_{t}+e_{1}'\Pi'\xi_{t}$
for some constant $\alpha$ (Plagborg-M\o{}ller and Wolf, 2021, Eq.17).
Note that $e_{l}'\Pi'\xi_{t}$ is the structural shock of variables
$\varepsilon_{t,l}$. Then, from (\ref{EQ_structuralEquationSystemRecursive})
we know that $\xi_{t}=\left(I-D\right)\Pi\varepsilon_{t}$. Substituting
the latter in the former equation, we have that 
\begin{equation}
\varepsilon_{t,1}=\alpha e_{l}'\Pi'\left(I-D\right)\Pi\varepsilon_{t}+e_{1}'\Pi'\xi_{t}.\label{EQ_assumption4_PM}
\end{equation}
Given that for Gaussian random variables zero correlation is equivalent
to independence, the above is a structural equation. In particular
it means that $\varepsilon_{t,1}$ is caused by all the variables
in $\varepsilon_{t}$ for which the $1\times K$ vector $e_{l}'\Pi'\left(I-D\right)\Pi$
has nonzero entries. The simplest case is when $\varepsilon_{t,l}$
is not caused by any other entry in $\varepsilon_{t}$. In graph language,
this means that $\varepsilon_{t,l}$ is a source node and in structural
equation notation it means that $\varepsilon_{t,l}=e_{l}'\Pi'\xi_{t}$.
From (\ref{EQ_structuralEquationSystemRecursive}), this can only
be the case if $e_{l}'\Pi'D\Pi$ is a zero row vector.

The above shows that the standard representation (\ref{EQ_assumption4_PM})
for the instrumental variable exclusion assumption for a SVAR has
non trivial implications in empirical work. In fact, given that (\ref{EQ_assumption4_PM})
is a structural equation, $\alpha\neq0$ means that $\varepsilon_{t,1}$,
the instrument conditioning on the past, must be caused by $\varepsilon_{t,l}$
and possibly by other variables. This is contradictory to the empirical
interpretation of an instrument. In one of our empirical illustrations,
we consider the oil supply shock identification methodology discussed
in K\"{a}nzig (2021). There the instrument is based on price changes
around OPEC announcements. The variable of interest for which we want
to measure the effect of a shock is real oil prices. When projections
and conditional expectations coincide, (\ref{EQ_assumption4_PM})
essentially implies that OPEC announcements ($\varepsilon_{t,1}$)
are contemporaneously caused by real oil price ($\varepsilon_{t,l}$)
and possibly other variables. This is contrary to what is usually
put forward as a justification for the use of this instrument. Of
course, projections and conditional expectations may be unrelated,
and more importantly the system may not be recursive. Nevertheless,
we shall show that an approach based on structural equations (and
equivalently causal graphs) can help us understanding the underlying
assumptions. 

Suppose that $\varepsilon_{t}$ satisfies (\ref{EQ_structuralEquationSystemRecursive}).
The exclusion restriction using an instrument $X_{t,K+1}=f_{K+1}\left(Z_{t,K+1}\right)$
where $Z_{t,K+1}$ is standard normal can instead be formulated as
$e_{l}'\Pi'\xi_{t,l}=\nu_{t,l}+\varepsilon_{t,K+1}$ where $\varepsilon_{t,K+1}$
is $Z_{t,K+1}$ conditioning on the past of $\left(Z_{t,1},Z_{t,2},...,Z_{t,K+1}\right)$
and $\nu_{t,l}$ is a structural shock independent of $\varepsilon_{t,K+1}$.
Then, $Z_{t,K+1}$ is a valid instrument if $\varepsilon_{t,K+1}=\xi_{t,K+1}$
is a structural shock. This means that $\xi_{t,l}$ is a structural
shock when we omit $Z_{t,K+1}$. To see this, note that $\varepsilon_{t,K+1}$
satisfies the IV exclusion restriction for the impact of $\varepsilon_{t,l}$
on the other variables, and it is compatible with a recursive structural
equation system. Assuming that (\ref{EQ_gaussianVAR}) holds for the
augmented $\left(K+1\right)\times1$ vector that also includes $Z_{t,K+1}$,
we can recover the joint distribution of $\left(\varepsilon_{t,1},\varepsilon_{t,2},...,\varepsilon_{t,K+1}\right)$
and $\left(Z_{t,1},Z_{t,2},...,Z_{t,K+1}\right)$, and apply any of
the projection methods used in the literature. Our methodology allows
us to do this. Moreover, relying on a sample version of the PC algorithm,
we can also estimate whether this exclusion restriction holds for
the augmented dataset. We shall illustrate this in Section \ref{Section_empiricalOilPriceShock}
with the dataset in K\"{a}nzig (2021). In summary, our framework
not only puts forward an alternative identification approach, but
also allows us to use existing methodologies. Relying on causal graphs
and structural equations systems can allow us to precisely define
assumptions, and its visual aspect may help our intuition.

Next, we introduce algorithms that will be shown to produce consistent
estimators, under assumptions stated in Section \ref{Section_assumptions}.

\section{Estimation Algorithms\label{Section_algos}}

For any positive integer $p$, $\left[p\right]:=\left\{ 1,2,...,p\right\} $.
For any matrix $Q$ of dimensions $p\times q$ and sets $\mathcal{A}\subseteq\left[p\right]$
and $\mathcal{B}\subseteq\left[q\right]$, $A_{\mathcal{A},\mathcal{B}}$
is the submatrix with rows in $\mathcal{A}$ and columns in $\mathcal{B}$.
In $A_{\mathcal{A},\mathcal{B}}$, when $\mathcal{A}=\left[p\right]$
we write $A_{\cdot,\mathcal{B}}$ and similarly if $\mathcal{B}=\left[q\right]$.
When $\mathcal{A}=\left[p\right]\setminus\left\{ i\right\} $ for
some $i\in\left[p\right]$, we write $A_{-i,\mathcal{B}}$ and similarly
for $\mathcal{B}$. When $A$ is a vector, it is always assumed that
it is a column vector and we shall use the same notation, but with
one single subscript. This notation will be used throughout the paper
with no further mention. 

The estimation methodology is based on a number of steps which extend
the methodology in Liu et al. (2012). First, we find an estimator
of the matrix $\Sigma$ in (\ref{EQ_copulaScalingMatrixW}), which
is the Gaussian copula scaling matrix of the vector $W_{t}=\left(X_{t}',X_{t-1}'\right)'$.
This is achieved using Algorithm \ref{Algo_scalingMatrix}. Once,
the estimator for $\Sigma$ is available, we identify the set of zero
entries in the precision matrix, i.e., the inverse of $\Sigma$. This
can be achieved using Lasso, as described in Algorithm \ref{Algo_LassoThresh}.
This algorithm follows the approach of Meinshausen and B\"{u}hlmann
(2006) to find the zeros in the inverse of (\ref{EQ_copulaScalingMatrixW}).
However, the algorithm also thresholds the resulting Lasso estimators
in order to achieve sign consistency. In this form, the algorithm
is equivalent to Gelato (Zhou et al., 2011). 

In Algorithm \ref{Algo_LassoThresh}, (\ref{EQ_LassoFOC}) is solved
by the $x$ that satisfies the first order conditions in a Lasso minimization
problem. The constraint $x_{i}=0$ is needed to avoid running the
regression of the $i^{th}$ variable on all the other covariates and
itself. We need the estimator to be in this form for later use. A
competing algorithm to find the zeros of the precision matrix is the
CLIME estimation algorithm with thresholding (Cai et al., 2011). The
procedure is described in Algorithm \ref{Algo_climeThresh}. The minimization
problem in Algorithm \ref{Algo_climeThresh} can be solved for one
column of $\Omega$ at the time, with $\Omega$ as defined there,
due to the use of the uniform norm. We shall show the validity of
both algorithms within the time series context of this paper. 

Algorithm \ref{Algo_sparseCopulaParametersEstimation} allows us to
estimate the parameters in (\ref{EQ_gaussianVAR}). In particular,
it uses the information on the zeros of the estimator for the precision
matrix $\Theta$ to construct a sparse estimator (Le and Zhong, 2021).
Using Lemma \ref{Lemma_identificationVarParms}, such sparse estimator
of the precision matrix is used to estimate the autoregressive matrix
$A$ and the covariance matrix of the innovations $\varepsilon_{t}$
in (\ref{EQ_gaussianVAR}). 

Finally, using Algorithm \ref{Algo_PCAlgo}, we identify the PCDAG.
Algorithm \ref{Algo_PCAlgo} makes reference to the PC algorithm.
We do not report the details in Algorithm \ref{Algo_PCAlgo}, as the
number of steps is relatively large and can be found in Spirtes et
al. (2000) among many other places. The aim of the PC algorithm is
to start with a dense graph with undirected edges for all variables.
It then aims at removing edges to obtain the skeleton of the graph.
Finally, it uses a set of rules to direct all possible edges based
on deterministic rules. It is not guaranteed that all edges can be
directed, of course. 

In order to delete edges, the PC algorithm uses the correlation coefficients
between two variables, conditional on subsets of other variables.
Note that the innovations in the latent model (\ref{EQ_gaussianVAR})
are Gaussian so that zero correlation implies independence. As soon
as we find a set of conditioning variables such that the two variables
are conditionally uncorrelated, we remove an edge between these two
variables. Given that the conditional correlations are unknown, Kalisch
and B\"{u}hlmann (2007) suggest to replace these with sample versions
as in Algorithm \ref{Algo_PCAlgo}. They define a parameter $\alpha$,
as in Algorithm \ref{Algo_PCAlgo}, and show that for $\alpha\rightarrow0$
at a certain speed we can obtain a consistent estimator of the PCDAG,
as if we knew the true conditional correlations. For this reason,
Algorithm \ref{Algo_PCAlgo} only gives details on the sample estimator
leaving out the deterministic steps, to avoid distracting details.

Identification of the SVAR requires that all edges are directed. Assuming
that Algorithm \ref{Algo_PCAlgo} can direct all the edges, for each
$i\in\left[K\right]$, we obtain estimators $\mathcal{\hat{V}}\left(i\right)$
for the set of parents of $\varepsilon_{t,i}$, using the notation
in Algorithm \ref{Algo_ImpulseResponse}. According to Lemma \ref{Lemma_SVAR_identification},
to find the matrix $D$, we need to find the regression coefficients
of the innovation $\varepsilon_{t,i}$ on $\varepsilon_{t,\mathcal{\hat{V}}\left(i\right)}$,
$i\in\left[K\right]$. Algorithm \ref{Algo_ImpulseResponse} finds
such regression coefficients and collects them into a $K\times K$
matrix $\hat{\Delta}$, $i=1,2,...,K$. In particular, the $i^{th}$
row of $\hat{\Delta}$ has entries $\mathcal{\hat{V}}\left(i\right)$
equal to the coefficients found regressing $\varepsilon_{t,i}$ on
$\varepsilon_{t,\mathcal{\hat{V}}\left(i\right)}$ and zeros elsewhere.
By the fact that the graph is a DAG, there is a permutation matrix
$\hat{\Pi}$ such that $\hat{\Pi}\hat{\Delta}\hat{\Pi}^{-1}$ is an
estimator for $D$ and is a lower triangular matrix with zeros along
the diagonal. The regression coefficients are obtained relying on
$\hat{\Sigma}_{\varepsilon}:=\hat{\Theta}_{11}^{-1}$. This is because
$\hat{\Theta}_{11}$ is a sparse estimator with good asymptotic properties.
Such properties are inherited by $\hat{\Sigma}_{\varepsilon}$ even
though $\Sigma_{\varepsilon}$ is not sparse. The estimator $\hat{\Sigma}_{\varepsilon}$
is not necessarily sparse. Moreover, regression coefficients are found
directly from $\hat{\Sigma}_{\varepsilon}$ with no need to estimate
the innovations. 

\begin{algorithm}
\caption{Copula Scaling Matrix Estimation.}
\label{Algo_scalingMatrix}

Define $W_{t}:=\left(X_{t}',X_{t-1}'\right)'$, $t\in\left[n\right]$.

For $1\leq i<j\leq2K$:

Let $\hat{\rho}_{i,j}$ be the sample Spearman's rho coefficient between
$\left(W_{s,i}\right)_{s\in\left[n\right]}$ and $\left(W_{s,j}\right)_{s\in\left[n\right]}$
(i.e. the sample correlation of their ranks).

Define the $2K\times2K$ matrix estimator $\hat{\Sigma}$ for (\ref{EQ_copulaScalingMatrixW})
with $i,j$ entry $\hat{\Sigma}_{i,j}=2\sin\left(\frac{\pi}{6}\hat{\rho}_{i,j}\right)$
and set $\hat{\Sigma}_{j,i}=\hat{\Sigma}_{i,j}$.

Ensure that the entries in $\hat{\Sigma}$ corresponding to $\Sigma_{11}$
and $\Sigma_{22}$ in (\ref{EQ_copulaScalingMatrixW}) are the same
by taking averages of the two estimators if needed.
\end{algorithm}
\begin{algorithm}
\caption{High Dimensional Causal Estimation with Lasso. Use Lasso (Meinshausen
and B\"{u}hlmann, 2006) to find the moral graph of $W_{t}$.}
\label{Algo_LassoThresh}

Set $\tau>\lambda>0$.

Run Algorithm \ref{Algo_scalingMatrix} to obtain $\hat{\Sigma}$.

For $i\in\left[K\right]$:

Denote by $\hat{\beta}^{\left(i\right)}\in\mathbb{R}^{2K}$ the solution
to 
\begin{equation}
\hat{\Sigma}_{\cdot,i}-\hat{\Sigma}x=\lambda{\rm sign}\left(x\right),\text{ s.t. }x_{i}=0,\,x\in\mathbb{R}^{K}\label{EQ_LassoFOC}
\end{equation}

Redefine $\hat{\beta}_{j}^{\left(i\right)}$ as $\hat{\beta}_{j}^{\left(i\right)}1_{\left\{ \left|\hat{\beta}_{j}^{\left(i\right)}\right|\geq\tau\right\} }$.

Let $j$ be a neighbour of $i$ if $\hat{\beta}_{j}^{\left(i\right)}\neq0$.

For each $i\in\left[K\right]$:

Set $\hat{\Omega}^{\left(i\right)}$ equal to $\hat{\beta}^{\left(i\right)}$,
but let $\hat{\Omega}_{i}^{\left(i\right)}=1$, where $\hat{\Omega}_{i}^{\left(i\right)}$
is the $i^{th}$ entry.
\end{algorithm}
\begin{algorithm}
\caption{High Dimensional Causal Estimation with CLIME. Use CLIME (Cai et al.,
2011) to find the moral graph of $W_{t}$.}
\label{Algo_climeThresh}

Set $\tau>\lambda>0$.

Run Algorithm \ref{Algo_scalingMatrix} to obtain $\hat{\Sigma}$.

Let $\hat{\Omega}\in\mathbb{R}^{2K\times2K}$ be the solution to $\min\left|\Omega\right|_{1,1}$
s.t. $\left|\hat{\Sigma}\Omega-I\right|_{\infty}\leq\lambda$.

Redefine $\hat{\Omega}_{i,j}$ as $\hat{\Omega}_{i,j}1_{\left\{ \left|\hat{\Omega}_{i,j}\right|\geq\tau\right\} }$
and denote by $\hat{\Omega}^{\left(i\right)}$ the $i^{th}$ column
of the redefined $\hat{\Omega}$.
\end{algorithm}

\begin{algorithm}
\caption{Estimation of the Parameters in (\ref{EQ_gaussianVAR}).}
\label{Algo_sparseCopulaParametersEstimation}

Run either Algorithm \ref{Algo_LassoThresh} or \ref{Algo_climeThresh}
to find $\hat{\Sigma}$ and $\hat{\Omega}^{\left(i\right)}$, $i=1,2,...,2K$.

Let $\tilde{\Omega}^{\left(i\right)}$ be the subvector obtained by
deleting the zero elements in $\hat{\Omega}^{\left(i\right)}$ and
denote by $\hat{s}_{i}$ its size.

Denote by $\hat{B}_{i}$ the $2K\times\hat{s}_{i}$ matrix such that
$\hat{\Omega}^{\left(i\right)}=\hat{B}_{i}\tilde{\Omega}^{\left(i\right)}$

Define $\hat{\Theta}^{\left(i\right)}=\hat{B}_{i}\left(\hat{B}_{i}'\hat{\Sigma}\hat{B}_{i}\right)^{-1}\hat{B}_{i}'e_{i}$
where $e_{i}$ is the $2K\times1$ vector with $i^{th}$ entry equal
to one and zero otherwise.

Let $\hat{\Theta}=\frac{1}{2}\left[\left(\hat{\Theta}^{\left(1\right)},\hat{\Theta}^{\left(2\right)},...,\hat{\Theta}^{\left(2K\right)}\right)+\left(\hat{\Theta}^{\left(1\right)},\hat{\Theta}^{\left(2\right)},...,\hat{\Theta}^{\left(2K\right)}\right)'\right]$.

Denote by $\hat{\Theta}_{11}$ the entries $\left(k,l\right)$ in
$\hat{\Theta}$, $k,l=1,2,...,K$.

Denote by $\hat{\Theta}_{12}$ the entries $\left(k,l\right)$ of
$\hat{\Theta}$ with $k=1,2,...,K$, and $l=K+1,K+2,...,2K$.

Define $\hat{A}=-\hat{\Theta}_{11}^{-1}\hat{\Theta}_{12}$ as an estimator
for $A$ in (\ref{EQ_copulaScalingMatrixW}).

Define $\hat{\Sigma}_{\varepsilon}=\hat{\Theta}_{11}^{-1}$ as an
estimator for $\Sigma_{\varepsilon}:=\mathbb{E}\varepsilon_{t}\varepsilon_{t}'$.
\end{algorithm}

\begin{algorithm}
\caption{Estimation of the PCDAG.}
\label{Algo_PCAlgo}

Run Algorithm \ref{Algo_sparseCopulaParametersEstimation} to find
$\hat{\Sigma}_{\varepsilon}$.

Use $\hat{\Sigma}_{\varepsilon}$ to find the estimator of the correlation
coefficient of $\varepsilon_{t,i}$ and $\varepsilon_{t,j}$ conditioning
on $\left\{ \varepsilon_{t,l}:l\in\mathbf{k}\right\} $ where $\mathbf{k}\subset\left[K\right]$
is a set that excludes $i,j$. Denote such correlation coefficient
by $\hat{\Xi}_{i,j|\mathbf{k}}$. 

Use the PC algorithm (Spirtes et al., 2000) and delete a node between
$\left(i,j\right)$ if $\sqrt{n-\left|\mathbf{k}\right|-3}\times g\left(\hat{\Xi}_{i,j|k}\right)\leq\Phi^{-1}\left(1-\frac{\alpha}{2}\right)$
where \textbf{$g\left(x\right)=2^{-1}\ln\left(\frac{1+x}{1-x}\right)$
}($x\in\left(-1,1\right)$) and $\alpha\in\left(0,1\right)$.
\end{algorithm}

\begin{algorithm}
\caption{Estimation of the impulse response.}
\label{Algo_ImpulseResponse}

Run Algorithm \ref{Algo_PCAlgo} and suppose that the PC algorithms
identifies the DAG in the sense that it produces and estimator $\mathcal{\hat{E}}\subseteq\mathcal{V}\times\mathcal{V}$
for the true edges $\mathcal{E}$, such that all elements in $\mathcal{\hat{E}}$
are directed.

For $i\in\left[K\right]$: 

Find all $j\in\mathcal{V}$ such that $\left(j,i\right)\in\hat{\mathcal{E}}$
so that conditioning on the $Z_{t-1}$, the $j$ covariate is a parent
of the $i$ one (i.e. $\varepsilon_{t,j}\rightarrow\varepsilon_{t,i}$).
Denote such set by $\mathcal{\hat{V}}\left(i\right)$. 

Find $\hat{d}_{i}=\hat{\Sigma}_{\varepsilon,\mathcal{\hat{V}}\left(i\right),\mathcal{\hat{V}}\left(i\right)}^{-1}\hat{\Sigma}_{\varepsilon,\mathcal{\hat{V}}\left(i\right),i}$.

Let $\hat{\Delta}$ be the matrix such that $\hat{\Delta}_{i,\mathcal{\hat{V}}\left(i\right)}=\hat{d}_{i}'$
and zero otherwise.

Find the matrix $\hat{\Pi}$ obtained from the least number of row
permutations of the identity matrix and such that $\hat{D}:=\hat{\Pi}\hat{\Delta}\hat{\Pi}^{-1}$
is lower diagonal with diagonal elements equal to zero. 
\end{algorithm}

The tuning parameters for Algorithms \ref{Algo_LassoThresh} and \ref{Algo_climeThresh}
are chosen using cross-validation (Section \ref{Section_ChoiceTuning}
in the Electronic Supplement, for details). 

In the Electronic Supplement, we also use simulations to investigate
the finite sample properties of the estimators in our algorithms (see
Section \ref{Section_summarySimulations} in the Electronic Supplement).The
simulation analysis show that our approach produces more reliable
results than methods that do not account for either sparsity or time
series dependence, i.e. setting $\lambda=0$ in Algorithms \ref{Algo_LassoThresh}
and \ref{Algo_climeThresh} or assuming $A=0$ in (\ref{EQ_gaussianVAR}).
Even when the persistence of the time series is reduced, our methodology
produces the best results for estimation of the causal structure and
the VAR parameters (for details, see Tables \ref{tab:SHD_HighDim}-\ref{tab:distance_LowDim}
in Section \ref{Section_summarySimulations} in the Electronic Supplement).
Although our approach is designed for a high dimensional setting,
it provides competitive results even in the low dimensional case.

\section{Asymptotic Analysis of the Algorithms\label{Section_asymptotics}}

The consistency of the algorithms relies on a set of conditions. Before
introducing our conditions, we introduce some additional notation.

\subsection{Additional Notation}

For any vector, the $\ell_{p}$ norm is denoted by $\left|\cdot\right|_{p}$,
$p\in\left[0,\infty\right]$. For any $I\times J$ dimensional matrix
$A$, $\left|A\right|_{p,q}=\left(\sum_{j=1}^{J}\left(\sum_{i=1}^{I}\left|A_{i,j}\right|^{p}\right)^{q/p}\right)^{1/q}$
is the elementwise norm. When $q=\infty$ we define $\left|A\right|_{p,\infty}=\max_{j\le J}\left(\sum_{i=1}^{I}\left|A_{i,j}\right|^{p}\right)^{1/p}$.
When both $p=q=\infty$ we simply write $\left|A\right|_{\infty}=\max_{i\leq I,j\le J}\left|A_{i,j}\right|$,
and this should not cause confusion with the $\ell_{\infty}$ norm.
For $p=0$, $\left|A\right|_{0,\infty}=\max_{j\le J}\sum_{i=1}^{I}1_{\left\{ \left|A_{i,j}\right|>0\right\} }$.
When $p=q=0$, this is just the total number of non-zero elements
in $A$. Finally, $\left|\cdot\right|_{{\rm op}}$ is used to define
the following operator norm: $\left|A\right|_{{\rm op}}=\max_{x:x'x\leq1}\left|Ax\right|_{2}$.
Then, $\left|A\right|_{{\rm op}}$ is the largest singular value of
$A$. For ease of reference, we call this norm the operator's norm.

Let 
\begin{equation}
\mathcal{U}\left(\omega,s\right)=\left\{ \Omega\in\mathbb{R}^{2K\times2K}:\Omega\succ0,\left|\Omega\right|_{1,\infty}\leq\omega,\left|\Omega\right|_{0,\infty}\leq s\right\} \label{EQ_setRestrictedMatrices}
\end{equation}
The symbol $\Omega\succ0$ is used to mean that $\Omega$ is a symmetric
strictly positive definite matrix. Then, $\mathcal{U}\left(\omega,s\right)$
is the set of symmetric strictly positive definite matrices whose
absolute sum of column entries is at most $\omega$, and with maximum
number of non-zero entries in each row equals $s$.

We shorten left and right and side with l.h.s. and r.h.s., respectively.
Finally, $\lesssim$ is used when the l.h.s. is bounded above by a
constant times the r.h.s.; $\gtrsim$ is bounded below by a constant
times the r.h.s.; $\asymp$ is used when the l.h.s. is bounded below
and above by constants times the r.h.s.. Finally, to avoid notational
trivialities, we assume that $K\geq2$. 

\subsection{Assumptions\label{Section_assumptions}}

\begin{condition}\label{Condition_model}(Model) There are monotone
functions $f_{k}$ such that $Z_{t,k}=f_{k}\left(X_{t,k}\right)$
is a standard Gaussian random variable such that (\ref{EQ_gaussianVAR})
holds. Moreover, $X_{t}$ has continuous marginal distributions.\end{condition}

\begin{condition}\label{Condition_dimension}(Dimension) The state
space is a subset of $\mathbb{R}^{K}$, where $K=O\left(n^{\eta_{K}}\right)$
for some $\eta_{K}<\infty$.\end{condition}

\begin{condition}\label{Condition_precisionMatrixSparsity}(Precision
matrix sparsity) The precision matrix $\Theta=\Sigma^{-1}$ is an
element of $\mathcal{U}\left(\omega,s\right)$ for $s=O\left(n^{\eta_{s}}\right)$
for some $\eta_{s}<1/2$. \end{condition}

\begin{condition}\label{Condition_thetaMin} (Identifiability) $\theta_{\min}\gtrsim n^{-\eta_{\theta}}$,
$\eta_{\theta}<1/2$, where $\theta_{\min}$ is the smallest absolute
value of the nonzero elements in $\Theta$. \end{condition}

\begin{condition}\label{Condition_eigenvals} (Eigenvalues) The singular
values of $A$ are in a compact interval inside $\left(0,1\right)$
and the eigenvalues of $\Sigma_{\varepsilon}$ are in a compact interval
inside $\left(0,\infty\right)$, uniformly in $K$. \end{condition}

Strictly speaking, if $K\rightarrow\infty$ as $n\rightarrow\infty$,
we should index both the process $X$ and its law by $n$ and think
in terms of a sequence of processes. We refrain to do so for notational
simplicity. No part in the proofs makes implicitly use of assumptions
that contradicts this.

\subsection{Remarks on the Assumptions\label{Section_remarksAssumptions}}

\paragraph{Assumption \ref{Condition_model}.}

The modelling assumption includes a Gaussian linear vector autoregressive
model as special case. However, it is clearly more general than that.
Once, we assume that the data satisfy a VAR model after a monotone
transformation, we do not need to impose any moment condition on the
original data. Hence the procedure is robust to fat tails. As discussed
in Section \ref{Section_Model}, we can view this assumption as a
Gaussian copula assumption for the cross-sectional and time series
dependence. Assumption \ref{Condition_model} can be viewed as a generalization
of the framework of Liu et al. (2012) in the time series direction
and has been recently exploited by Fan et al. (2022) to test for Granger
causality in high dimensional models. 

The continuity of the marginal distribution of $X_{t}$ is not needed.
As shown in Fan et al. (2017) we can recover the parameters of the
latent Gaussian process even for mixed data types (see also Section
\ref{Section_remarksGaussianTransform} in the Appendix). In this
case, we would modify Algorithm \ref{Algo_scalingMatrix} accordingly.

Our results apply to a VAR($p$) for fixed and finite $p$, if we
redefine $W_{t}:=\left(X_{t}',X_{t-1}'\right)'$ to be $W_{t}:=\left(X_{t}',X_{t-1}',...,X_{t-p}'\right)'$;
here $p$ is defined locally and not related to the same symbol in
other parts of the paper. Then, we just need to change the dimension
of the set of matrices in (\ref{EQ_setRestrictedMatrices}) from $2K\times2K$
to $K_{p}\times K_{p}$ where $K_{p}=\left(p+1\right)K$. The conditions,
will then apply to these new quantities. Clearly, the dimension of
the matrix $\Theta$ is $K_{p}\times K_{p}$ while the dimension of
the submatrix $\Theta_{11}$ in (\ref{EQ_precisionMatrixPartition})
is still $K\times K$. 

\paragraph{Assumption \ref{Condition_precisionMatrixSparsity}.}

The precision matrix is supposed to have maximum absolute sum of each
column bounded by a constant $\omega$. Our bounds make explicit the
dependence on $\omega$ so that we can have $\omega\rightarrow\infty$
if needed. This constant is only used in Algorithms \ref{Algo_LassoThresh}
and \ref{Algo_climeThresh}. The total number of non zero elements
in each row is supposed to be bounded by a constant $s$. This is
allowed to grow to infinity with the sample size at a certain rate.
This assumption is different from Fan et al. (2022) who assumes that
the autoregressive matrix $A$ in (\ref{EQ_gaussianVAR}) is sparse.
This is not the case here. By Lemma \ref{Lemma_identificationVarParms},
sparsity of $\Theta$ does not imply sparsity of either $A$ or $\Sigma_{\varepsilon}$.
In order to see this, we recall that $\Theta_{i,j}=0$ if and only
if $W_{t,i}$ and $W_{t,j}$ are independent, conditioning on all
the other remaining variables Lauritzen (1996, Proposition 5.2), where
$W_{t}:=\left(X_{t}',X_{t-1}'\right)'$. 

\begin{example}\label{Example_sparseThetaNoSparseA}For random variables
$Y_{1},Y_{2},Y_{3}$, let $Y_{1}\perp Y_{2}|Y_{3}$ mean that $Y_{1}$
and $Y_{2}$ are independent given $Y_{3}$. Now, suppose that for
all $k\in\left[K\right]$ and $l\neq k$, 
\[
X_{t,k}\perp X_{t,l}|\left\{ X_{t,k+1},X_{t,k-1}\right\} \cap\left\{ X_{t,i}:i\in\left[K\right]\right\} 
\]
 
\[
X_{t,k}\perp X_{t,l}|\left\{ X_{t-1,k+1},X_{t-1,k},X_{t-1,k-1}\right\} \cap\left\{ X_{t-1,i}:i\in\left[K\right]\right\} 
\]
and 
\[
X_{t-1,k}\perp X_{t-1,l}|\left\{ X_{t,k+1},X_{t,k},X_{t,k-1}\right\} \cap\left\{ X_{t,i}:i\in\left[K\right]\right\} 
\]
and such that $\Theta_{11}=\Theta_{22}$. Intuitively, this means
that variables that are not close to each other in terms of index
are conditionally independent. The intersection with $\left\{ X_{t,i}:i\in\left[K\right]\right\} $
is to avoid conditioning on $X_{t,K+1}$ for example, as we only have
$K$ variables. Given our modelling assumption (\ref{EQ_gaussianVAR}),
and the previous remarks about $\Theta$, this means that $\Theta_{11}$
and $\Theta_{12}$ are tridiagonal. Moreover, $\Theta_{11}=\Theta_{22}$
means that the partial correlation between $Z_{t,k}$ and $Z_{t,k+i}$
given all other covariates (including $Z_{t-1}$) is the same as the
partial correlation between $Z_{t-1,k}$ and $Z_{t-1,k+i}$ given
all other covariates (including $Z_{t}$) is the same. From Lemma
\ref{Lemma_identificationVarParms} and the fact that the inverse
of a tridiagonal matrix is not sparse, we deduce that both $A$ and
$\Sigma_{\varepsilon}$ are not sparse. \end{example}

Clearly, we can obtain non-sparse $A$ from sparse $\Theta$ under
more general setups than Example \ref{Example_sparseThetaNoSparseA}.
This is just chosen as a simple illustration for the sake of conciseness. 

\paragraph{Assumption \ref{Condition_thetaMin}.}

This assumption is only used to ensure that we can identify the zero
entries in $\Theta$. It is necessary in order to ensure the validity
of post selection asymptotic, though the rate can be arbitrarily slow
when $\theta_{\min}\rightarrow0$ (Leeb and P\"{o}tscher, 2005, p.29ff). 

\paragraph{Assumption \ref{Condition_eigenvals}.}

The eigenvalues condition means that the variables are linearly independent
in the population. This could be weakened, but at the cost of technical
complexity. This assumption also implies the following.

\begin{lemma}\label{Lemma_implicationsConditionEigenvals}Under Assumption
\ref{Condition_eigenvals} the following statements hold uniformly
in $K$:
\begin{enumerate}
\item The eigenvalues of $\Gamma=Var\left(Z_{t}\right)$ are bounded away
from zero and infinity; 
\item There are constants $\sigma_{\min},\sigma_{\max}\in\left(0,\infty\right)$
such that the eigenvalues of $\Sigma$ in (\ref{EQ_copulaScalingMatrixW})
are in the interval $\left[\sigma_{\min},\sigma_{\max}\right]$;
\item There is a $\nu>0$ such that $\left|\Theta_{i,i}\right|\geq\nu^{2}$;
\item The partial correlations of $\varepsilon_{t,i}$ and $\varepsilon_{t,j}$
conditioning on any other subset of remaining innovations is bounded
above by a constant $\bar{\sigma}<1$.
\end{enumerate}
\end{lemma}

\subsection{Uniform Convergence of the Scaling Matrix Estimator}

The uniform consistency of the covariance estimator from Algorithm
\ref{Algo_scalingMatrix} is well known (Liu et al., 2012). It is
still consistent for dependent data.

\begin{theorem}\label{Theorem_covMatConvergence}Under the Assumptions,
$\left|\hat{\Sigma}-\Sigma\right|_{\infty}=O_{P}\left(\sqrt{\frac{\ln K}{n}}\right)$.\end{theorem}

Fan et al. (2022) show a similar result using Kendall's tau instead
of Spearman's rho with a different method of proof. 

\subsection{Estimation of the Undirected Graph}

\subsubsection{Consistency for Algorithm \ref{Algo_LassoThresh} \label{Section_lassoResults}}

The reader is referred to the Assumptions and Algorithm \ref{Algo_LassoThresh}
for the notation. Let $\beta^{\left(i\right)}$ be the population
regression coefficient including a zero in the $i^{th}$ entry, i.e.
the solution to $\Sigma_{\cdot,i}x-\Sigma=0$ s.t. $x_{i}=0$.

\begin{theorem}\label{Theorem_lassoL1Consistency}Suppose that the
Assumptions hold. There is a finite constant $c$ large enough such
that in Algorithm \ref{Algo_LassoThresh}, choosing $\lambda=\lambda_{n}=c\omega\sqrt{\frac{\ln K}{n}}$,
with $\omega$ is as in Assumption \ref{Condition_precisionMatrixSparsity}
we have that $\max_{i\in\left[K\right]}\left|\hat{\beta}^{\left(i\right)}-\beta^{\left(i\right)}\right|_{1}=O_{P}\left(\omega s\sqrt{\frac{\ln K}{n}}\right)$.\end{theorem}

One could choose $c\rightarrow\infty$ slowly enough, in which case
the bound would be $O_{P}\left(c\times\omega s\sqrt{\frac{\ln K}{n}}\right)$
instead of $O_{P}\left(\omega s\sqrt{\frac{\ln K}{n}}\right)$. The
proof of this result shows that we could have stated the results as
finite sample one with high probability. However, such statement would
still depend on an unknown constant. Hence, for simplicity, we have
chosen not to do so. 

Using appropriate thresholding, with threshold constant greater than
the noise level, but smaller than $\theta_{\min}$, the absolute value
of the smallest nonzero entry in $\Theta$, leads to set identification.
In what follows ${\rm sign}\left(x\right)$ is the sign of the real
variable $x$ with ${\rm sign}\left(0\right)=0$. 

\begin{theorem}\label{Theorem_lassoSignConsistency}Suppose that
the Assumptions hold. In Algorithm \ref{Algo_LassoThresh}, set $\tau=\tau_{n}=o\left(\theta_{\min}\right)$
such that $\lambda=\lambda_{n}=o\left(\tau_{n}\right)$ with $\lambda$
as in Theorem \ref{Theorem_lassoL1Consistency}. If $\omega s\sqrt{n^{-1}\ln K}\rightarrow0$,
then, 
\[
\Pr\left({\rm sign}\left(\hat{\beta}_{j}^{\left(i\right)}\right)\neq{\rm sign}\left(\beta_{j}^{\left(i\right)}\right)\text{ for at least one }i\in\left[K\right],j\in\left[2K\right]\right)\rightarrow0.
\]
\end{theorem}

\subsubsection{Consistency Results for Algorithm \ref{Algo_climeThresh}}

The reader is referred to the Assumptions and Algorithm \ref{Algo_climeThresh}
for the notation.

\begin{theorem}\label{Theorem_ClimeOmegaConvergence}Suppose that
the Assumptions hold. There is a finite constant $c$ large enough
such that in Algorithm \ref{Algo_climeThresh}, $\lambda=\lambda_{n}=c\omega\sqrt{\frac{\ln K}{n}}$,
where $\omega$ is as in Assumption \ref{Condition_precisionMatrixSparsity},
implies that $\left|\hat{\Omega}-\Theta\right|_{\infty}=O_{P}\left(\omega^{2}\sqrt{\frac{\ln K}{n}}\right)$.\end{theorem}

The same remark we made about $c$ in Theorem \ref{Theorem_lassoL1Consistency}
applies here. Also here, we could have stated the result as a finite
sample one with high probability. 

Using the appropriate level of thresholding, Theorem \ref{Theorem_ClimeOmegaConvergence}
implies the following.

\begin{theorem}\label{Theorem_ClimeSignConvergence}Suppose that
the Assumptions hold. In Algorithm \ref{Algo_climeThresh}, set $\tau=\tau_{n}=o\left(\theta_{\min}\right)$
and $\lambda=\lambda_{n}=o\left(\tau_{n}/\omega\right)$ with $\lambda$
as in Theorem \ref{Theorem_ClimeOmegaConvergence}. If $\omega^{2}\sqrt{n^{-1}\ln K}\rightarrow0$,
then, 
\[
\Pr\left({\rm sign}\left(\hat{\Omega}_{i,j}\right)\neq{\rm sign}\left(\Theta_{i,j}\right)\text{ for some }i,j\in\left[2K\right]\right)\rightarrow0.
\]
\end{theorem}

\subsection{Estimation of the Process Parameters and Causal Graph}

In what follows, we suppose that the conditions of either Theorem
\ref{Theorem_lassoSignConsistency} or Theorem \ref{Theorem_ClimeSignConvergence}
hold, depending on which algorithm is used. For short we generically
refer to these as the Assumptions $\left(\lambda,\tau\right)$ as
they also involve restrictions on the choice of penalty $\lambda$
and threshold $\tau$. 

\subsubsection{Consistency of Precision Matrix Estimation}

The estimator for the precision matrix is elementwise uniformly consistent
under sparseness conditions.

\begin{theorem}\label{Theorem_precisionMatrixConsistency}Suppose
that the Assumptions $\left(\lambda,\tau\right)$ hold. Then, the
estimator $\hat{\Theta}$ from Algorithm \ref{Algo_sparseCopulaParametersEstimation}
satisfies $\left|\hat{\Theta}-\Theta\right|_{\infty}=O_{P}\left(\sqrt{\frac{\ln K}{n}}\right)$.\end{theorem}

While the quantity $s=\left|\Theta\right|_{0,\infty}$ does not enter
the bound, a constraint on its growth rate, as prescribed by Assumption
\ref{Condition_precisionMatrixSparsity}, is required for Theorem
\ref{Theorem_precisionMatrixConsistency} to hold. 

\subsubsection{Consistency of the Estimators for the Autoregressive Matrix and Innovation
Covariance Matrix}

Recall that by Lemma \ref{Lemma_identificationVarParms}, using the
notation in (\ref{EQ_gaussianVAR}) and (\ref{EQ_precisionMatrixPartition}),
$A=$ $-\Theta_{11}^{-1}\Theta_{12}$ and $\Sigma_{\varepsilon}=\Theta_{11}^{-1}$.
Hence, we need consistency of $\Theta_{12}$ and the inverse of $\Theta_{11}$,
which is the case under sparseness. Recall that $s=\left|\Theta\right|_{0,\infty}$
as in Assumption \ref{Condition_precisionMatrixSparsity}. We have
the following bounds in terms of the operator's norm.

\begin{theorem}\label{Theorem_InnovationsCovAutoregressiveMatrix}Suppose
that the Assumptions $\left(\lambda,\tau\right)$ hold. Then, $\left|\hat{\Sigma}_{\varepsilon}-\Sigma_{\varepsilon}\right|_{{\rm op}}=O_{P}\left(s\sqrt{\frac{\ln K}{n}}\right)$
and $\left|\hat{A}-A\right|_{{\rm op}}=O_{P}\left(s\sqrt{\frac{\ln K}{n}}\right)$.\end{theorem}

\subsubsection{PC Algorithm}

Let $\hat{G}$ be the estimated PCDAG from Algorithm \ref{Algo_sparseCopulaParametersEstimation}
and $G$ the true PCDAG. The next result requires faithfulness of
the distribution of the data to the graph, as defined in Section \ref{Section_preliminaryConcepts}.
In what follows, $\Phi\left(\cdot\right)$ is the cumulative distribution
function of a standard normal random variable. 

\begin{theorem}\label{Theorem_PcAlgoConsistency}Suppose that the
Assumptions $\left(\lambda,\tau\right)$ hold and that the joint distribution
of the innovations $\varepsilon_{t}$ in (\ref{EQ_gaussianVAR}) is
faithful to the DAG for all $K$. Run the PC algorithm as referenced
in Algorithm \ref{Algo_PCAlgo} with $\alpha=\alpha_{n}$ such that
$\alpha_{n}=2\left(1-\Phi\left(n^{1/2}c_{n}/2\right)\right)$ for
$c_{n}\asymp n^{-\eta_{c}}$ where $2\eta_{c}+3\eta_{s}<1$ with $\eta_{s}$
as in Assumption \ref{Condition_precisionMatrixSparsity}. Then, $\Pr\left(\hat{G}\neq G\right)\lesssim n^{-p}$
for any constant $p<\infty$.\end{theorem}

Theorem \ref{Theorem_PcAlgoConsistency} says that the estimator for
the PCDAG converges to the true one at an arbitrarily fast polynomial
rate. This is worse that the exponential rate obtained by Kalisch
and B\"{u}hlmann (2007) for causal discovery using independent identically
distributed data. 

\subsubsection{Consistency of Structural Model Parameters}

We show that $\hat{D}$ from Algorithm \ref{Algo_ImpulseResponse}
is consistent for $D$, with $D$ as in Lemma \ref{Lemma_SVAR_identification}.
When the PC algorithms in Algorithm \ref{Algo_PCAlgo} produces edges
that are all directed, we interpret $D$ to be the one corresponding
to the permutation matrix $\Pi$ that is obtained by the least number
of row permutations of the identity. Then, $D$ is unique. 

In the following, we state the consistency of $\hat{D}$ for $D$,
and the consistency of an estimator $\hat{H}$ for $H$, in (\ref{EQ_maInfinitySVAR}),
with convergence rates. We shall denote by $\kappa$ the maximum number
of direct descendants among all parents. It is not difficult to show
that this is the same as the maximum number of nonzero elements among
the columns of $D$. Such number is bounded above by $s$, which corresponds
to the maximum number of adjacent variables across all the nodes.

\begin{theorem}\label{Theorem_impulseResponse}Suppose that the Assumptions
$\left(\lambda,\tau\right)$ hold, that the joint distribution of
the innovations $\varepsilon_{t}$ in (\ref{EQ_gaussianVAR}) is faithful
to the DAG for all $K$, and that all the estimated edges resulting
from Algorithm \ref{Algo_PCAlgo} are directed. Then, using Algorithm
\ref{Algo_ImpulseResponse}, $\left|\hat{D}-D\right|_{{\rm op}}=O_{P}\left(s\sqrt{\frac{\kappa\ln K}{n}}\right)$,
where $D$ is as in (\ref{EQ_SVAR}) with $\Pi$ obtained by the least
number of row permutations of the identity. Moreover, we also have
that $\hat{H}=\left(I-\hat{D}\right)^{-1}$ satisfies $\left|\hat{H}-H\right|_{{\rm op}}=O_{P}\left(s\sqrt{\frac{\kappa\ln K}{n}}\right)$.\end{theorem}

\section{Empirical Illustrations\label{Section_empirics}}

To showcase the methodology presented in this paper we consider two
illustrations. The first considers a supply side oil price shock.
This problem has recently been considered by K\"{a}nzig (2021). While
the baseline model used in K\"{a}nzig (2021) only includes 6 variables,
this is still a high dimensional problem due to the fact that the
selected number of lags is 12. Our aim is to highlight the features
of our methodology and how it can be used to gain additional insights
on the role of an external instrument. This application should clarify
some of the language used in the paper and draw a clear parallel between
the more common language used in economics and causal DAG's. We hope
to convince the reader that the use of the DAG has much to offer,
once its role is understood. 

The second application focuses on the causal relation between information
in the order book in high frequency trading. For this application,
the latent model is a VAR(1), however, the number $K$ of variables
is large: $K=60$. Among other things, there we highlight how the
information from the impulse response functions produces a net effect
that is different from the causal information flow represented by
the structural equation model or equivalently the causal graph. 

\subsection{The Effect of Oil Price Shocks\label{Section_empiricalOilPriceShock}}

Shocks in real oil price can be caused by either demand or supply
shocks. K\"{a}nzig (2021) uses oil futures price changes around OPEC
announcements as an instrument to identify supply side shocks. We
use our methodology to show how we identify the latent structural
VAR model without an instrument for this specific dataset. We then
include the instrument as an additional variable to our model and
show that the resulting DAG suggests that this is a valid instrument,
though unnecessary for the purpose of identifying the structural parameters.
We stress that the goal of this application is to refer to a state
of the art approach for a well known problem and show what we can
achieve with our methodology, assuming that the system is recursive. 

\subsubsection{The Data and the Covariates}

We consider the same dataset used in K\"{a}nzig (2021). The data
consists of real oil price, U.S. CPI, U.S. industrial production,
world industrial production, world oil inventories and world oil production.
We also include a Crude Oil Shock variable constructed in K\"{a}nzig
(2021), as additional variable. This variable can be used as either
internal or external instrument for identification. Here, it will
be used as an internal instrument to show that it is a valid instrument,
relying on the graphical method of the paper. The sample period is
from February 1975 to December 2017. The data is at monthly frequencies.
Due to either the persistency or nonstationarity of the data, we first
difference all variables except for the oil supply news shock. The
covariates are listed in Table \ref{Table_covariatesListOilShock}.

\begin{table}
\caption{List of Covariates Used in the Model of Oil Price Shocks. The covariates
are listed together with their short name for ease of reference. The
covariate OilShock is an instrument; see K\"{a}nzig (2021) for details
on the covariates.}

\label{Table_covariatesListOilShock}
\centering{}%
\begin{tabular}{llc}
 &  & \tabularnewline
Name & Short Name & \tabularnewline
\cline{1-2} \cline{2-2} 
 &  & \tabularnewline
Crude Oil News Shock & OilShock & \tabularnewline
Real Oil Price & OilPrice & \tabularnewline
World Oil Production & WorldOilProd & \tabularnewline
World Oil Inventories & WorldOilInv & \tabularnewline
World Industrial Production & WorldIndProd & \tabularnewline
U.S. Industrial Production & UsIndProd & \tabularnewline
U.S. CPI & UsCpi & \tabularnewline
 &  & \tabularnewline
\end{tabular}
\end{table}

\subsubsection{Estimation}

We estimate the causal graph using our proposed methodology and the
six variables introduced in the previous section. We allow for lags
greater than one, by the minor modification discussed in Section \ref{Section_remarksAssumptions}.
We choose a lag length of 12 as in K\"{a}nzig (2021). This is in
line with the choice by Akaike's information criterion as implemented
in Section \ref{Section_AICLagLength} of the Electronic Supplement.
We use both Lasso (Algorithm \ref{Algo_LassoThresh}) and CLIME (Algorithm
\ref{Algo_climeThresh}) for the estimation of the sparse precision
matrix. For these algorithms, the penalization parameter $\lambda$
and the threshold parameter $\tau$ are selected using cross-validation
(see Section \ref{Section_ChoiceTuning} in the Electronic Supplement
for details). We then apply Algorithms \ref{Algo_sparseCopulaParametersEstimation},
\ref{Algo_PCAlgo}, and \ref{Algo_ImpulseResponse} to estimate the
Gaussian copula VAR parameters, recover the contemporaneous causal
structure and possibly identify the matrix of contemporaneous relations
$D$. The latter can then be used for estimation of the impulse response
functions.\textbf{ }

\subsubsection{Summary of Results}

The results for Lasso and CLIME were very similar. In the interest
of space, we report and discuss only the results when Lasso (Algorithm
\ref{Algo_LassoThresh}) is used as intermediate step, with no further
mention. 

\begin{figure}
\begin{centering}
\noindent\begin{minipage}[c]{1\textwidth}%
\begin{center}
\includegraphics[scale=0.12]{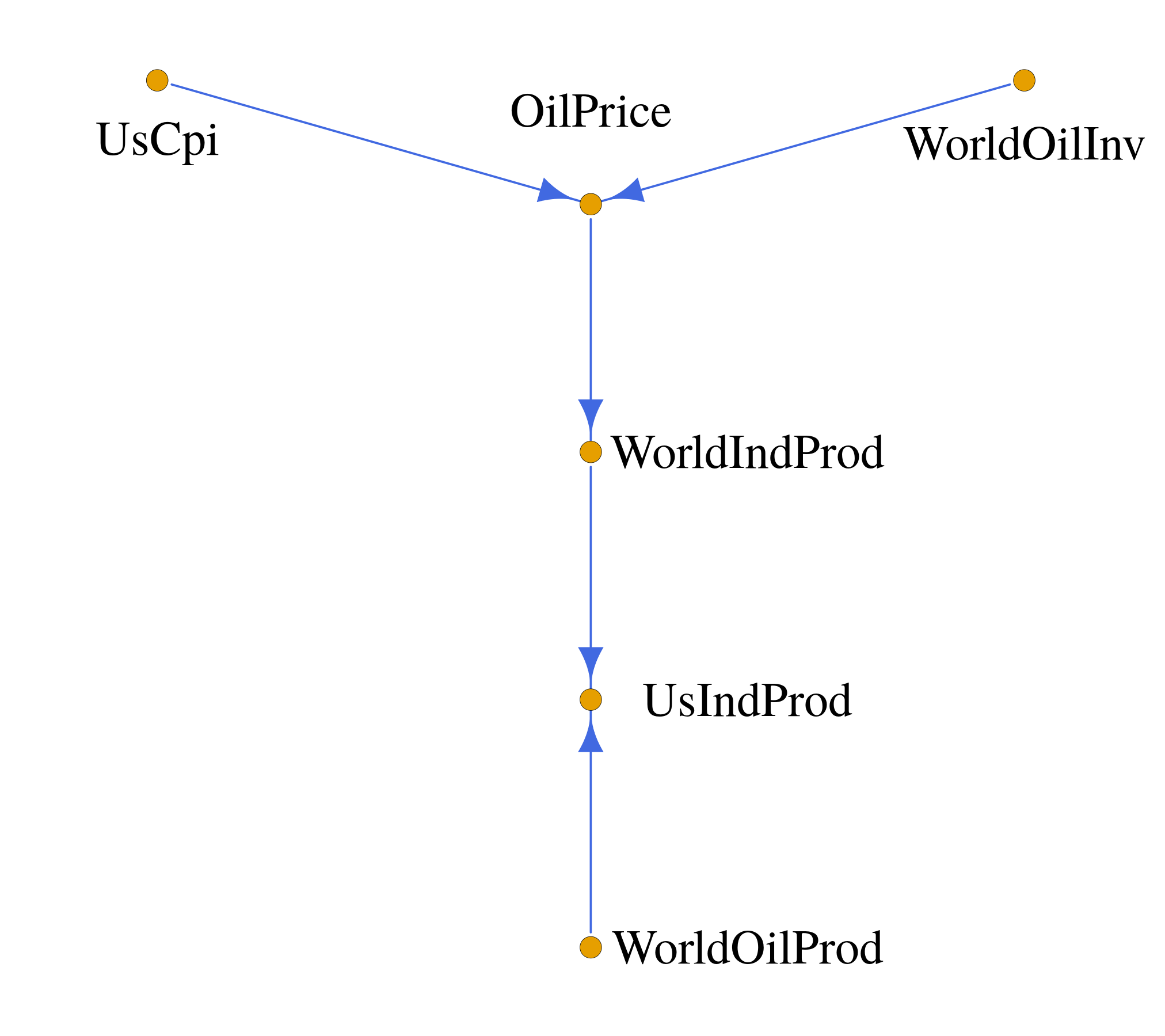}
\par\end{center}
\begin{center}
(a)
\par\end{center}%
\end{minipage} \vfill{}
\noindent\begin{minipage}[c]{1\textwidth}%
\begin{center}
\includegraphics[scale=0.12]{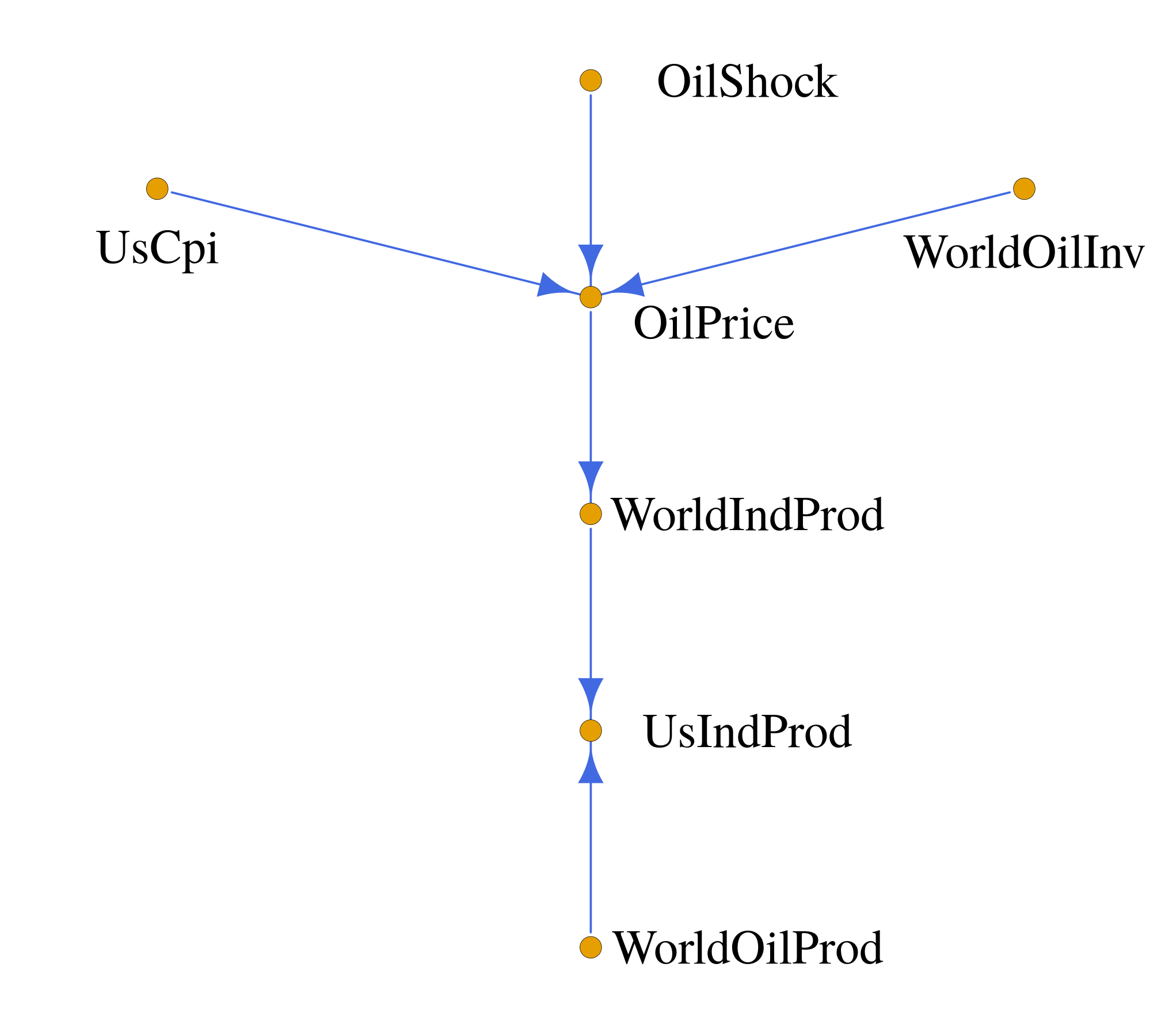}
\par\end{center}
\begin{center}
(b)
\par\end{center}%
\end{minipage}
\par\end{centering}
\caption{Contemporaneous Causal Graph for or Real Oil Price and Related Macro
Variables. The graph is estimated using the methodology of the paper
with Lasso. The penalty parameters $\lambda$ and $\tau$ are chosen
by cross-validation. The results are robust to parameters choice locally
around the cross-validation ones. The graph is estimated using six
variables in Panel (a), and adding also the supply side news shock
in Panel (b).}
\label{Figure_macroDag}
\end{figure}

Using our methodology, we estimate a model with 12 lags, in line with
K\"{a}nzig (2021). This means that the number of relevant parameters
to be estimated is $12K^{2}+\frac{K\cdot(K+1)}{2}=453$, where $K=6$
is the number of covariates in the model without instrument. Given
a sample size of $n=503$ we clearly are in a high dimensional setting.
We found that all the edges of the causal graph were directed. This
means that we are able to identify the permutation matrix $\Pi$ and
the matrix $D$ in Lemma \ref{Lemma_SVAR_identification}. In consequence,
the SVAR parameters are identified without the need of an instrument,
under the assumption that the system is recursive. We report the DAG
in Figure \ref{Figure_macroDag}. 

To shed further light on the value of OilShock as instrument used
in K\"{a}nzig (2021) and confirm that it satisfies the exclusion
restriction for an instrument, we estimate the model including the
latter as an additional variable. We now have $K=7$ variable with
same number of lags. The result shows that OilShock is a source node,
it impacts real oil prices directly and is not connected to any of
the other variables. According to the discussion in Section \ref{Section_identificationWithExternalInstruments}
this means that it is a valid instrument. 

The results also highlight a challenge. We find that a shock to UsCpi
leads to a contemporaneous effect on OilPrice. This appears inconsistent
with logic and economic theory. One explanation is that the underlying
assumption that the system is recursive, after proper permutation,
is not satisfied. However, the direction of this particular causal
relation is identified using the example at the end of Section \ref{Section_preliminaryConcepts},
which does not rely on recursivity. To see this, note that from the
data we have that OilShock and UsCpi are unconditionally independent,
but dependent when we condition on OilPrice. Then, this implies that
OilShock and UsCpi are unrelated common causes to OilPrice (Section
\ref{Section_preliminaryConcepts}). Hence, the answer needs to be
found elsewhere. To check that the results are not statistical artifacts
specific to our methodology, we estimate the same model as in K\"{a}nzig
(2021): a VAR for the levels of the observed variables with 12 lags.
We then check whether the residuals of OilShock and UsCpi are unconditionally
independent of each other and all other variables, but are dependent
when conditioning on OilPrice. We found that with 95\% confidence,
this is the case. Hence, we rule out that this is a statistical artifact
specific to our methodology. The answer requires work beyond the scope
of this paper. 

There are two take away from this empirical illustration. First, for
this specific data set, we are able to identify the latent SVAR parameters,
under the assumption of a recursive structure with no need for an
instrument. Note even assuming that the structure is recursive, the
set of possible solutions has cardinality that grows exponentially
with the number $K$ of variables. Hence, this is a nontrivial exercise.
Second, by including the potential instrument as one of the variables,
we are able to confirm the validity of the instrument. Again the underlying
assumption is that the system is recursive. However, given the structure
of the graph, recursivity is not used to orient the edges in the subgraph
OilPrice, OilShock and UsCpi.

\subsection{Causal Relations in the Limit Order Book}

Large orders to buy or sell stocks in a financial market are usually
broken into smaller ones and executed over a fixed time frame, where
time can be measured in clock time or volume time (Donnelly, 2022
for a review). One important dilemma when executing orders is to decide
whether to execute crossing the spread or posting passive orders.
The latter does not guarantee immediate execution. However, it is
believed to reduce market impact. Empirical evidence shows that passive
orders that skew the limit order book also cause market impact. Hence,
it is of interest to understand the causal implication of an algorithm
that keeps posting limit orders resulting on an order book imbalance
(Table \ref{Table_covariatesListLOB} for definition) versus an algorithm
that continuously trade crossing the spread. Moreover, limit orders
may be posted at deeper levels in to book to gain queue priority.
Such actions tend to have a persistent impact on the book as many
orders need to be executed over a finite number of time. We are interested
in understanding the implications of such persistent actions. 

To this end, we apply our methodology to study the causal relations
between aggregated order book and trades variables in high frequency
electronic trading. Aggregation allows us to reduce noise and extract
information that is concealed at high frequency. We aggregate the
information in volume time (Section \ref{Section_dataCovariatesLOB},
for more details). This is different from the analysis of order book
tick data which has been studied extensively in the literature (Cont
et al., 2014, Kercheval and Zhang, 2015, Sancetta, 2018, Mucciante
and Sancetta, 2022a, 2022b). It is well known that market participants
look at the order book to extract market information (MacKenzie, 2017).
We want to extract average causal relations. The underlying assumption
is that (\ref{EQ_structuralEquationSystemRecursive}) holds, i.e.
the causal structure can be represented in terms of a DAG. 

We shall estimate a model with 5 stocks to investigate the direction
of information dissemination within each stock, via the order book
and trades, as well intra stocks. This requires the estimation of
a large dimensional model. Our results will also show how the methodology
of this paper allows us to disentangle contemporaneous causal effects
from time series effects.

\subsubsection{The Data and the Covariates\label{Section_dataCovariatesLOB}}

We consider four stocks constituents of the S\&P500 traded on the
NYSE: Amazon (AMZN), Cisco (CSCO), Disney (DIS) and Coca Cola (KO).
We also consider the ETF on the S\&P500 (SPY). The stock tickers are
given inside the parenthesis. The sample period is from 01/March/2019
to 30/April/2019, from 9:30am until 4:30pm on every trading day. The
data were collected from the LOBSTER data provider (Huang and Polak,
2011)\footnote{https://lobsterdata.com/.}. This is a Level 3 dataset,
meaning that it contains all limit orders and cancellations for the
first 10 levels of the order book as well as trades, all in a sequential
order.

We construct a set of covariates related to the ones that are commonly
found in the studies of high frequency order book and trades. However,
we use aggregated data in volume time in intervals of 10\% of daily
volume of SPY. Volume time means that instead of clock time, we use
cumulated trades as measure of time. We choose SPY as common time
for all the instruments as this is the asset that replicates the S\&P500
index. Aggregated data allow us to estimate an average propensity
of each covariate to cause the other. For example, an order book where
limit orders to buy tend to be much higher than limit orders to sell
could drive the price up over. The covariates are the book imbalance
up to ten levels, a geometric average return, and the trade imbalance,
often termed order flow imbalance. The covariates are listed in Table
\ref{Table_covariatesListLOB}, where their definition can be found.
In Table \ref{Table_covariatesListLOB}, ${\rm Mid}=\left({\rm AskPrice}_{1}+{\rm BidPrice}_{1}\right)/2$
and ${\rm LagMid}$ is the ${\rm Mid}$ from the previous minute bucket,
where ${\rm AskPrice}_{i}$ is the ask price at level $i$ and similarly
for ${\rm BidPrice}_{i}$. The operator ${\rm avg}\left(\cdot\right)$
takes the data from the same one minute bucket and computes the average
value. In case of much market activity, the exchange will use the
same timestamp for a number of messages at different levels. In the
case of the orderbook, we use the last book snapshot of the many with
the same time stamp. We do not apply this logic to trades. These covariates
are directional ones. For this reason, we have omitted other interesting
ones, such as the spread. Moreover, the instruments we use are all
very liquid and the spread does not change much in this case.\textbf{ }

For ease of reference, in what follows, we shall use the convention
of merging the ticker and covariate short name.

\begin{table}
\caption{List of Covariates Derived from the Order Book. The covariates are
listed together with their definition.}

\label{Table_covariatesListLOB}
\centering{}%
\begin{tabular}{lllc}
 &  &  & \tabularnewline
Name & Short Name & Definition & \tabularnewline
\cline{1-3} \cline{2-3} \cline{3-3} 
 &  &  & \tabularnewline
Book imbalance & ${\rm BookImb}_{i}$ & $\frac{{\rm avg}\left({\rm BidSize}_{i}-{\rm AskSize}_{i}\right)}{{\rm avg}\left({\rm BidSize}_{i}+{\rm AskSize}_{i}\right)}$ & \tabularnewline
at level $i\in\left[10\right]$ &  &  & \tabularnewline
 &  &  & \tabularnewline
Return & ${\rm Ret}$ & $100\times\left[{\rm avg}\left(\ln\left({\rm Mid}\right)\right)-{\rm avg}\left(\ln\left({\rm LagMid}\right)\right)\right]$ & \tabularnewline
 &  &  & \tabularnewline
Trade Imbalance & ${\rm TradeImb}$ & $\frac{{\rm avg}\left({\rm SignedTrdSize}\right)}{{\rm avg}\left({\rm TrdSize}\right)}$ & \tabularnewline
\end{tabular}
\end{table}

\subsubsection{Estimation}

The estimation is the same as in Section \ref{Section_empiricalOilPriceShock},
but constraining the analysis to one lag only. We shall also compute
the impulse response functions for a subset of the variables using
the methodology discussed in Section \ref{Section_impulseResponseMCIntegration}
in the Appendix.

\subsubsection{Summary of Results}

The results for Lasso and CLIME were very similar. We discuss only
the results when Lasso (Algorithm \ref{Algo_LassoThresh}) is used
as intermediate step. Our results show that the causal structure of
the order book of each instrument exhibits a dense network structure.
Within each instrument, the first level of order book imbalance is
not contemporaneously caused by any other variable (this is called
a source node). In general we observe how the causal structure goes
from top levels of the book to deeper ones. Usually, the return is
affected directly by the deeper levels of the order book imbalance.
For all instruments the return is a cause of the trade imbalance variable
that does not happen to cause any other variable (this is called a
sink node). We also observe cross-causal effects across instruments.
We observe how in general the return of an instrument could be affected
by other instrument returns, e.g., AMZN return impacts CSCO and the
SPY return. In particular, the SPY return is affected by the other
returns. We also observe that the trade imbalance of an instrument
may directly affect the top levels of the book of other instruments
impacting so on all the order book structure, e.g., the AMZN trade
imbalance directly affects the first level of CSCO and SPY book imbalance
as well as the respective trade imbalance together with trade imbalance
of DIS and the eighth level of the SPY book imbalance.

\begin{sidewaysfigure}
\centering{}\includegraphics[scale=0.15]{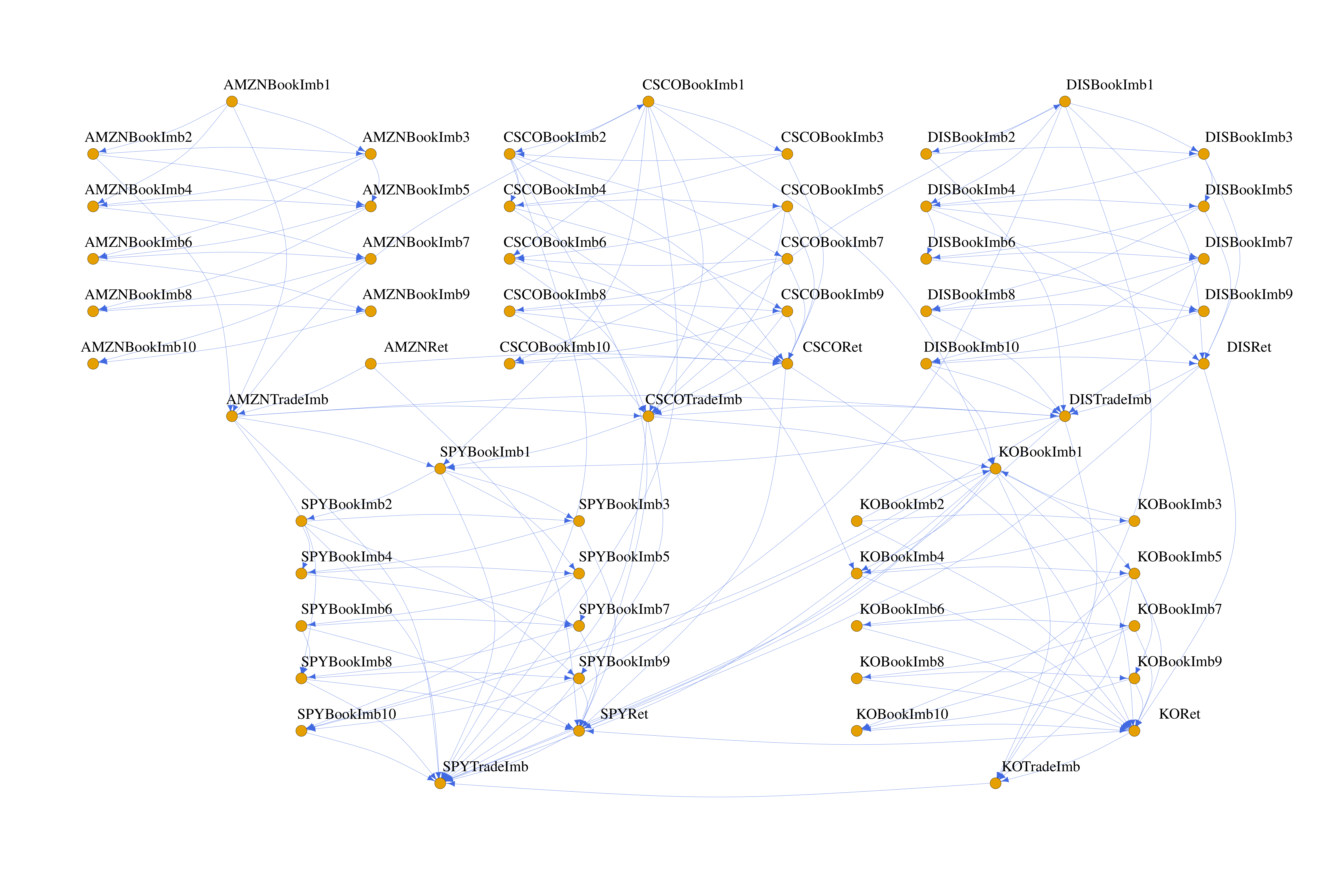}\caption{Contemporaneous Causal Graph for the Aggregated Oderbook Information.
The graph is estimated using the methodology of the paper with Lasso.
The penalty parameters $\lambda$ and $\tau$ are chosen by cross-validation.
The results are robust to parameters chosen locally around the cross-validation
ones.}
\label{Figure_lassoDAGResults}
\end{sidewaysfigure}

The details can be found in Figure \ref{Figure_lassoDAGResults} that
shows the DAG of contemporaneous causal relations obtained from our
estimation procedure. 

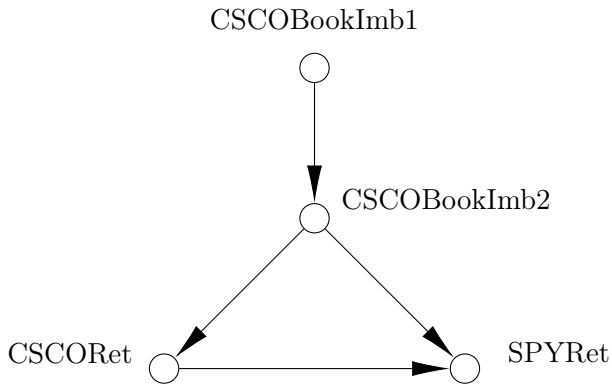
\begin{figure}
\centering
\begin{tikzpicture}[scale=1,every node/.style={draw=black,circle}]
\tikzstyle{EdgeStyle} = [->,>=stealth']   

 \node[label={[xshift=0.0cm, yshift=-1.1cm]\small CSCOBookImb1}] (a) at (0,0) {};      \node[label={[xshift=+1.75cm, yshift=-1.5cm]\small CSCOBookImb2}] (b) at (0,-2) {};  
\node[label={[xshift=-1.25cm, yshift=-1cm]\small CSCORet}] (c) at (-2,-4) {};      
\node[label={[xshift=1.25cm, yshift=-0.85cm]\small SPYRet}] (d) at (2,-4) {};  \draw[-{Latex[length=5mm, width=2mm]}] (a) edge (b); 
\draw[-{Latex[length=5mm, width=2mm]}] (c) edge (d); \draw[-{Latex[length=5mm, width=2mm]}] (b) edge (d);
\draw[-{Latex[length=5mm, width=2mm]}] (b) edge (c);
\end{tikzpicture}\caption{Subgraph of Estimated Graph in Figure \ref{Figure_lassoDAGResults}.
The subgraph only considers the contemporaneous causal relations between
${\rm CSCOBookImb}_{1}$, ${\rm CSCOBookImb}_{2}$, ${\rm CSCORet}$
and ${\rm SPYRet}$.}
\label{Figure_subgraphDagCSCOImb}
\end{figure}

\begin{figure}
\noindent\begin{minipage}[t]{1\columnwidth}%
\begin{center}
{\includegraphics[width=0.45\textwidth]{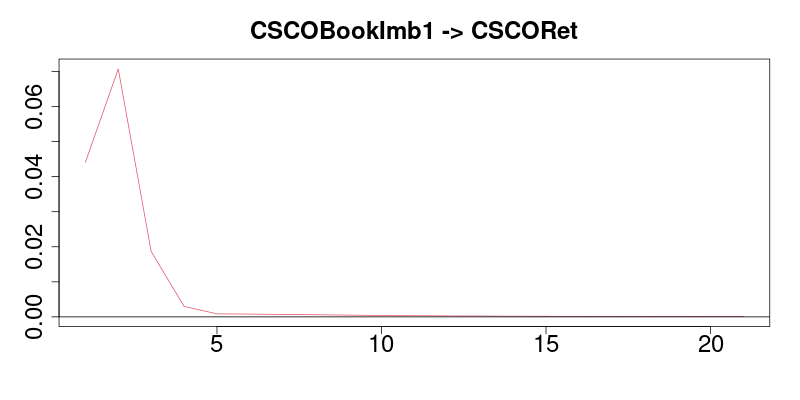}}
\quad{}{\includegraphics[width=0.45\textwidth]{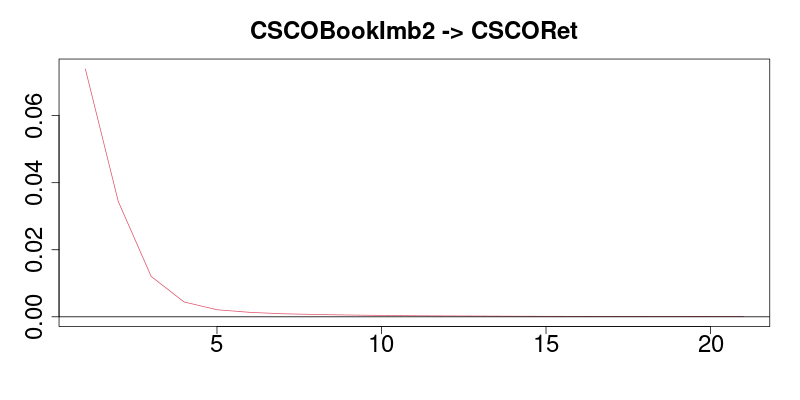}}\\
\par\end{center}
\begin{center}
{\includegraphics[width=0.45\textwidth]{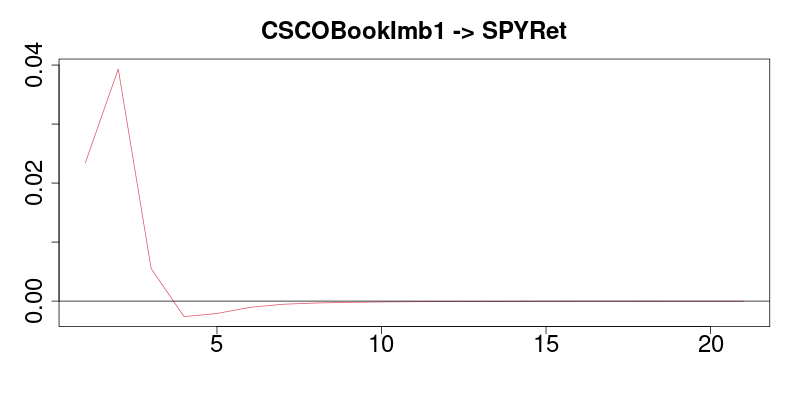}}
\quad{}{\includegraphics[width=0.45\textwidth]{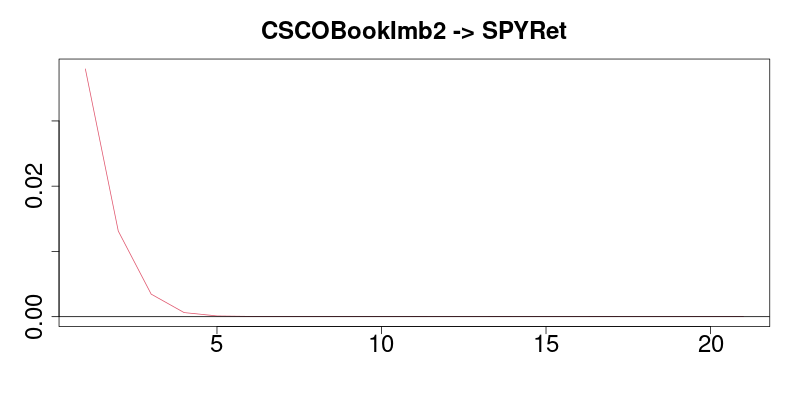}}
\par\end{center}
\begin{center}
(a)
\par\end{center}%
\end{minipage}

\noindent\begin{minipage}[t]{1\columnwidth}%
\begin{center}
{\includegraphics[width=0.45\textwidth]{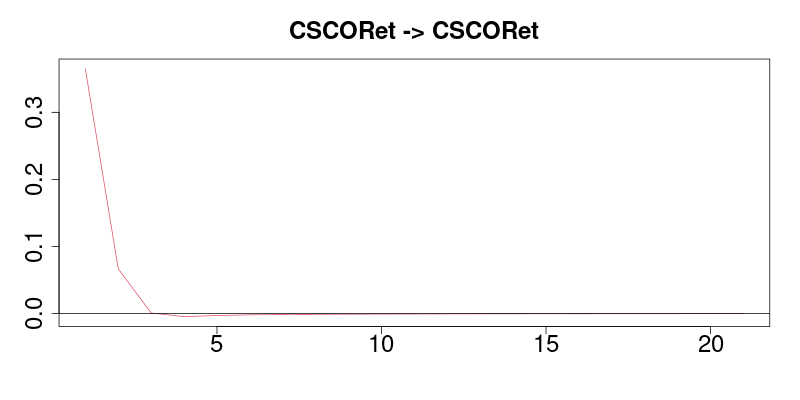}}
\quad{}{\includegraphics[width=0.45\textwidth]{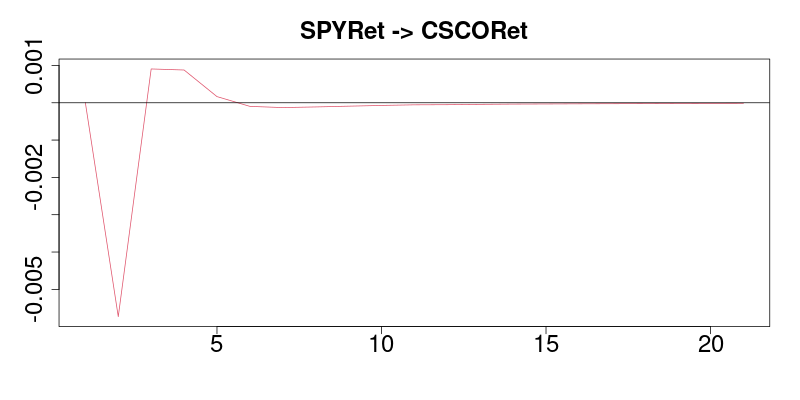}}\\
\par\end{center}
\begin{center}
{\includegraphics[width=0.45\textwidth]{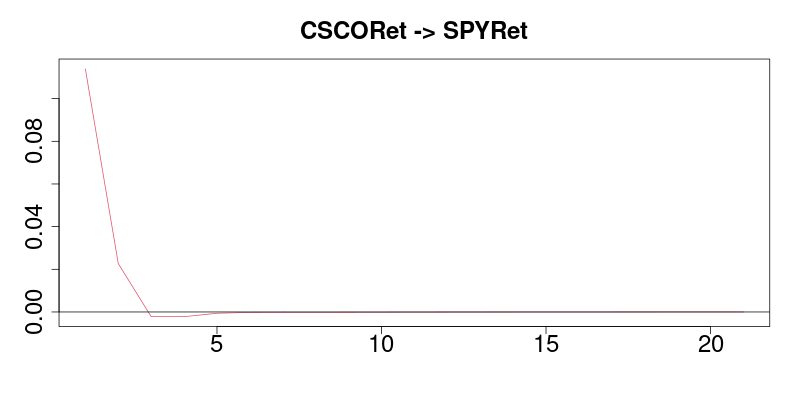}}
\quad{}{\includegraphics[width=0.45\textwidth]{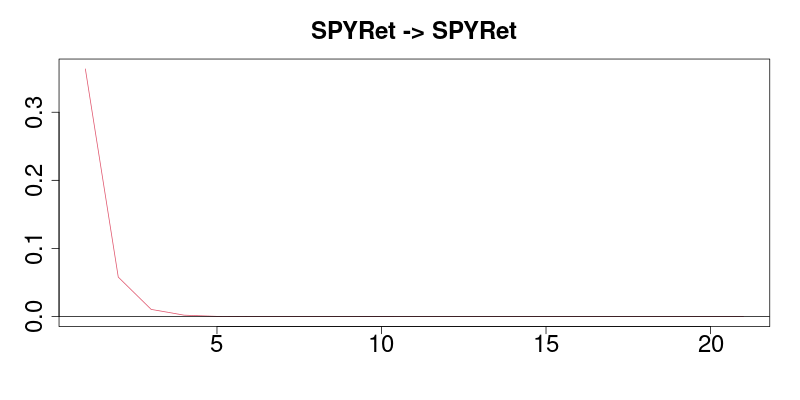}}
\par\end{center}
\begin{center}
(b)
\par\end{center}%
\end{minipage}

\caption{Impulse Response Functions for a Subset of the Covariates. Panel (a)
shows the impulse response functions for ${\rm CSCORet}$, ${\rm SPYRet}$
as a result of a unitary shock on ${\rm CSCOBookImb}_{1}$, ${\rm CSCOBookImb}_{2}$.
Panel (b) show the same information for ${\rm CSCORet}$ and ${\rm SPYRet}$
on each other. The x-axis represents time lag shifted by one unit,
so that $1$ is the effect of the shock at lag $0$.}
\label{Figure_subfig_imb1_2-1}
\end{figure}

We also show how the contemporaneous impulse response function $\Pi'H\Pi$
may fail to show the direct contemporaneous causal relations defined
via $\Pi'D\Pi$ in (\ref{EQ_structuralEquationSystemRecursive}).
Consider the subgraph composed by ${\rm CSCOBookImb}_{1}$, ${\rm CSCOBookImb}_{2}$,
${\rm CSCORet}$ and ${\rm SPYRet}$ as shown in Figure \ref{Figure_subgraphDagCSCOImb}.
The related impulse response functions are plotted in Figure \ref{Figure_subfig_imb1_2-1}.\footnote{Bootstrap confidence intervals were very tight due to the large sample
size, so they are not plotted.} By looking at the impulse response functions, we may conclude that
${\rm CSCOBookImb}_{1}$ and ${\rm CSCOBookImb}_{2}$ are directly
affecting ${\rm CSCORet}$ and ${\rm SPYRet}$. However, this effect
is mediated as shown in Figure \ref{Figure_subgraphDagCSCOImb}. There,
we observe that a shock on ${\rm CSCOBookImb}_{1}$ will first impact
the ${\rm CSCOBookImb}_{2}$ and ${\rm CSCORet}$ and then it propagates
to the ${\rm SPYRet}$. Only the causal graph or equivalently the
structural equations system allows us to understand the information
flow in the order book. The impulse response functions only represent
the contemporaneous net effect of a shock. 

\section{Conclusion\label{Section_conclusion}}

This paper has introduced a novel approach for the estimation of causal
relations in time series. It essentially uses a Gaussian copula VAR
model. Such causal relations differ from Granger causality. Our methodology,
allows us to identify causal relations in high dimensional models.
Using a sparsity condition we are able to consistently estimate the
model parameters. Our sparsity condition does not impose sparsity
of the autoregressive matrix and of the covariance matrix of the innovations
implied by the Gaussian copula VAR model. Our sparsity conditions
can be viewed as weak assumptions on conditional independence. We
are then able to identify the related directed acyclic graph of causal
relations, using observational data, as if we knew the true distribution
of the data. 

Asymptotic results and finite sample investigation confirm the viability
of our methodology and its practical usefulness for high dimensional
problems. A finite sample analysis, carried out using simulation (Section
\ref{Section_summarySimulations} in the Electronic Supplement), confirms
the asymptotic results of the paper. Moreover, the simulations show
that not accounting for time series dependence leads to wrong causal
inference. Failing to exploit sparsity leads to suboptimal results,
even in low dimensions. 

We also relied on two empirical applications to highlight the methodology
of the paper. We considered the effect of oil price shocks to the
economy as studied in K\"{a}nzig (2021). We showed how our methodology
can be used to verify whether an instrument is needed and whether
the instrument is a valid one. Then, we applied our methodology to
the analysis of the conditional contemporaneous causal relations of
order book data aggregated in volume time. To the best of our knowledge
this has not been done before and has important implications for understanding
the aetiology of electronic trading. The applications also showed
how causal inference provides the path followed by a shock via a system
of structural equations that has a graphical representation. On the
other hand the contemporaneous impulse response functions show the
net effect with no information on the actual causal path.

There are a number of areas that have been overlooked and require
further research in the future. For example, the methodology assumes
that the system of structural innovations of the latent VAR process
is recursive. In this case, strategies for partial identifications
within our framework need to be devised. However, we showed that methods
based on instruments can still be used in our setup. Moreover, the
literature has put forward the possibility of models that exhibit
some form of time variation. This time variation is then exploited
for identification via heteroskedasticity. Our framework does not
cover this, yet. This extension requires careful study, as it has
nontrivial implications for the meaning of causality, as used in this
paper. For example, time variation may result from omitted variables/causes.
In this case, a nonlinear framework, such as ours, can be a suitable
starting point to address the problem. To conclude, our approach provides
an opportunity for much new research building on the existing contributions
in the literature.

\section*{Appendix \label{Section_Appendix}}

\setcounter{figure}{0} \renewcommand{\thefigure}{A.\arabic{figure}}

\setcounter{equation}{0} \renewcommand{\theequation}{A.\arabic{equation}} 

\setcounter{lemma}{0} \renewcommand{\thelemma}{A.\arabic{lemma}}

\setcounter{section}{0} \renewcommand{\thesection}{A.\arabic{section}}

\section{Remarks on the Gaussian Transformation\label{Section_remarksGaussianTransform}}

We provide some remarks on the model in order to clarify its applicability.
For simplicity of exposition, suppose that the number of covariates
$K=2$ and that there is no time series dependence (i.e. $A$ is a
zero matrix). For any univariate random variable $X_{t,k}$, there
is always a monotonic increasing function $f_{k}$ such that $f_{k}\left(X_{t,k}\right)$
is standard normal. For example, let $F_{k}\left(x\right)=\Pr\left(X_{t,k}\leq x\right)$
and $F_{k}\left(x-\right)=\Pr\left(X_{t,k}<x\right)$, $x\in\mathbb{R}$.
By stationarity, the probability is independent of $t\geq1$. Define 

\begin{equation}
\tilde{F}_{k}\left(x,v\right)=\left(1-v\right)F_{k}\left(x-\right)+vF_{k}\left(x\right).\label{EQ_uniformTransformFunction}
\end{equation}
If the variables are continuous with density with no atoms, $\tilde{F}_{k}\left(x,v\right)=F_{k}\left(x\right)$.
The purpose of this section is to consider the case where $F_{k}$
is not necessarily continuous and show that our methodology still
applies. Once we state the following result, it will become clear
how to discuss the case where the variables are time dependent. 

\begin{lemma}\label{Lemma_transformsProperties} Let $V_{t,1}$ and
$V_{t,2}$ be uniform random variables in $\left[0,1\right]$ independent
of $X_{t,1}$ and $X_{t,2}$. The following hold. 
\begin{enumerate}
\item $\tilde{U}_{t,k}:=\tilde{F}_{k}\left(X_{t,k},V_{t,k}\right)$ is a
uniform random variable in $\left[0,1\right]$ and $X_{t,k}=F_{k}^{-1}\left(\tilde{U}_{t,k}\right)$
almost surely, $k=1,2$.
\item Let $\Phi^{-1}:\left[0,1\right]\rightarrow\mathbb{R}$ be the quantile
function of the standard normal distribution. Then, $\tilde{Z}_{t,k}:=\Phi^{-1}\left(\tilde{U}_{t,k}\right)$
is a standard normal random variable, $k=1,2$.
\item Define $\pi_{V}:=\mathbb{E}V_{t,1}V_{t,2}$. Then, $\tilde{\rho}\left(\pi_{V}\right):=12Cov\left(\tilde{F}_{1}\left(X_{t,1},V_{t,1}\right),\tilde{F}_{2}\left(X_{t,1},V_{t,2}\right)\right)$
is a function of $\pi_{V}$. If $V_{t,1}$ and $V_{t,2}$ are independent,
$\pi_{V}=1/4$ and we have that 
\begin{align*}
\tilde{\rho}\left(\frac{1}{4}\right)= & 12\mathbb{E}\left[\frac{F_{1}\left(X_{t,1}-\right)+F_{1}\left(X_{t,1}\right)}{2}\right]\left[\frac{F_{2}\left(X_{t,2}-\right)+F_{2}\left(X_{t,2}\right)}{2}\right]-3.
\end{align*}
\item Let $U_{t,1}$ and $U_{t,2}$ be uniform random variables in $\left[0,1\right]$
with Gaussian copula, and $Z_{t,k}=\Phi^{-1}\left(U_{t,k}\right)$,
$k=1,2$. Let $\rho:=12Cov\left(U_{t,1},U_{t,2}\right)$, then, $r_{Z}:=Cov\left(Z_{t,1},Z_{t,2}\right)=2\sin\left(\frac{\pi}{6}\rho\right)$. 
\item Let $Z_{t,1}$ and $Z_{t,2}$ be standard Gaussian with correlation
coefficient $r_{Z}$. Let $U_{t,k}=\Phi\left(Z_{t,k}\right)$ and
$X_{t,k}=F_{k}^{-1}\left(U_{t,k}\right)$, $k=1,2$, and 
\[
h\left(r\right)=\mathbb{E}^{X_{t,1}}\mathbb{E}^{X_{t,2}}\Phi\left(\Phi^{-1}\left(1-F_{1}\left(X_{t,1}\right)\right),\Phi^{-1}\left(1-F_{2}\left(X_{t,2}\right)\right);r\right),
\]
where $\mathbb{E}^{X_{t,k}}$ is expectation w.r.t. the marginal law
of $X_{t,k}$, $k=1,2$. Then, the function $h\left(r\right)$ is
strictly increasing w.r.t. $r\in\left[-1,1\right]$. Moreover $\mathbb{E}F_{1}\left(X_{t,1}\right)F_{2}\left(X_{t,2}\right)=h\left(r_{Z}\right)$,
where $\Phi\left(\cdot,\cdot;r_{Z}\right)$ is the bivariate distribution
of two standard normal random variables with correlation equal to
$r_{Z}$.
\item Suppose that $\hat{F}_{k}$ is an estimator for $F_{k}$ satisfying
$\sup_{x\in\mathbb{R}}\left|\hat{F}_{k}\left(x\right)-F_{k}\left(x\right)\right|\rightarrow0$
in probability, $k=1,2$ and that the data is ergodic. Then 
\[
\frac{1}{n}\sum_{t=1}^{n}\hat{F}_{1}\left(X_{t,1}\right)\hat{F}_{2}\left(X_{t,2}\right)\rightarrow\mathbb{E}F_{1}\left(X_{t,1}\right)F_{2}\left(X_{t,2}\right)
\]
in probability. 
\end{enumerate}
\end{lemma}

Lemma \ref{Lemma_transformsProperties} show how the transformation
in (\ref{EQ_uniformTransformFunction}) can be used construct uniform
random variables in $\left[0,1\right]$ (Point 1). Once variables
are uniform, we can obtain standard Gaussian random variables (Point
2). Points 1 and 2 also mean that we can choose $f_{k}$ in the definition
of (\ref{EQ_gaussianVAR}) such that $f_{k}^{-1}\left(Z_{t,k}\right):=F_{k}^{-1}\left(\Phi\left(Z_{t,k}\right)\right)$.
Spearman's rho is a commonly used measure of dependence which is invariant
under strictly monotone transformation. However, the population Spearman's
rho for the transformed variables depends on the dependence between
$V_{t,1}$ and $V_{t,2}$, i.e. $\pi_{V}$. Hence, it is no unique.
When $\pi_{V}=1/4$, the transformation produces a Spearman's rho
through independent linear interpolations between the discontinuity
points of the distribution functions of the two variables (Point 3).
On the other hand, if the variables are continuous, the transformation
produces produces uniform random variables with dependence structure
that maps into the dependence structure of the latent process via
a closed for expression (Point 4). However, discontinuities do not
preclude us from identification of the correlation coefficient $r_{Z}$
of the latent Gaussian variables (Points 5). All we need to do is
to replace the map $\rho\mapsto2\sin\left(\frac{\pi}{6}\rho\right)$
with $\mathbb{E}F_{1}\left(X_{t,1}\right)F_{2}\left(X_{t,2}\right)\mapsto h^{-1}\left(\mathbb{E}F_{1}\left(X_{t,1}\right)F_{2}\left(X_{t,2}\right)\right)$.
Mutatis mutandis, this observation has been made in Fan et al. (2017)
and can be used for identification of the distributional parameters
of the latent process when $F_{k}$ is not continuous. 

\begin{example}Suppose that $X_{t,1}$ and $X_{t,2}$ are binary
random variables with values in $\left\{ 0,1\right\} $ and such that
$\Pr\left(X_{t,k}=1\right)=p_{k}$, $k=1,2$. Suppose that their joint
dependence is captured by a Gaussian copula. Then from Lemma \ref{Lemma_transformsProperties},
$X_{t,1}=1_{\left\{ U_{t,1}\geq p_{1}\right\} }$ and $X_{t,2}=1_{\left\{ U_{t,2}\geq p_{2}\right\} }$
where $\left(U_{t,1},U_{t,2}\right)$ are uniform random variables
in $\left[0,1\right]$ with Gaussian copula with scaling matrix $\Sigma$
such that the $\left(1,2\right)$ entry is $\Sigma_{1,2}=r_{Z}=2\sin\left(\frac{\pi}{6}\rho\right)$
where $\rho=12Cov\left(U_{t,1},U_{t,2}\right)$. Moreover, from Lemma
\ref{Lemma_transformsProperties} (Point 5), we have that, 
\begin{align*}
\mathbb{E}F_{1}\left(X_{t,1}\right)F_{2}\left(X_{t,2}\right)= & \sum_{x_{1}\in\left\{ 0,1\right\} }\sum_{x_{2}\in\left\{ 0,1\right\} }\Phi\left(\Phi^{-1}\left(\Pr\left(X_{t,1}\geq x_{1}\right)\right),\Phi^{-1}\left(\Pr\left(X_{t,2}\geq x_{2}\right)\right);r_{Z}\right)\\
 & \times\Pr\left(X_{t,1}=x_{1}\right)\Pr\left(X_{t,2}=x_{2}\right).
\end{align*}
By strict monotonicity w.r.t. $r_{Z}$, if we know $\mathbb{E}F_{1}\left(X_{t,1}\right)F_{2}\left(X_{t,2}\right)$
and $p_{1}$, $p_{2}$, we can uniquely identify $r_{Z}$.\end{example}

Knowledge of $r_{Z}$ essentially hinges on knowledge of $\mathbb{E}F_{1}\left(X_{t,1}\right)F_{2}\left(X_{t,2}\right)$.
A uniformly consistent estimator of the distribution function of the
data assures that such quantity is consistently estimated (Point 6).
A natural estimator $\hat{F}_{k}$ for $F_{k}$ is the empirical distribution
function based on a sample of size $n$. In this case, the uniform
convergence is exponentially fast (Lemma \ref{Lemma_exponentialIneqEDF}
in the Electronic Supplement). We also note that with an additive
error $O\left(\frac{1}{n}\right)$,
\[
\frac{12}{n}\sum_{t=1}^{n}\left(\hat{F}_{1}\left(X_{t,1}\right)\hat{F}_{2}\left(X_{t,2}\right)-\frac{1}{4}\right)
\]
is equal to the sample rank correlation coefficient (sample Spearman's
rho) because $n\hat{F}_{k}\left(X_{t,k}\right)=\sum_{s=1}^{n}1_{\left\{ X_{s,k}\leq X_{t,k}\right\} }$
is the rank of variable $X_{t,k}$, $k=1,2$. This also means that
if $F_{k}$ is discontinuous, we can use $h^{-1}\left(\hat{\rho}+3\right)$
as an estimator of $r_{Z}$, where $\hat{\rho}$ is the sample Spearman's
rho. See Fan et al (2017) for consistency when $h$ needs to be estimated.

\section{Impulse Response Functions\label{Section_impulseResponseFunctions}}

Given that our model is Markovian, we define 
\begin{equation}
\mathbb{E}\left[X_{t+s,k}|X_{t-1}=x,\xi_{t,l}=\delta\right]-\mathbb{E}\left[X_{t+s,k}|X_{t-1}=x,\xi_{t,l}=0\right]\label{EQ_impulseResponseConditional}
\end{equation}
to be the impulse response of $X_{t+s}$ to a shock in $\xi_{t,l}$
equal to $\delta$ and conditioning on a fixed value of $X_{t-1}=x\in\mathbb{R}^{K}$.
Recall that $\xi_{t,l}$ is not necessarily the shock corresponding
to $Z_{t,l}$. The latter is given by the $l^{th}$ entry in $\Pi\xi_{t}$.
Integrating out $x$ w.r.t. the marginal distribution of $X_{t}$
(\ref{EQ_impulseResponseConditional}) gives an unconditional impulse
response function. We introduce some notation to simplify the statement
of the details in what follows. For any matrix $B$, let $\left[B\right]_{k,l}$,
$\left[B\right]_{k,\cdot}$, $\left[B\right]_{\cdot,l}$ be the $k,l$
entry, the $k^{th}$ row and $l^{th}$ column respectively. If $B$
is a column vector, write $\left[B\right]_{k}$ for its $k^{th}$
entry. 

\begin{lemma}\label{Lemma_impulseResponse} Under the conditions
of Lemma \ref{Lemma_SVAR_identification}, for any scalar $\delta$,
in (\ref{EQ_impulseResponseConditional}) we have that 
\begin{equation}
\mathbb{E}\left[X_{t+s,k}|X_{t-1}=x,\xi_{t,l}=\delta\right]=\mathbb{E}f_{k}^{-1}\left(\left[A^{s+1}z+\sum_{r=0}^{s-1}\Upsilon_{r}\xi_{t+s-r}+\Upsilon_{s}\xi_{t}\left(\delta,l\right)\right]_{k}\right)\label{EQ_impulseResponseConditionalExplicit}
\end{equation}
where $z\in\mathbb{R}^{K}$ has $l^{th}$ entry $z_{l}=f_{l}\left(x_{l}\right)$,
$l=1,2,...,K$ and $\xi_{t}\left(a,l\right)$ equals $\xi_{t}$ except
for the $l^{th}$ entry which is fixed to a value equal to $a\in\mathbb{R}$.

Let $f_{k}'\left(x\right)=df_{k}\left(x\right)/dx$. Then, for $\delta\rightarrow0$,
(\ref{EQ_impulseResponseConditional}) equals 
\[
\mathbb{E}\left[\frac{\partial X_{t+s,k}}{\partial\xi_{t,l}}|X_{t-1}=x,\xi_{t,l}=\delta\right]=\mathbb{E}\left[f_{k}'\left(\left[A^{s+1}z+\sum_{r=0}^{s-1}\Upsilon_{r}\xi_{t+s-r}+\Upsilon_{s}\xi_{t}\left(0,l\right)\right]_{k}\right)\right]^{-1}\left[\Upsilon_{s}\right]_{\cdot,l}\delta
\]

\end{lemma}

Despite the possibly involved notation, the conclusions of Lemma \ref{Lemma_impulseResponse}
are simple. To find the impulse response, we need to find the inverse
of $f_{k}$. This function is unknown. Our methodology to estimate
the parameters of the latent SVAR does not require explicit knowledge
of $f_{k}$. However, if we want to compute (\ref{EQ_impulseResponseConditional})
such knowledge is needed. An estimator can be based on the truncated
inverse of the empirical distribution function (Liu et al., 2009).
Given that the latent model is Gaussian with i.i.d. innovations, the
expectation can be simply computed by Monte Carlo integration (Section
\ref{Section_impulseResponseMCIntegration} for more details). According
to Lemma \ref{Lemma_transformsProperties}, using the notation therein,
we have that $f_{k}^{-1}\left(\cdot\right):=F_{k}^{-1}\left(\Phi\left(\cdot\right)\right)$.
Finally, Lemma \ref{Lemma_impulseResponse} says that if we are interested
in the infinitesimal effect of a shock, we can linearize (\ref{EQ_impulseResponseConditional}).
In this case, up to a proportionality constant, (\ref{EQ_impulseResponseConditional})
is equal to $\left[\Upsilon_{s}\right]_{\cdot,l}\delta$. Hence, we
are only interested in the shape of the impulse response, knowledge
of $\Upsilon_{s}$ is sufficient. 

\subsection{Monte Carlo Integration\label{Section_impulseResponseMCIntegration}}

From Lemma \ref{Lemma_SVAR_identification}, $\xi_{t}=H^{-1}\Pi\varepsilon_{t}$.
Define $\Sigma_{\xi}:=\mathbb{E}\xi_{t}\xi_{t}'$ so that $\Sigma_{\xi}=H^{-1}\Pi\Sigma_{\varepsilon}\left(H^{-1}\Pi\right)'$.
For each of the variables $\xi_{t+s-r}$ simulate $m$ i.i.d. Gaussian
random vectors with covariance matrix $\Sigma_{\xi}$. Use a superscript
to denote these simulated data, i.e. $\left\{ \xi_{t+s-r}^{\left(v\right)}:v=1,2,...,m\right\} $.
The expectation in (\ref{EQ_impulseResponseConditionalExplicit})
is approximated by 
\[
\frac{1}{m}\sum_{v=1}^{m}f_{k}^{-1}\left(\left[A^{s+1}z+\sum_{r=0}^{s-1}\Upsilon_{r}\xi_{t+s-r}^{\left(v\right)}+\Upsilon_{s}\xi_{t}^{\left(v\right)}\left(\delta,l\right)\right]_{k}\right).
\]
To compute an unconditional impulse response function, we need to
integrate out $z$. To do so, we replace the above with 
\[
\frac{1}{m}\sum_{v=1}^{m}f_{k}^{-1}\left(\left[A^{s+1}Z^{\left(v\right)}+\sum_{r=0}^{s-1}\Upsilon_{r}\xi_{t+s-r}^{\left(v\right)}+\Upsilon_{s}\xi_{t}^{\left(v\right)}\left(\delta,l\right)\right]_{k}\right)
\]
where the random vectors $Z^{\left(v\right)}$ are Gaussian mean zero
with covariance matrix $\Gamma$ as in (\ref{EQ_copulaScalingMatrixW}). 

\newpage{}

\section*{Supplementary Material to ``Consistent Causal Inference for High
Dimensional Time Series'' by F. Cordoni and A. Sancetta\label{Section_Supplement}}

\setcounter{figure}{0} \renewcommand{\thefigure}{S.\arabic{figure}}

\setcounter{table}{0} \renewcommand{\thetable}{S.\arabic{table}}

\setcounter{equation}{0} \renewcommand{\theequation}{S.\arabic{equation}} 

\setcounter{page}{1}

\setcounter{lemma}{0} \renewcommand{\thelemma}{S.\arabic{lemma}} 

\setcounter{section}{0} \renewcommand{\thesection}{S.\arabic{section}}

\section{Proofs \label{Section_Proofs}}

Throughout, we use $c_{0},c_{1},c_{2},...$ to denote constants. 

We also recall a property of symmetric strictly positive definite
partitioned matrices. Let $\Sigma=\left(\begin{array}{cc}
A_{11} & A_{12}\\
A_{12}' & A_{22}
\end{array}\right)$ where $A_{i,j}$ $i,j\in\left\{ 1,2\right\} $ is a partition of
$\Sigma$. Then, $\Sigma^{-1}=\Theta=\left(\begin{array}{cc}
B_{11} & B_{12}\\
B_{12}' & B_{22}
\end{array}\right)$ where {\small{}
\begin{align}
B_{11}=\left(A_{11}-A_{12}A_{22}^{-1}A_{21}\right)^{-1},\,B_{12}=-B_{11}A_{12}A_{22}^{-1},\,B_{22}=\left(A_{22}-A_{21}A_{11}^{-1}A_{12}\right)^{-1}\label{EQ_partitionedInverse}
\end{align}
}(e.g. Lauritzen, 1996, eq. B.2).

The conclusions from Lemma \ref{Lemma_implicationsConditionEigenvals}
will be used in a number of places. Hence, we prove this first. 

\subsection{Proof of Lemma \ref{Lemma_implicationsConditionEigenvals}}

We prove one point at the time. 

\paragraph{Proof of Point 1. }

From the condition on $A$, we have that $Var\left(X_{t}\right)=\sum_{i=0}^{\infty}A^{i}\Sigma_{\varepsilon}\left(A'\right)^{i}$.
We note that 
\[
\sum_{i=0}^{\infty}{\rm eig}_{\min}\left(A^{i}\Sigma_{\varepsilon}\left(A'\right)^{i}\right)\leq{\rm eig}_{j}\left(\sum_{i=0}^{\infty}A^{i}\Sigma_{\varepsilon}\left(A'\right)^{i}\right)\leq\sum_{i=0}^{\infty}{\rm eig}_{\max}\left(A^{i}\Sigma_{\varepsilon}\left(A'\right)^{i}\right)
\]
$j=1,2,...,K$, where ${\rm eig}_{j}\left(\cdot\right)$, ${\rm eig}_{\min}\left(\cdot\right)$
and ${\rm eig}_{\max}\left(\cdot\right)$ are the $j^{th}$ eigenvalue,
the minimum and the maximum eigenvalue of the argument (Bhatia, 1996,
eq. III.13, using induction). Moreover, we have that 
\[
{\rm eig}_{\min}\left(\Sigma_{\varepsilon}\right){\rm eig}_{\min}\left(A^{i}\left(A'\right)^{i}\right)\leq{\rm eig}_{\min}\left(A^{i}\Sigma_{\varepsilon}\left(A'\right)^{i}\right)
\]
and 
\[
{\rm eig}_{\max}\left(A^{i}\Sigma_{\varepsilon}\left(A'\right)^{i}\right)\leq{\rm eig}_{\max}\left(A^{i}\left(A'\right)^{i}\right){\rm eig}_{\max}\left(\Sigma_{\varepsilon}\right).
\]
To see this note that
\[
\max_{x:x'x=1}x'A\Sigma_{\varepsilon}A'x\leq\max_{y:y'y=x'A'Ax}y'\Sigma_{\varepsilon}y={\rm eig}_{\max}\left(A'A\right){\rm eig}_{\max}\left(\Sigma_{\varepsilon}\right)
\]
and similarly for the lower bound and for $i>1$. Given that the eigenvalues
${\rm eig}_{j}\left(A'A\right)$ are in $\left(0,1\right)$ and the
eigenvalues ${\rm eig}_{j}\left(\Sigma_{\varepsilon}\right)$ are
in $\left(0,\infty\right)$ by assumption, we conclude that the eigenvalues
of $Var\left(X_{t}\right)$ are bounded away from zero and infinity,
uniformly in $K$.

\paragraph{Proof of Point 2. }

From the definition in (\ref{EQ_copulaScalingMatrixW}), we have the
following equality, \textbf{
\[
\Sigma=\left[\left(\begin{array}{cc}
I & \boldsymbol{0}\\
\boldsymbol{0} & I
\end{array}\right)+\left(\begin{array}{cc}
\boldsymbol{0} & A\\
A' & \boldsymbol{0}
\end{array}\right)\right]\left(\begin{array}{cc}
\Gamma & \boldsymbol{0}\\
\boldsymbol{0} & \Gamma
\end{array}\right),
\]
}where, here, $\boldsymbol{0}$ represents a $K\times K$ matrix of
zeros. From the assumption on $A$ and the fact that $\Gamma=Var\left(X_{t}\right)$,
we can use the definition of eigenvalues and, mutatis mutandis, the
previous inequalities, from the proof of Point 1, to deduce the result.

\paragraph{Proof of Point 3. }

From (\ref{EQ_partitionedInverse}) and the definition of $\Sigma$
as variance of $\left(Z_{t}',Z_{t-1}'\right)'$, we deduce that the
$\left(i,i\right)$ element in $\Theta_{11}$ is the inverse of the
variance of $Z_{t,i}$ conditioning on $Z_{t-1,i}$, all the other
variables and their first lag. Given that the eigenvalues of $\Sigma$
are bounded away from zero, uniformly in $K$, the random variables
are not perfectly correlated. Hence there must be a constant $\nu>0$
as in the statement of the lemma.

\paragraph{Proof of Point 4.}

The eigenvalues of $\Sigma_{\varepsilon}$ are in some compact interval
inside $\left(0,\infty\right)$, uniformly in $K$, by assumption.
Hence, the innovation vector has entries that are not perfectly dependent.
This means that no conditional correlation between any two variables
can be equal to one, uniformly in $K$.

\subsection{Proof of Proposition \ref{Proposition_strongMixing}}

It is clear that the process $X$ is a stationary Markov chain. The
mixing coefficients are invariant of monotone transformations of the
random variables. Hence, we can consider the mixing coefficients of
$Z$ in (\ref{EQ_gaussianVAR}). For the Gaussian VAR model in (\ref{EQ_gaussianVAR}),
Theorem 3.1 in Han and Wu (2019) says that the strong mixing coefficient
$\alpha\left(k\right)$ for variables $k$ periods apart satisfies
$\alpha\left(k\right)\leq c\left|A\right|_{{\rm op}}^{k}$ where $c$
is the square root of the ratio between the largest and smallest eigenvalue
of $Var\left(Z_{t}\right)$. This ratio is bounded by Lemma \ref{Lemma_implicationsConditionEigenvals}.
On the other hand, $\left|A\right|_{{\rm op}}$ is the largest singular
value of $A$, which is smaller than one, uniformly in $K$, by assumption.
Hence, the strong mixing coefficients decay exponentially fast.

\subsection{Proof of Lemmas \ref{Lemma_identificationVarParms} and \ref{Lemma_SVAR_identification}}

The conditions in Proposition \ref{Proposition_strongMixing} ensure
that the model is stationary. We use this with no explicit mention
in the following. 

\subsubsection{Proof of Lemma \ref{Lemma_identificationVarParms}}

This follows from (\ref{EQ_copulaScalingMatrixW}) and Lauritzen (1996,
eq. C3-C4) or from (\ref{EQ_partitionedInverse}).

\subsubsection{Proof of Lemma \ref{Lemma_SVAR_identification}}

By the assumption of the lemma, all edges of the graph of $\varepsilon_{t}$
are directed. There are also no cycles. Hence, there must be a permutation
matrix $\Pi$ of the elements in $\varepsilon_{t}$ such that the
$i$ element in $\Pi\varepsilon_{t}$ is not a parent of the $i-1$
element. This implies the structure $\Pi\varepsilon_{t}=H\xi_{t}$
where $H$ is a lower triangular matrix with diagonal entries equal
to one. Note that $H$ can have diagonal elements equal to one because
we are not assuming that $\mathbb{E}\xi_{t}\xi_{t}'$ is the identity.
The fact that the graph is acyclic means that $H$ is full rank. Otherwise,
we would have a descendant that is an ancestor of itself. Now note
that the inverse of a lower triangular matrix is also lower triangular.
Moreover, if the matrix has diagonal elements equal to one, also the
inverse has diagonal elements equal to one. Hence, we can write $H^{-1}=I-D$
where $D$ is as in the statement of the lemma and obtain (\ref{EQ_SVAR}).
To find the infinite moving average representation, rewrite (\ref{EQ_SVAR})
as $H^{-1}\Pi\left(I-AL\right)Z_{t}=\xi_{t}$ where, here, $L$ is
the lag operator. By assumption, $\left(I-AL\right)$ can be inverted
and has an infinite convergent series representation. Hence, we deduce
(\ref{EQ_maInfinitySVAR}) by standard algebra and the aforementioned
remarks on $H$.

\subsection{Exponential Inequality for Spearman's Rho}

For simplicity, we use notation that is local to this section only.
In this section we assume that $\left(\left(X_{t,i},X_{t,j}\right)\right)_{t\geq1}$
are real valued stationary random variables with exponentially decaying
strong mixing coefficients. We also assume that the variables have
continuous distribution function $F_{i}$ and $F_{j}$. As usual,
we denote by $\hat{F}_{i}$ and $\hat{F}_{j}$ the empirical distribution.
Let $R_{t,i}:=\sum_{t=1}^{n}1_{\left\{ X_{t,i}\leq X_{t,i}\right\} }$
be the rank of variable $X_{t,j}$ and similarly for $R_{t,j}$. In
Hoeffding (1948, p.318) we have that the sample version of Spearman's
rho is defined to be 
\begin{equation}
\hat{\rho}_{i,j}=\frac{12}{n^{3}-n}\sum_{t=1}^{n}\left(R_{t,i}-\frac{n+1}{2}\right)\left(R_{t,j}-\frac{n+1}{2}\right).\label{EQ_sampleSpearmanRho}
\end{equation}
This same statistic is also used in Liu et al. (2012, proof of Theorem
4.1). Note that other versions of of sample Spearman's rho can be
defined. These would essentially be equal to the above up to an additive
$O\left(n^{-1}\right)$ term. Their analysis can be treated in a way
similar to what follows. For simplicity, we only focus on the above. 

We recall the following Bernstein inequality from Merlev\`{e}de et
al. (2009) which we shall use twice. 

\begin{lemma}\label{Lemma_BernsteinInequ}Let $\left(Y_{t}\right)_{t\geq1}$
be a sequence of mean zero, stationary random variables whose absolute
value is uniformly bounded by $\bar{y}<\infty$, and with exponentially
decaying strong mixing coefficients. Then, for $n\geq4$ and $z\geq0$,
there is a constant $c_{1}>0$, depending on the mixing coefficients
only and such that 
\[
\Pr\left(\left|\frac{1}{n}\sum_{t=1}^{n}Y_{t}\right|\geq z\right)\leq\exp\left\{ -\frac{c_{1}nz^{2}}{\bar{y}^{2}+z\bar{y}\ln n\left(\ln\ln n\right)}\right\} .
\]

\end{lemma}

A general main ingredient for our derivation of an exponential inequality
for (\ref{EQ_sampleSpearmanRho}) is the following. 

\begin{lemma}\label{Lemma_exponentialIneqEDF} Under the assumptions
of this section, choose a $c_{2}\in\left(0,\infty\right)$, and let
$z:=y-n^{-c_{2}}$ for any $y\geq n^{-c_{2}}$. Then, there is a constant
$c_{1}>0$ such that

\[
\Pr\left(\sup_{x\in\mathbb{R}}\left|\hat{F}_{i}\left(x\right)-F_{i}\left(x\right)\right|\geq y\right)\leq2\exp\left\{ -\frac{c_{1}nz^{2}}{1+z\ln n\left(\ln\ln n\right)}+c_{2}\ln n\right\} .
\]

\end{lemma}

\begin{proof}We can always find a continuous monotone transformation
$x\mapsto g\left(x\right)\in\left[0,1\right]$ for $x$ in the range
of $X_{t,i}$. Hence, given that $1_{\left\{ X_{t,i}\leq x\right\} }=1_{\left\{ g\left(X_{t,i}\right)\leq g\left(x\right)\right\} }$,
we can assume that $X_{t,i}\in\left[0,1\right]$ for the purpose of
the proof. Note that continuity of $g$ does not mean that $g\left(X_{t,i}\right)$
is a continuous random variable. Using standard techniques, we replace
the supremum by the maximum over a finite number of elements. We then
apply Lemma \ref{Lemma_BernsteinInequ}.

To do so, for fixed but arbitrary $\epsilon>0$, we construct intervals
$\left[x_{l}^{L},x_{l}^{U}\right]$, $l=1,2,...,N\left(\epsilon\right)$,
such that $\left|F_{i}\left(x\right)-F_{i}\left(z\right)\right|\leq\epsilon$
for $x,z\in\left[x_{l}^{L},x_{l}^{U}\right]$. The construction is
as follows and similar to the one of the Lebesgue integral. Fix an
arbitrary $\epsilon>0$ and divide the interval $\left[0,1\right]$
into $N\left(\epsilon\right)$ intervals $\left[t_{l-1},t_{l}\right]$
where $0=t_{0}<t_{1}<\cdots<t_{N\left(\epsilon\right)}=1$ such $t_{l}-t_{l-1}\leq\epsilon^{-1}$
. Then, $N\left(\epsilon\right)$ is the smallest integer greater
than or equal to $\epsilon^{-1}$. Define variables $0\leq x_{1}^{L}\leq x_{2}^{L}\leq\cdots\leq x_{N\left(\epsilon\right)}^{L}=1$
as $x_{l}^{L}:=\inf\left\{ x>0:F_{i}\left(x\right)\geq t_{l-1}\right\} $.
Similarly, define variables $0\leq x_{1}^{U}\leq x_{2}^{U}\leq\cdots\leq x_{N\left(\epsilon\right)}^{U}=1$
as $x_{l}^{U}:=\sup\left\{ x\leq1:F_{i}\left(x\right)\leq t_{l}\right\} $.
It is not difficult to see that this construction has the aforementioned
properties. Note that we can have $\left[x_{l}^{L},x_{l}^{U}\right]$
equal to a singleton, i.e. $x_{l}^{L}=x_{l}^{U}$, if there are discontinuities
in $F_{i}$ and such discontinuities are larger than $\epsilon$. 

The following is a standard argument in the proof of the Glivenko-Cantelli
Theorem (van der Vaart and Wellener, 2000, proof of Theorem 2.4.1).
From the fact that $F_{i}\left(x\right)$ and $\hat{F}_{i}\left(x\right)$
are monotonically increasing, we have that $F_{i}\left(x_{l}^{L}\right)\leq F_{i}\left(x\right)\leq F_{i}\left(x_{l}^{U}\right)$
and $\hat{F}_{i}\left(x_{l}^{L}\right)\leq\hat{F}_{i}\left(x\right)\leq\hat{F}_{i}\left(x_{l}^{U}\right)$
for $x\in\left[x_{l}^{L},x_{l}^{U}\right]$. Also recall that $\mathbb{E}\hat{F}_{i}\left(x\right)=F_{i}\left(x\right)$.
In consequence, 
\begin{align*}
\max_{x\in\left[x_{l}^{L},x_{l}^{U}\right]}\left(\hat{F}_{i}\left(x\right)-F_{i}\left(x\right)\right)= & \max_{x\in\left[x_{l}^{L},x_{l}^{U}\right]}\left(1-\mathbb{E}\right)\hat{F}_{i}\left(x\right)\leq\left(1-\mathbb{E}\right)\hat{F}_{i}\left(x_{l}^{U}\right)\\
 & +\max_{x\in\left[x_{l}^{L},x_{l}^{U}\right]}\mathbb{E}\left(\hat{F}_{i}\left(x_{l}^{U}\right)-\hat{F}_{i}\left(x\right)\right).\\
\leq & \left(1-\mathbb{E}\right)\hat{F}_{i}\left(x_{l}^{U}\right)+\epsilon
\end{align*}
using monotonicity and the fact that $\left|F_{i}\left(x_{l}^{U}\right)-F_{i}\left(x_{l}^{L}\right)\right|\leq\epsilon$
by construction. In consequence, 
\[
\max_{x\in\left[0,1\right]}\left(\hat{F}_{i}\left(x\right)-F_{i}\left(x\right)\right)\leq\max_{l\in\left\{ 1,2,...,N\left(\epsilon\right)\right\} }\left(1-\mathbb{E}\right)\hat{F}_{i}\left(x_{l}^{U}\right)+\epsilon.
\]
Hence, using the union bound, 
\[
\Pr\left(\max_{x\in\left[0,1\right]}\left(\hat{F}_{i}\left(x\right)-F_{i}\left(x\right)\right)\geq y\right)\leq N\left(\epsilon\right)\max_{x\in\left[0,1\right]}\Pr\left(\left(1-\mathbb{E}\right)\hat{F}_{i}\left(x\right)\geq y-\epsilon\right).
\]
Set $\epsilon=n^{-c_{2}}$. Apply Lemma \ref{Lemma_BernsteinInequ}
with $Y_{t}=\left(1-\mathbb{E}\right)1_{\left\{ X_{t,i}\leq x\right\} }$
for arbitrary, but fixed $x$, and $z:=y-\epsilon=y-n^{-c_{2}}$.
A similar inequality holds for $\min_{x\in\left[x_{l}^{L},x_{l}^{U}\right]}$.
Hence, we deduce the final result.\end{proof}

Control of the quantity below will be shown to be essentially equivalent
to control of (\ref{EQ_sampleSpearmanRho}).

\begin{lemma}\label{Lemma_sampleJointExpectationECDF}Under the assumptions
of this section, choose a $c_{2}\in\left(0,\infty\right)$, and let
$z:=y-n^{-c_{2}}$ for any $y\geq n^{-c_{2}}$. Then, there is a constant
$c_{1}>0$ such that
\begin{align*}
 & \Pr\left(\left|\frac{1}{n}\sum_{i=1}^{n}\left(\hat{F}_{i}\left(X_{t,i}\right)\hat{F}_{j}\left(X_{t,i}\right)-\mathbb{E}F_{i}\left(X_{t,i}\right)F_{j}\left(X_{t,i}\right)\right)\right|\geq y\right)\\
\leq & 5\exp\left\{ -\frac{c_{1}nz^{2}}{1+z\ln n\left(\ln\ln n\right)}+c_{2}\ln n\right\} .
\end{align*}
\end{lemma}

\begin{proof}By the triangle inequality and the uniform boundedness
of the empirical distribution function, 
\begin{align*}
 & \left|\frac{1}{n}\sum_{i=1}^{n}\left(\hat{F}_{i}\left(X_{t,i}\right)\hat{F}_{j}\left(X_{t,i}\right)-\mathbb{E}F_{i}\left(X_{t,i}\right)F_{j}\left(X_{t,i}\right)\right)\right|\\
 & \leq\left|\frac{1}{n}\sum_{i=1}^{n}\left(\hat{F}_{i}\left(X_{t,i}\right)-F_{i}\left(X_{t,i}\right)\right)\right|+\left|\frac{1}{n}\sum_{i=1}^{n}\left(\hat{F}_{j}\left(X_{t,j}\right)-F_{j}\left(X_{t,j}\right)\right)\right|\\
 & \left|\frac{1}{n}\sum_{i=1}^{n}\left(F_{i}\left(X_{t,i}\right)F_{j}\left(X_{t,i}\right)-\mathbb{E}F_{i}\left(X_{t,i}\right)F_{j}\left(X_{t,i}\right)\right)\right|.
\end{align*}
We apply Lemma \ref{Lemma_exponentialIneqEDF} to the first two terms
on the r.h.s. and Lemma \ref{Lemma_BernsteinInequ} to the last one
to deduce the result. \end{proof}

The definition of the population version of Spearman's rho (e.g.,
Joe, 1997, p.32) between two random variables with joint distribution
$F_{i,j}$ and marginals $F_{i}$ and $F_{j}$ is $\rho_{i,j}=12\int\int F_{i}\left(x\right)F_{j}\left(x\right)dF_{i,j}\left(x,y\right)-3$.
Hence, we have the following.

\begin{lemma}\label{Lemma_expIneqHanModified}Under the assumptions
of this section, there is a constant $c_{1}>0$ such that for $n$
large enough and any $x\geq6/n$,
\[
\Pr\left(\max_{i,j\leq K}\left|\hat{\rho}_{i,j}-\rho_{i,j}\right|\geq x\right)\leq5\exp\left\{ -\frac{c_{1}nx^{2}}{4\left(1+x\ln n\left(\ln\ln n\right)\right)}+\ln n+2\ln K\right\} .
\]
\end{lemma}

\begin{proof}Dividing and multiplying by $n^{2}$, (\ref{EQ_sampleSpearmanRho})
is equal to 
\[
\frac{12n}{n^{2}-1}\sum_{t=1}^{n}\left(\hat{F}_{i}\left(X_{t,i}\right)-\frac{n+1}{2n}\right)\left(\hat{F}_{i}\left(X_{t,i}\right)-\frac{n+1}{2n}\right).
\]
Again, by simple algebra, the triangle inequality and the fact that
$\hat{F}_{i}$ has range in $\left[0,1\right]$, we have that for
$n$ large enough, e.g. $n\geq24$,
\[
\left|\hat{\rho}_{i,j}-\frac{12}{n}\sum_{i=1}^{n}\left(\hat{F}_{i}\left(X_{t,i}\right)\hat{F}_{j}\left(X_{t,i}\right)-\frac{1}{4}\right)\right|\leq\frac{24}{n}.
\]
In consequence, 
\[
\Pr\left(\left|\hat{\rho}_{i,j}-\rho_{i,j}\right|\geq x\right)\leq\Pr\left(\frac{1}{n}\sum_{i=1}^{n}\left(\hat{F}_{i}\left(X_{t,i}\right)\hat{F}_{j}\left(X_{t,i}\right)-\mathbb{E}F_{i}\left(X_{t,i}\right)F_{j}\left(X_{t,i}\right)\right)\geq x-\frac{2}{n}\right).
\]
We can then apply Lemma \ref{Lemma_sampleJointExpectationECDF} with
$c_{2}=1$ and $y=x-2n^{-1}$ to the r.h.s. of the above display.
In Lemma \ref{Lemma_sampleJointExpectationECDF} for $x\geq6/n$,
we have $z=\left(x-\frac{2}{n}\right)-\frac{1}{n}$ which implies
that $z\in\left[x/2,x\right]$. In Lemma \ref{Lemma_sampleJointExpectationECDF},
replace $z$ its lower bound and upper bound in the numerator and
denominator of the exponential function to deduce the result.\end{proof}

\subsection{Lemmas on Control of the Sample Covariance Estimator and Related
Quantities\label{Section_LemmasSampleCov}}

To avoid notational trivialities, suppose that $K\geq n$. If not,
replace $K$ with $n$ in what follows. Recall that $\rho_{i,j}$
is the rank correlation between $W_{t,i}$ and $W_{t,j}$. By stationarity,
this does not depend on $t$. We have the following.

\begin{lemma}\label{Lemma_SpearmanBound}Under the Assumptions, for
$n$ large enough, there is a finite constant $c_{0}$ such that
\[
\Pr\left(\max_{i,j\leq K}\left|\hat{\rho}_{i,j}-\rho_{i,j}\right|\geq c_{0}\sqrt{\frac{\ln K}{n}}\right)\leq K^{-1}.
\]

\end{lemma}

\begin{proof}This follows from the inequality in Lemma \ref{Lemma_expIneqHanModified}.
There, we set $x^{2}=32\ln\left(K\right)/\left(c_{1}n\right)$ to
deduce that for $c_{0}=\sqrt{32/c_{1}}$, 
\[
\Pr\left(\max_{i,j\leq K}\left|\hat{\rho}_{i,j}-\rho_{i,j}\right|\geq c_{0}\sqrt{\frac{\ln K}{n}}\right)\leq5\exp\left\{ -\frac{8\left(\ln K\right)-3\left(1+\epsilon\right)\ln K}{1+\epsilon}\right\} 
\]
for $\epsilon=\sqrt{32\ln\left(K\right)/\left(c_{1}n\right)}\left(\ln n\right)\left(\ln\ln n\right)$.
Under the Assumptions, for $n$ large enough, $\epsilon\le1$. Substituting
in the above display we find that the r.h.s. is bounded above by $K^{-1}$
and this proves the lemma.\end{proof}

We now show that the correlation matrix obtained from Spearman's rho
converges.

\begin{lemma}\label{Lemma_CorrelationBound}Under the Assumptions,
for $n$ large enough, there is a constant $c_{0}$ (the same as in
Lemma \ref{Lemma_SpearmanBound}), such that, 
\[
\Pr\left(\max_{i,j\leq K}\left|\hat{\Sigma}_{i,j}-\Sigma_{i,j}\right|\geq\frac{3c_{0}}{\pi}\sqrt{\frac{\ln K}{n}}\right)\leq K^{-1}.
\]

\end{lemma}

\begin{proof}From Lemma \ref{Lemma_transformsProperties} we have
that $\hat{\Sigma}_{i,j}-\Sigma_{i,j}=2\sin\left(\frac{\pi}{6}\hat{\rho}_{i,j}\right)-2\sin\left(\frac{\pi}{6}\rho_{i,j}\right)$.
If the variables were not continuous, we would need to use another
transformation (see the remarks in Section \ref{Section_remarksGaussianTransform}).
Given that $\sin\left(x\right)$ is Lipschitz with constant one, the
result follows from Lemma \ref{Lemma_SpearmanBound}. \end{proof}

\begin{lemma}\label{Lemma_CorrExpBound}Suppose that the Assumptions
hold. Then, there is a constant $c_{3}>0$, such that, for $n$ large
enough,
\[
\max_{i,j\leq K}\Pr\left(\left|\hat{\Sigma}_{i,j}-\Sigma_{i,j}\right|\geq x\right)\leq\exp\left\{ -nc_{3}x^{2}\right\} 
\]
for any $x$ satisfying $xn^{1/2}\rightarrow\infty$ and $x\left(\ln n\right)\left(\ln\ln n\right)\rightarrow0$.

\end{lemma}

\begin{proof}This follows from the remarks in the proof of Lemma
\ref{Lemma_SpearmanBound} and then an application of Lemma \ref{Lemma_expIneqHanModified}using
the constraints on $x$.\end{proof}

\subsection{Lemmas for the Control of the Precision Matrix Estimator}

The following result for the control of the operator norm will be
used in the proofs.

\begin{lemma}\label{Lemma_inverseMatrixConvergence}Suppose that
$\hat{Q}$ and $Q$ are symmetric matrices such that $Q$ has eigenvalues
bounded away from zero an infinity. If $\left|\hat{Q}-Q\right|_{{\rm op}}=\epsilon$,
then $\left|\hat{Q}^{-1}-Q^{-1}\right|_{{\rm op}}=O\left(\left|Q^{-1}\right|_{{\rm op}}^{2}\epsilon\right)$
as long as $\left|Q^{-1}\right|_{{\rm op}}<\epsilon^{-1}$.\end{lemma}

\begin{proof}With the present notation, Lemma 4 Le and Zhong (2021)
says that 
\begin{equation}
\left|\hat{Q}^{-1}-Q^{-1}\right|_{{\rm op}}\leq\left|Q^{-1}\right|_{{\rm op}}\frac{\left|Q^{-1}\left(\hat{Q}-Q\right)\right|_{{\rm op}}}{1-\left|Q^{-1}\left(\hat{Q}-Q\right)\right|_{{\rm op}}}.\label{EQ_LeZhongLemma4}
\end{equation}
Then, the result follows from the fact that $\left|Q^{-1}\left(\hat{Q}-Q\right)\right|_{{\rm op}}\leq\left|Q^{-1}\right|_{{\rm op}}\left|\hat{Q}-Q\right|_{{\rm op}}$
together with the condition of the lemma to ensure that the denominator
is greater than zero.\end{proof}

The operator norm can be bounded by the uniform norm of the elements
using the following.

\begin{lemma}\label{Lemma_OperatorNormUniformNormBound}Suppose that
$\hat{Q}$ and $Q$ are symmetric matrices. Then, $\left|\hat{Q}-Q\right|_{{\rm op}}\leq\left|\hat{Q}-Q\right|_{0,\infty}\left|\hat{Q}-Q\right|_{\infty}$.\end{lemma}

\begin{proof}First, note that $\left|\hat{Q}-Q\right|_{{\rm op}}\leq\left|\hat{Q}-Q\right|_{1,\infty}$
because $\hat{Q}-Q$ is symmetric. This is well known because, for
any matrix $A$ (not to be confused with the autoregressive matrix
in (\ref{EQ_gaussianVAR})), $A'Ax=\sigma^{2}x$ where $\sigma^{2}$
is the maximum eigenvalue of $A'A$ and $x$ is the corresponding
eigenvector. Hence, $\sigma^{2}\left|x\right|_{\infty}=\left|A'Ax\right|_{\infty}$.
By a special case of Holder inequality, $\left|A'Ax\right|_{\infty}\leq\left|A'\right|_{\infty,1}\left|A\right|_{\infty,1}\left|x\right|_{\infty}$.
This implies that $\sigma^{2}=\left|A\right|_{{\rm op}}^{2}\leq\left|A\right|_{1,\infty}\left|A\right|_{\infty,1}$.
Then, using the fact that, in our case, $A=\hat{Q}-Q$ is symmetric,
we deduce the inequality at the start of the proof. Moreover, $\left|\hat{Q}-Q\right|_{1,\infty}\leq\left|\hat{Q}-Q\right|_{0,\infty}\left|\hat{Q}-Q\right|_{\infty}$
because $\left|\hat{Q}-Q\right|_{0,\infty}$ is the maximum number
of nonzero elements across the columns of $\hat{Q}-Q$. \end{proof}

Define the event 
\begin{equation}
E:=\left\{ 1_{\left\{ \hat{\Theta}_{i,j}>0\right\} }=1_{\left\{ \Theta_{i,j}>0\right\} }\right\} \label{EQ_eventSetIdentification}
\end{equation}
We shall derive a number of results conditional on such event. The
event $E$ means that $\left\{ \hat{B}_{i}:i\in\left[2K\right]\right\} $
in Algorithm \ref{Algo_sparseCopulaParametersEstimation} correctly
identifies all the nonzero entries in $\Theta$. The next result can
be found in the proof of Theorem 3 in Le and Zhong (2021).

\begin{lemma}\label{Lemma_LeZhongTh3} Suppose that the Assumptions
hold. On the event (\ref{EQ_eventSetIdentification}), there is a
constant $c_{4}$ such that 
\begin{equation}
\Pr\left(\left|\hat{\Theta}-\Theta\right|_{\infty}\geq z\right)\leq2K\Pr\left(\left|\hat{\Sigma}-\Sigma\right|_{\infty}\geq zc_{4}\right).\label{EQ_lemmaLeZhongTh3}
\end{equation}
 \end{lemma}

We can now use the lemmas from Section \ref{Section_LemmasSampleCov}.

\begin{lemma}\label{Lemma_ExpThetaIneq}Suppose that the Assumptions
hold. On the event (\ref{EQ_eventSetIdentification}), there is a
constant $c_{5}>0$, such that, for $n$ large enough,
\[
\Pr\left(\left|\hat{\Theta}-\Theta\right|_{\infty}\geq z\right)\leq2\exp\left\{ -nc_{5}z^{2}+3\ln K\right\} 
\]
for any $z$ satisfying $zn^{1/2}\rightarrow\infty$ and $z\left(\ln n\right)\left(\ln\ln n\right)\rightarrow0$.
Moreover, $\left|\hat{\Theta}-\Theta\right|_{\infty}=O_{P}\left(\sqrt{\frac{\ln K}{n}}\right)$.\end{lemma}

\begin{proof}We bound the r.h.s. in the display of Lemma \ref{Lemma_LeZhongTh3}
using Lemma \ref{Lemma_CorrExpBound} and the union bound. We can
then deduce that the r.h.s. of (\ref{EQ_lemmaLeZhongTh3}) is bounded
above by $2K^{3}\exp\left\{ -nc_{3}c_{4}^{2}z^{2}\right\} $. Defining
$c_{5}:=c_{3}c_{4}^{2}$ and rearranging we deduce the first statement.
The second statement follows by choosing $z$ large enough and proportional
to a quantity $O\left(\sqrt{\frac{\ln K}{n}}\right)$ so that the
first statement immediately gives that $\left|\hat{\Theta}-\Theta\right|_{\infty}=O_{P}\left(\sqrt{\frac{\ln K}{n}}\right)$.
Such choice of $z$ is consistent with the constraint given in the
lemma.\end{proof}

We also need an exponential inequality for $\hat{\Theta}_{11}^{-1}-\Theta_{11}^{-1}$.
For simplicity, we state the result for $\hat{\Theta}^{-1}$ rather
than $\hat{\Theta}_{11}^{-1}$.

\begin{lemma}\label{Lemma_ExpThetaInverseIneq}Suppose that the Assumptions
hold and that $s\sqrt{\ln K/n}=o\left(1\right)$. On the event (\ref{EQ_eventSetIdentification}),
there is a constant $c_{6}>0$ such that, for $n$ large enough, 
\[
\Pr\left(\left|\hat{\Theta}^{-1}-\Theta^{-1}\right|_{\infty}\geq z\right)\leq2\exp\left\{ -ns^{-2}c_{6}z^{2}+3\ln K\right\} 
\]
for any $z$ satisfying $zn^{1/2}\rightarrow\infty$ and $z\left(\ln n\right)\left(\ln\ln n\right)\rightarrow0$.\end{lemma}

\begin{proof}First, we note that for any symmetric matrix $Q$, $\left|Q\right|_{\infty}\leq\left|Q\right|_{{\rm op}}$.
This is because $\left|Q\right|_{{\rm op}}=\max_{x,y}x'Qy$ where
the maximum is over vectors with unit Euclidean norm. By this remark
and (\ref{EQ_LeZhongLemma4}) we deduce that the set $\left\{ \left|\hat{\Theta}^{-1}-\Theta^{-1}\right|_{\infty}\geq z\right\} $
is contained in the set 
\[
\left\{ \left|\Theta^{-1}\right|_{{\rm op}}\frac{\left|\Theta^{-1}\left(\hat{\Theta}-\Theta\right)\right|_{{\rm op}}}{1-\left|\Theta^{-1}\left(\hat{\Theta}-\Theta\right)\right|_{{\rm op}}}\geq z\right\} .
\]
For arbitrary events $A$ and $B$, we shall use the trivial decomposition
$A=\left\{ A\cap B\right\} \cup\left\{ A\cap B^{c}\right\} \subseteq\left\{ A\cap B\right\} \cup B^{c}$,
where $B^{c}$ is the complement of $B$. Then, we deduce that the
event in the above display is contained in the event 
\begin{equation}
\left\{ \left|\Theta^{-1}\left(\hat{\Theta}-\Theta\right)\right|_{{\rm op}}\geq1/2\right\} \cup\left\{ \left|\Theta^{-1}\right|_{{\rm op}}\left|\Theta^{-1}\left(\hat{\Theta}-\Theta\right)\right|_{{\rm op}}\geq z/2\right\} \label{EQ_eventsBoundInverse}
\end{equation}
For $z/\left|\Theta^{-1}\right|_{{\rm op}}\rightarrow0$, the above
union of two events is contained in the second event. This is the
case because the eigenvalues of $\Theta$ are bounded away from zero
and infinity by Lemma \ref{Lemma_implicationsConditionEigenvals}.
Hence, it is sufficient to bound the latter. Using a standard inequality
for operator norms, and then Lemma \ref{Lemma_OperatorNormUniformNormBound},
we deduce that 
\[
\left|\Theta^{-1}\left(\hat{\Theta}-\Theta\right)\right|_{{\rm op}}\leq\left|\Theta^{-1}\right|_{{\rm op}}\left|\left(\hat{\Theta}-\Theta\right)\right|_{0,\infty}\left|\left(\hat{\Theta}-\Theta\right)\right|_{\infty}.
\]
On the event $E$ in (\ref{EQ_eventSetIdentification}), $\left|\left(\hat{\Theta}-\Theta\right)\right|_{0,\infty}\leq\left|\Theta\right|_{0,\infty}\leq s$.
We assume $E$ holds without making it explicit in the notation. In
consequence, recalling that, by Lemma \ref{Lemma_implicationsConditionEigenvals},
$\sigma_{\max}$ is the largest singular value of $\Theta^{-1}=\Sigma$,
which is bounded uniformly in $K$, we have that
\[
\Pr\left(\left|\Theta^{-1}\right|_{{\rm op}}\left|\Theta^{-1}\left(\hat{\Theta}-\Theta\right)\right|_{{\rm op}}\geq z/2\right)\leq\Pr\left(\left|\left(\hat{\Theta}-\Theta\right)\right|_{\infty}\geq z/\left(2\sigma_{\max}^{2}s\right)\right).
\]
By Lemma \ref{Lemma_ExpThetaIneq} and the conditions of the present
lemma, the r.h.s. is bounded above by $2\exp\left\{ -nc_{5}z^{2}/\left(2\sigma_{\max}^{2}s\right)^{2}+3\ln K\right\} $.
Setting $c_{6}=c_{5}/\left(4\sigma_{\max}^{4}\right)$, which is strictly
positive, gives the result. \end{proof}

The following result will be used in due course.

\begin{lemma}\label{Lemma_IneqCorrelationsFromCov} Suppose that
$U$, $V_{1}$, $V_{2}$ and $\hat{U}$, $\hat{V}_{1}$, $\hat{V}_{2}$
are random variables. Then, the event $\left\{ \left|\frac{\hat{U}}{\hat{V}_{1}\hat{V}_{2}}-\frac{U}{V_{1}V_{2}}\right|\geq x\right\} $
is contained in the union of the following three events: $\left\{ \left|\frac{\hat{U}\left(\hat{V}_{1}-V_{1}\right)}{\hat{V}_{1}V_{1}V_{2}}\right|\ge x/4\right\} $,
$\left\{ \left|\frac{\hat{U}\left(\hat{V}_{2}-V_{2}\right)}{\hat{V}_{1}\hat{V}_{2}V_{2}}\right|\geq x/4\right\} $
and $\left\{ \left|\frac{\hat{U}-U}{V_{1}V_{2}}\right|\geq x/2\right\} $.
\end{lemma}

\begin{proof}Add and subtract $\frac{\hat{U}}{V_{1}V_{2}}$ to find
that 
\[
\frac{\hat{U}}{\hat{V}_{1}\hat{V}_{2}}-\frac{U}{V_{1}V_{2}}=\left(\frac{\hat{U}}{\hat{V}_{1}\hat{V}_{2}}-\frac{\hat{U}}{V_{1}V_{2}}\right)+\left(\frac{\hat{U}}{V_{1}V_{2}}-\frac{U}{V_{1}V_{2}}\right).
\]
The first term on the r.h.s. can be written as 
\[
\left(\frac{\hat{U}}{\hat{V}_{1}\hat{V}_{2}}-\frac{\hat{U}}{V_{1}V_{2}}\right)=\left(\frac{\hat{U}}{\hat{V}_{1}\hat{V}_{2}V_{1}V_{2}}\right)\left[\hat{V}_{2}\left(\hat{V}_{1}-V_{1}\right)+V_{1}\left(\hat{V}_{2}-V_{2}\right)\right].
\]
We can then deduce the statement of the lemma by basic set inequalities.\end{proof}

Let $\hat{\Xi}_{i,j}=\hat{\Sigma}_{\varepsilon,i,j}/\sqrt{\hat{\Sigma}_{\varepsilon,i,i}\hat{\Sigma}_{\varepsilon,j,j}}$
and similarly for $\Xi_{i,j}$ using $\Sigma_{\varepsilon}$ in place
of $\hat{\Sigma}_{\varepsilon}$. These are estimated and population
correlation coefficients between $\varepsilon_{t,i}$ and $\varepsilon_{t,j}$.

\begin{lemma}\label{Lemma_corrEpsilonExpIneq} Suppose that the Assumptions
hold. There is a constant $c_{7}>0$, such that, for $n$ large enough,
\[
\max_{i,j\leq K}\Pr\left(\left|\hat{\Xi}_{i,j}-\Xi_{i,j|k}\right|\geq z\right)\leq16\exp\left\{ -ns^{-2}c_{7}z^{2}+3\ln K\right\} 
\]
for any $z$ satisfying $zn\rightarrow\infty$ and $z\left(\ln n\right)\left(\ln\ln n\right)\rightarrow0$.\end{lemma}

\begin{proof}We apply Lemma \ref{Lemma_IneqCorrelationsFromCov}
to deduce that we need to bound the following probabilities
\[
\Pr\left(E_{1}\right):=\Pr\left(\left|\frac{\hat{\Sigma}_{\varepsilon,i,j}\left(\hat{\Sigma}_{\varepsilon,i,i}-\Sigma_{\varepsilon,i,i}\right)}{\sqrt{\hat{\Sigma}_{\varepsilon,i,i}\Sigma_{\varepsilon,i,i}\Sigma_{\varepsilon,j,j}}}\right|\geq z/4\right),
\]
\[
\Pr\left(E_{2}\right):=\Pr\left(\left|\frac{\hat{\Sigma}_{\varepsilon,i,j}\left(\hat{\Sigma}_{\varepsilon,j,j}-\Sigma_{\varepsilon,j,j}\right)}{\sqrt{\hat{\Sigma}_{\varepsilon,i,i}\hat{\Sigma}_{\varepsilon,j,j}\Sigma_{\varepsilon,j,j}}}\right|\geq z/4\right)
\]
and
\[
\Pr\left(E_{3}\right):=\Pr\left(\left|\frac{\hat{\Sigma}_{\varepsilon,i,j}\left(\hat{\Sigma}_{\varepsilon,i,j}-\Sigma_{\varepsilon,i,j}\right)}{\sqrt{\Sigma_{\varepsilon,i,i}\Sigma_{\varepsilon,j,j}}}\right|\geq z/2\right).
\]
We further define the following events: $E_{4}:=\left\{ \max_{i,j\leq K}\left|\hat{\Sigma}_{\varepsilon,i,j}\right|\leq3/2\right\} $,
and $E_{5}:=\left\{ \min_{i\leq K}\hat{\Sigma}_{\varepsilon,i,i}\geq\sigma_{\min}/2\right\} $
where $\sigma_{\min}>0$ is the minimum eigenvalue of $\Sigma$, by
Lemma \ref{Lemma_implicationsConditionEigenvals}. Then, $\Pr\left(E_{1}\right)\leq\Pr\left(E_{1}\cap E_{4}\cap E_{5}\right)+\Pr\left(E_{4}^{c}\right)+\Pr\left(E_{5}^{c}\right)$
where, as usual, the superscript $c$ is used to denote the complement
of a set. Before bounding each term separately, we note that by the
Cauchy interlacing theorem (Bhatia, 1996, Corollary III. 1.5), the
smallest eigenvalue of $\Sigma_{\varepsilon}$ is no smaller than
$\sigma_{\min}$. Moreover, $\Sigma_{\varepsilon,i,i}\geq\sigma_{\min}$.
To see this note that the l.h.s. is equal to $e_{i}'\Sigma_{\varepsilon}e_{i}$,
where $e_{i}$ is the vector with $i^{th}$ entry equal to one and
all other entries equal to zero. On the other hand the r.h.s. is smaller
than $\min_{x:x'x=1}x'\Sigma_{\varepsilon}x$ by the definition of
minimum eigenvalue and the Cauchy's interlacing theorem. Now, 
\begin{align}
\Pr\left(E_{1}\cap E_{4}\cap E_{5}\right)\leq & \Pr\left(\left|3\sigma_{\min}^{-3/2}\left(\hat{\Sigma}_{\varepsilon,i,i}-\Sigma_{\varepsilon,i,i}\right)\right|\geq z/4\right)\nonumber \\
\leq & 2\exp\left\{ -ns^{-2}12^{-2}\sigma_{\min}^{3}c_{6}z^{2}+3\ln K\right\} \label{EQ_corrEpsilonIneq1}
\end{align}
using the bounds implied by the events $E_{4}$ and $E_{5}$, the
aforementioned remarks on $\Sigma_{\varepsilon,i,i}$, and then Lemma
\ref{Lemma_ExpThetaInverseIneq}. Noting that $\hat{\Sigma}_{\varepsilon,i,j}\leq\Sigma_{\varepsilon,i,j}+\left|\text{\ensuremath{\hat{\Sigma}_{\varepsilon,i,j}}-\ensuremath{\Sigma_{\varepsilon,i,j}}}\right|$
and that $\left|\Sigma_{\varepsilon,i,j}\right|\leq1$ because $\varepsilon_{t}$
is the innovation of the variable $Z_{t}$ with entries having variance
one, we deduce that $\Pr\left(E_{4}^{c}\right)\leq\Pr\left(\left|\text{\ensuremath{\hat{\Sigma}_{\varepsilon,i,j}}-\ensuremath{\Sigma_{\varepsilon,i,j}}}\right|\geq1/2\right)$
and this probability is eventually bounded by (\ref{EQ_corrEpsilonIneq1})
as long as $z\rightarrow0$. By the same argument used to bound $\Pr\left(E_{4}^{c}\right)$,
we deduce that $\Pr\left(E_{5}^{c}\right)$ is eventually less than
(\ref{EQ_corrEpsilonIneq1}). Hence, $\Pr\left(E_{1}\right)$ is bounded
by three times the r.h.s. of (\ref{EQ_corrEpsilonIneq1}) for $n$
large enough. By similar arguments, we also note that $\Pr\left(E_{2}\right)$
and $\Pr\left(E_{3}\right)$ are bounded by three and two times, respectively,
the r.h.s. of (\ref{EQ_corrEpsilonIneq1}). Putting everything together,
and setting $c_{7}:=12^{-2}\sigma_{\min}^{3}c_{6}$, the result follows.\end{proof}

For any set $\mathbf{k}\subset\left[K\right]$ we let $\hat{\Xi}_{i,j|\mathbf{k}}$
be the correlation of $\varepsilon_{t,i}$ with $\varepsilon_{t,j}$
conditioning on $\left\{ \varepsilon_{t,l}:l\in\mathbf{k}\right\} $.

\begin{lemma}\label{Lemma_ParCorrExpBound} Under the Assumptions,
there is a constant $c_{7}>0$ (same as in Lemma \ref{Lemma_corrEpsilonExpIneq}),
such that, for $n$ large enough, 
\[
\max_{i,j\leq K,\mathbf{k}\in\mathcal{K}_{i,j}}\Pr\left(\left|\hat{\Xi}_{i,j|\mathbf{k}}-\Xi_{i,j|\mathbf{k}}\right|\geq z\right)\leq16\exp\left\{ -\left(n-m\right)s^{-2}c_{7}z^{2}+3\ln K\right\} 
\]
for $\mathcal{K}_{i,j}\subseteq\left\{ \left[K\right]\setminus\left\{ i,j\right\} \right\} $
of cardinality $m$ and $z$ satisfying 
\[
z\left(n-m\right)\rightarrow\infty\text{ and }z\left(\ln\left(n-m\right)\right)\left(\ln\ln\left(n-m\right)\right)\rightarrow0.
\]
\end{lemma}

\begin{proof}By Lemma 2 in Kalisch and B\"{u}hlmann (2007) if the
distribution of the sample correlation coefficient is $f\left(x;n\right)$
where $n$ is the sample size, the distribution of the partial correlation
coefficient is the same with $n$ replaced by $n-m$, i.e. $f\left(x;n-m\right)$.
Hence, we can use Lemma \ref{Lemma_CorrExpBound} with $n$ replaced
by $n-m$ everywhere and the lemma is proved. \end{proof}

The next is a trivial variation of lemma 3 in Kalisch and B\"{u}hlmann
(2007) adapted to our inequalities.

\begin{lemma}\label{Lemma_LogTransParCorrExpBound} Suppose that
the Assumptions hold. Define $L:=1/\left(1-2^{-2}\left[1+\bar{\sigma}\right]^{2}\right)$
where $\bar{\sigma}$ is as in Lemma \ref{Lemma_implicationsConditionEigenvals}.
For $g\left(x\right)=2^{-1}\ln\left(\frac{1+x}{1-x}\right)$, $x\in\left(-1,1\right)$,
there is a constant $c_{7}>0$ (same as the one in Lemma \ref{Lemma_ParCorrExpBound}),
such that, for $n$ large enough, 
\[
\max_{i,j\leq K,\mathbf{k}\in\mathcal{K}_{i,j}}\Pr\left(\left|g\left(\hat{\Xi}_{i,j|\mathbf{k}}\right)-g\left(\Xi_{i,j|\mathbf{k}}\right)\right|\geq z\right)\leq32\exp\left\{ -\left(n-m\right)s^{-2}c_{8}\left(z/L\right)+3\ln K\right\} 
\]
for $\mathcal{K}_{i,j}\subseteq\left\{ \left[K\right]\setminus\left\{ i,j\right\} \right\} $
of cardinality $m$ and for $z$ satisfying $z\left(n-m\right)\rightarrow\infty$
and $z\left(\ln\left(n-m\right)\right)\left(\ln\ln\left(n-m\right)\right)\rightarrow0$.\end{lemma}

\begin{proof}By the mean value theorem $g\left(x\right)-g\left(y\right)=\partial g\left(\tilde{y}\right)\left(x-y\right)$
for $\tilde{y}$ is in the convex hull of $\left\{ x,y\right\} $,
$x,y\in\left(-1,1\right)$; here, $\partial g\left(\tilde{y}\right)=1/\left(1-\tilde{y}^{2}\right)$
is the derivative of $g$ evaluated at $\tilde{y}$. Suppose $\left|x-y\right|\leq\left(1-\bar{\sigma}\right)/2$
and $y\in\left[-\bar{\sigma},\bar{\sigma}\right]$ for some $\bar{\sigma}<1$.
Note that $\tilde{y}^{2}\leq\left(y+\left|x-y\right|\right)^{2}$,
so that $\partial g\left(\tilde{y}\right)\leq L$ and substituting
the aforementioned upper bound for $y$ and $\left|x-y\right|$ in
terms of $\bar{\sigma}$, and using the definition of $L$. Set $V:=\hat{\Xi}_{i,j|\mathbf{k}}-\Xi_{i,j|\mathbf{k}}$
and $U:=\partial g\left(\tilde{\Xi}_{i,j|\mathbf{k}}\right)$ where
$\tilde{\Xi}_{i,j|\mathbf{k}}$ is in the convex hull of $\left\{ \hat{\Xi}_{i,j|\mathbf{k}},\Xi_{i,j|\mathbf{k}}\right\} $.
The event $\left\{ UV\geq z\right\} $ is contained in the union of
the events $\left\{ V\geq z/L\right\} $ and $\left\{ U>L\right\} $.
From Lemma \ref{Lemma_ParCorrExpBound} we have that $\Pr\left(V\geq z/L\right)\leq16\exp\left\{ -\left(n-m\right)s^{-2}c_{7}\left(z/L\right)+3\ln K\right\} $
for $z$ satisfying the conditions of that lemma. The lemma then follows
if we show that $\left\{ U\geq L\right\} \subseteq\left\{ V\geq z/L\right\} $
for $z\rightarrow0$, as in the statement of the lemma. To this end,
note that $\left\{ U\geq L\right\} $ is contained in the union of
the events $\left\{ U>L,V\leq\left(1-\bar{\sigma}\right)/2\right\} $
and $\left\{ V>\left(1-\bar{\sigma}\right)/2\right\} $. The latter
event is eventually contained in $\left\{ V\geq z/L\right\} $ when
$z\rightarrow0$. Finally, the event $\left\{ U>L,V\leq\left(1-\bar{\sigma}\right)/2\right\} $
has probability zero because, by the remarks at the beginning of the
proof, we know that $U\leq L$ when $V\leq\left(1-\bar{\sigma}\right)/2$
and $\left|\Xi_{i,j|\mathbf{k}}\right|\leq\bar{\sigma}$, which is
the case by Lemma \ref{Lemma_implicationsConditionEigenvals}, uniformly
in $K$, for any $\mathbf{k}\in\mathcal{K}_{i,j}$. Hence, the lemma
is proved.\end{proof}

\subsection{Technical Lemmas for Lasso}

For $S\subseteq\left[2K\right]$ and some constant $L>0$, recall
that the square of the compatibility constant is $\phi_{{\rm comp}}^{2}\left(L,S,\Sigma\right):=\min\left\{ \frac{sb'\Sigma b}{\left|b\right|_{1}^{2}}:b\in\mathcal{R}\left(L,S\right)\right\} $
where $\mathcal{R}\left(L,S\right):=\left\{ b:\left|b_{S^{c}}\right|_{1}\leq L\left|b_{S}\right|_{1}\neq0\right\} $
(van de Geer and B\"{u}hlmann, 2009) . Here $S^{c}$ is the complement
of $S$ in $\left[2K\right]$. Throughout this section, the notation
is as in Algorithm \ref{Algo_LassoThresh} and Section \ref{Section_lassoResults}
and $\sigma_{\min}$ is as in Lemma \ref{Lemma_implicationsConditionEigenvals}.
We have the following.

\begin{lemma}\label{Lemma_compConditionFromEigs}Under the Assumptions,
for any $S\subseteq\left[2K\right]$ of cardinality $s$, and $L>0$,
$\phi_{{\rm comp}}\left(L,S,\hat{\Sigma}\right)\geq\sigma_{\min}^{1/2}-\left(L+1\right)\sqrt{s\left|\hat{\Sigma}-\Sigma\right|_{\infty}}$.\end{lemma}

\begin{proof}Note that the square root of the minimum eigenvalue
of a matrix is a lower bound for the compatibility constant. To see
this, note that $sb'\Sigma b/\left|b_{S}\right|_{1}^{2}\geq s\sigma_{\min}\left|b\right|_{2}^{2}/\left|b_{S}\right|_{1}^{2}\geq\sigma_{\min}$
because $s\left|b\right|_{2}^{2}\geq s\left|b_{S}\right|_{2}^{2}\geq\left|b_{S}\right|_{1}^{2}$.
Then, the lemma is special case of Corollary 10.1 in van de Geer and
B\"{u}hlmann (2009). \end{proof}

We now derive a basic bound for the Lasso procedure computed across
$2K$ response variables, one at the time, using the sufficient statistic
$\hat{\Sigma}$.

\begin{lemma}\label{Lemma_lassoBasicBound} Define 
\begin{equation}
\lambda_{0}=2\left(1+\max_{i\in\left[2K\right]}\sum_{j\in\left[2K\right]:j\neq i}\left|\Theta_{i,j}/\Theta_{i,i}\right|\right)\left|\hat{\Sigma}-\Sigma\right|_{\infty}.\label{EQ_lambda0}
\end{equation}
Under the Assumptions, on the event $E_{{\rm Lasso}}:=\left\{ \lambda\geq2\lambda_{0}\right\} $,
we have that $\max_{i\in\left[K\right]}\left|\hat{\beta}^{\left(i\right)}-\beta^{\left(i\right)}\right|_{1}=O_{P}\left(s\lambda/\sigma_{\min}\right)$.\end{lemma}

\begin{proof}We prove first the result for a fixed $i$. We shall
then see that the bound is uniform in $i\in\left[K\right]$. To avoid
notational complexities, we use a notation that is only local to this
proof. Set $\Gamma=\Sigma_{-i,-i}$ , $\gamma=\Sigma_{-i,i}$, $b=\beta_{-i}^{\left(i\right)}$
and $\hat{b}=\hat{\beta}_{-i}^{\left(i\right)}$. Note that $b=\Gamma^{-1}\gamma$
by definition. As in the text we use the hat for estimators of various
quantities. Write $\delta=\hat{b}-b$. Given that the Lasso estimator
minimises the Lasso objective function we have that 
\[
-2\hat{\gamma}'\hat{b}+\hat{b}'\hat{\Gamma}\hat{b}+\lambda\left|\hat{b}\right|_{1}\leq-2\hat{\gamma}'b+b'\hat{\Gamma}b+\lambda\left|b\right|_{1}.
\]
This can be rearranged to give the following inequality
\[
\delta'\hat{\Gamma}\delta\leq2\left(\hat{\gamma}'-b'\hat{\Gamma}\right)\delta+\lambda\left(\text{\ensuremath{\left|b\right|_{1}}}-\left|\hat{b}\right|_{1}\right)
\]
(Loh and Wainwright, 2012, eq. 5.1). Adding and subtracting $b'\Gamma$,
we write $\left(\hat{\gamma}'-b'\hat{\Gamma}\right)=\left(\hat{\gamma}'-b'\Gamma\right)+b'\left(\Gamma-\hat{\Gamma}\right)$.
Given that $b'\Gamma=\gamma'$, by definition of $\gamma$ and $\hat{\gamma}$,
we have that $\left|\hat{\gamma}-\Gamma b\right|_{\infty}\leq\left|\hat{\Sigma}-\Sigma\right|_{\infty}$.
By definition of $\Gamma$ and $\hat{\Gamma}$ and a basic inequality,
$\left|\left(\Gamma-\hat{\Gamma}\right)b\right|_{\infty}\leq\left|b\right|_{1}\left|\hat{\Sigma}-\Sigma\right|_{\infty}$.
However, $\left|b\right|_{1}=\sum_{j\in\left[2K\right]:j\neq i}\left|\Theta_{i,j}/\Theta_{i,i}\right|$
because the regression coefficients can be obtained from the precision
matrix: $\beta_{j}^{\left(i\right)}=-\Theta_{i,j}/\Theta_{i,i}$.
Hence, by definition of $\lambda_{0}$ as in the statement of the
lemma and the last display, we deduce that $\delta'\hat{\Gamma}\delta\leq\lambda_{0}\left|\delta\right|_{1}+\lambda\left(\text{\ensuremath{\left|b\right|_{1}}}-\left|\hat{b}\right|_{1}\right)$.
This is in the form of the basic inequality in van de Geer and B\"{u}hlmann
(2009, last display on p.1387). On the set $\left\{ \lambda\geq2\lambda_{0}\right\} $,
the r.h.s. of the previous inequality is bounded above by $2^{-1}\lambda\left|\delta\right|_{1}+\lambda\left(\text{\ensuremath{\left|b\right|_{1}}}-\left|\hat{b}\right|_{1}\right)$.
Then, by arguments in van de Geer and B\"{u}hlmann (2009, second
and third display on p.1388, replacing $\lambda_{0}$ with $2^{-1}\lambda$
in their definition of $L$, so that here $L=3$), we deduce that
\[
\left|\delta\right|_{1}\leq4\sqrt{s\delta'\hat{\Gamma}\delta/\hat{\phi}_{{\rm comp}}^{2}}
\]
where $\hat{\phi}_{{\rm comp}}:=\phi_{{\rm comp}}\left(L,S,\hat{\Sigma}\right)$
is the compatibility constant, which we shall show to be strictly
positive. Lemma 11.2 in van de Geer and B\"{u}hlmann (2009) says
that $\sqrt{\delta'\hat{\Gamma}\delta}=O\left(\frac{\lambda\sqrt{s}}{\hat{\phi}_{{\rm comp}}}\right)$
once we replace $\lambda_{0}$ with $\lambda/2$ in their lemma. By
Lemmas \ref{Lemma_compConditionFromEigs} and \ref{Lemma_CorrelationBound},
$\hat{\phi}_{{\rm comp}}=\sigma_{\min}^{1/2}-O_{P}\left(\sqrt{s\frac{\ln K}{n}}\right)$
choosing $L=3$ in Lemma \ref{Lemma_compConditionFromEigs}. We also
have that $\sqrt{s\frac{\ln K}{n}}=o\left(\sigma_{\min}^{1/2}\right)$.
By these remarks and the above display, we deduce $\left|\delta\right|_{1}=O_{P}\left(\frac{s\lambda}{\sigma_{\min}}\right)$.
The bound is uniform in $i\in\left[K\right]$ because Lemma \ref{Lemma_implicationsConditionEigenvals}.
Hence, the result follows.\end{proof}

\begin{lemma}\label{Lemma_lassoPenaltyDominance} Suppose that the
Assumptions hold. Then, for $\lambda_{0}$ as in (\ref{EQ_lambda0}),
$\lambda_{0}=O_{P}\left(\left(\omega/\nu^{2}\right)\sqrt{\frac{\ln K}{n}}\right)$
where $\nu$ is as in Lemma \ref{Lemma_implicationsConditionEigenvals}.\end{lemma}

\begin{proof}Under the Assumptions, an upper bound for (\ref{EQ_lambda0})
is given by $2\left(1+\omega/\nu^{2}\right)\left|\hat{\Sigma}-\Sigma\right|_{\infty}$.
This is $O_{P}\left(\left(\omega/\nu^{2}\right)\sqrt{\frac{\ln K}{n}}\right)$
using Lemma \ref{Lemma_CorrelationBound}. Hence, the result follows.\end{proof}

\subsection{Proof of Theorem \ref{Theorem_covMatConvergence}}

This follows from Lemma \ref{Lemma_CorrelationBound}.

\subsection{Proof of Theorem \ref{Theorem_lassoL1Consistency}}

An upper bound for (\ref{EQ_lambda0}) is given by $2\left(1+\omega/\nu^{2}\right)\left|\hat{\Sigma}-\Sigma\right|_{\infty}$.
Then, in Lemma \ref{Lemma_lassoBasicBound}, the set $\Pr\left(E_{{\rm Lasso}}\right)\rightarrow1$
as $K\rightarrow\infty$, for $\lambda=4\left(1+\omega/\nu^{2}\right)\times\frac{3c_{0}}{\pi}\sqrt{\frac{\ln K}{n}}$,
by Lemma \ref{Lemma_CorrelationBound}. Therefore, by Lemma \ref{Lemma_lassoBasicBound},
$\max_{i\in\left[K\right]}\left|\hat{\beta}^{\left(i\right)}-\beta^{\left(i\right)}\right|_{1}=O_{P}\left(\omega s\sqrt{\frac{\ln K}{n}}\right)$
and we can choose $c=12\left(1+\nu^{-2}\right)c_{0}/\pi$ in the statement
of the theorem. Hence, the result follows.

\subsection{Proof of Theorem \ref{Theorem_lassoSignConsistency}}

Note that $\theta_{\min}$ is a lower bound on $\min_{i,j}\left\{ \left|\beta_{j}^{\left(i\right)}\right|:\left|\beta_{j}^{\left(i\right)}\right|>0\right\} $.
This is because $\left|\beta_{j}^{\left(i\right)}\right|=\left|\Theta_{i,j}/\Theta_{i,i}\right|$.
Note that $-\Theta_{i,i}$ is the variance of $Z_{t,i}$ conditioning
on all other covariates. Hence, $\left|\Theta_{i,i}\right|\leq1$
because $Var\left(Z_{ti}\right)=1$ so that $\left|\beta_{j}^{\left(i\right)}\right|$
is either zero or greater than $\theta_{\min}$. Then, the event in
the probability of the theorem is contained in the event $\max_{i\in\left[K\right]}\left|\hat{\beta}^{\left(i\right)}-\beta^{\left(i\right)}\right|_{1}>\tau$,
because $\tau=o\left(\theta_{\min}\right)$. The latter event has
probability going to zero according to Theorem \ref{Theorem_lassoL1Consistency}.

\subsection{Proof of Theorem \ref{Theorem_ClimeOmegaConvergence}}

By Theorem 6 in Cai et al. (2011), $\left|\hat{\Omega}-\Theta\right|_{\infty}\leq4\left|\Theta\right|_{1,\infty}\lambda_{n}$,
on the event $E_{{\rm Clime}}:=\left\{ \lambda_{n}\geq\left|\Theta\right|_{1,\infty}\left|\hat{\Sigma}-\Sigma\right|_{\infty}\right\} $.
Choosing $\lambda_{n}=\omega\left(\frac{3c_{0}}{\pi}\sqrt{\frac{\ln K}{n}}\right)$
, by Lemma \ref{Lemma_CorrelationBound}, $\Pr\left(E_{{\rm Clime}}\right)\rightarrow1$
as $K\rightarrow\infty$.

\subsection{Proof of Theorem \ref{Theorem_ClimeSignConvergence}}

Due to the fact that $\left|\Theta_{i,j}\right|\in\left\{ 0\right\} \cup\left[\theta_{\min},\infty\right)$
and $\left|\hat{\Omega}_{i,j}\right|\in\left\{ 0\right\} \cup\left[\tau,\infty\right)$
uniformly in $i,j\in\left[2K\right]$, the event in the probability
of the theorem is eventually contained in $\left\{ \left|\hat{\Omega}-\Theta\right|_{\infty}\geq\tau\right\} $.
This goes to zero by Theorem \ref{Theorem_ClimeOmegaConvergence}
because $\tau$ is of larger order of magnitude than $\left|\hat{\Omega}-\Theta\right|_{\infty}$.

\subsection{Proof of Theorem \ref{Theorem_precisionMatrixConsistency}}

Under the event $E$ in (\ref{EQ_eventSetIdentification}), we are
within the framework of the results in Le and Zhong (2021). When such
event is true, the result follows from Theorem 3 in Le and Zhong (2021).
The proof of their result requires a bound in probability for $\text{\ensuremath{\left|\hat{\Sigma}-\Sigma\right|}}_{\infty}$;
see the third display on their page 12. In their proof this is denoted
by the symbol $\left|W_{X,nj}\right|_{\infty}$. We control this quantity
using Lemma \ref{Lemma_CorrelationBound}. To finish the proof note
that $\Pr\left(E\right)\rightarrow1$ using either Theorem \ref{Theorem_lassoSignConsistency}
or Theorem \ref{Theorem_ClimeSignConvergence}.

\subsection{Proof of Theorem \ref{Theorem_InnovationsCovAutoregressiveMatrix}}

From Lemma \ref{Lemma_identificationVarParms}, recall that $\Sigma_{\varepsilon}=\Theta_{11}^{-1}$
and $A=$ $-\Theta_{11}^{-1}\Theta_{12}$. By Lemmas \ref{Lemma_inverseMatrixConvergence}
and \ref{Lemma_OperatorNormUniformNormBound}, the Assumptions and
Theorem \ref{Theorem_precisionMatrixConsistency}, we deduce that
$\left|\hat{\Theta}_{11}^{-1}-\Theta_{11}^{-1}\right|_{{\rm op}}=O_{P}\left(s\sqrt{\frac{\ln K}{n}}\right)$
on the event $E$ in (\ref{EQ_eventSetIdentification}); note that
$\left|\Theta_{11}\right|_{0,\infty}\leq s$. The event $E$ has probability
going to one by either Theorem \ref{Theorem_lassoSignConsistency}
or Theorem \ref{Theorem_ClimeSignConvergence}. This proves the first
bound in the theorem. To prove the convergence of the autoregressive
matrix estimator, we note that $A-\hat{A}=\hat{\Theta}_{11}^{-1}\hat{\Theta}_{12}-\Theta_{11}^{-1}\Theta_{12}$.
The r.h.s. can be rewritten as $\hat{\Theta}_{11}^{-1}\left(\hat{\Theta}_{12}-\Theta_{12}\right)+\left(\hat{\Theta}_{11}^{-1}-\Theta_{11}^{-1}\right)\Theta_{12}$.
The first term in the sum is equal to 
\[
\Theta_{11}^{-1}\left(\hat{\Theta}_{12}-\Theta_{12}\right)+\left(\hat{\Theta}_{11}^{-1}-\Theta_{11}^{-1}\right)\left(\hat{\Theta}_{12}-\Theta_{12}\right).
\]
Then, by standard inequalities and the previous bounds, it is not
difficult to deduce that its operator norm is $O_{P}\left(s\sqrt{\frac{\ln K}{n}}\right)$.
The same follows for the operator norm of $\left(\hat{\Theta}_{11}^{-1}-\Theta_{11}^{-1}\right)\Theta_{12}$.
This concluded the proof of the theorem.

\subsection{Proof of Theorem \ref{Theorem_PcAlgoConsistency}}

The assumptions in Kalisch and B\"{u}hlmann (2007) are satisfied
by our Assumptions together with the faithfulness condition stated
in the theorem. In particular, from Kalisch and B\"{u}hlmann (2007,
proof of Lemma 4), it is sufficient to bound the probability of a
Type I and Type II error, as given by the following
\[
\Pr\left(\left|g\left(\hat{\Xi}_{i,j|\mathbf{k}}\right)-g\left(\Xi_{i,j|\mathbf{k}}\right)\right|\geq z\right)\leq32\exp\left\{ -\left(n-m\right)s^{-2}c_{7}\left(z/L\right)^{2}+3\ln K\right\} 
\]
where $m$ is the cardinality of $\mathbf{k}$, $g$ is as defined
in Lemma \ref{Lemma_LogTransParCorrExpBound}, and setting $z=c_{n}$
where $c_{n}$ is as in Kalisch and B\"{u}hlmann (2007): $c_{n}\asymp n^{-\eta_{c}}$.
Choosing $m$ equal to the maximal number of adjacent nodes, there
are $O\left(K^{m}\right)$ hypotheses to test. By Lemma 5 in Kalisch
and B\"{u}hlmann (2007), we can assume $m\leq s$ with probability
going to one. By this remark and the union bound we need the following
to go to zero: $K^{s}32\exp\left\{ -\left(n-s\right)s^{-2}c_{7}\left(c_{n}/L\right)^{2}+3\ln K\right\} $.
By the Assumptions, $s=O\left(n^{\eta_{s}}\right)=o\left(n^{1/2}\right)$
and $K^{s}=O\left(n^{s\eta_{K}}\right)$ for some finite $\eta_{K}$.
Hence we must have $n^{\eta_{s}}\ln n=o\left(n^{1-2\left(\eta_{s}+\eta_{c}\right)}\right)$.
This is the case if $2\eta_{c}+3\eta_{s}<1$, as stated in the theorem.
The theorem is then proved following the steps in the proof of Lemma
4 in Kalisch and B\"{u}hlmann (2007).

\subsection{Proof of Theorem \ref{Theorem_impulseResponse}}

Define the set $E_{G}:=\left\{ \hat{G}=G\right\} $, where $\hat{G}$
is the PCDAG estimated using Algorithm \ref{Algo_PCAlgo} and $G$
is the true PCDAG. Hence, on $E_{G}$ we have that that $\mathcal{\hat{V}}\left(i\right)=\mathcal{V}\left(i\right)$.
By Theorem \ref{Theorem_PcAlgoConsistency}, the event $E_{G}$ has
probability going to one. Hence, in what follows, we shall replace
$\mathcal{\hat{V}}\left(i\right)$ with $\mathcal{V}\left(i\right)$.
By the assumption of the present theorem, $G$ has all edges that
are directed. Let 
\[
\hat{\Psi}:=\left[\begin{array}{cccc}
\hat{\Sigma}_{\varepsilon,\mathcal{\hat{V}}\left(1\right),\mathcal{\hat{V}}\left(1\right)} & \mathbf{0} & \cdots & \mathbf{0}\\
\mathbf{0} & \hat{\Sigma}_{\varepsilon,\mathcal{\hat{V}}\left(2\right),\mathcal{\hat{V}}\left(2\right)} & \ddots & \vdots\\
\vdots & \mathbf{0} & \ddots & \mathbf{0}\\
\mathbf{0} & \cdots & \mathbf{0} & \hat{\Sigma}_{\varepsilon,\mathcal{\hat{V}}\left(K\right),\mathcal{\hat{V}}\left(K\right)}
\end{array}\right]
\]
and 
\[
\hat{\Phi}:=\left[\begin{array}{cccc}
\hat{\Sigma}_{\varepsilon,\mathcal{\hat{V}}\left(1\right),1} & \mathbf{0} & \cdots & \mathbf{0}\\
\mathbf{0} & \hat{\Sigma}_{\varepsilon,\mathcal{\hat{V}}\left(2\right),2} & \ddots & \vdots\\
\vdots & \mathbf{0} & \ddots & \mathbf{0}\\
\mathbf{0} & \cdots & \mathbf{0} & \hat{\Sigma}_{\varepsilon,\mathcal{\hat{V}}\left(K\right),K}
\end{array}\right];
\]
where the symbol $\mathbf{0}$ denotes a generic conformable matrix
of zeros. Then, the nonzero consecutive entries in the $i^{th}$ column
of $\hat{\Psi}^{-1}\hat{\Phi}$ is equal to $\hat{d}_{i}$ as defined
in Algorithm \ref{Algo_ImpulseResponse}. Here, we shall define the
population version of the above by $\Psi$ and $\Phi$. We define
a matrix $R$ such that $\Delta=\left(R\hat{\Psi}^{-1}\hat{\Phi}\right)'$.
The matrix $R$ reshapes $\hat{\Psi}^{-1}\hat{\Phi}$ so that we can
find $\Delta$. We write such matrix $R$ as 
\[
R:=\left[\begin{array}{cccc}
R_{1}^{\left(1\right)} & R_{1}^{\left(2\right)} & \cdots & R_{1}^{\left(K\right)}\\
R_{2}^{\left(1\right)} & R_{2}^{\left(2\right)} & \cdots & R_{2}^{\left(K\right)}\\
\vdots & \vdots & \ddots & \vdots\\
R_{K}^{\left(1\right)} & R_{K}^{\left(2\right)} & \cdots & R_{K}^{\left(K\right)}
\end{array}\right],
\]
where $R_{k}^{\left(i\right)}$ is a $1\times\mathcal{V}\left(i\right)$
vector defined as follows. If $k\notin\mathcal{V}\left(i\right)$,
then, $R_{k}^{\left(i\right)}$ is a row vector of zeros; for example
$R_{k}^{\left(k\right)}=0$, $k\in\left[K\right]$. If $k\in\mathcal{V}\left(i\right)$,
$R_{k}^{\left(i\right)}$ will have a one in the position such that
$R_{k}^{\left(i\right)}\hat{d}_{i}'\varepsilon_{t,\mathcal{V}\left(i\right)}=\hat{d}_{i,j}\varepsilon_{t,k}$,\textbf{
}where $j$ is the position of the element in $\mathcal{V}\left(i\right)$
that is equal to $k$; $\hat{d}_{i,j}$ is the estimated regression
coefficient of $\varepsilon_{t,k}$ in the regression of $\varepsilon_{t,i}$
on $\varepsilon_{t,\mathcal{V}\left(i\right)}$. This also means that
the number of ones in the $k^{th}$ row of $R$ is equal to the number
of direct descendants of the variable $\varepsilon_{t,k}$. We denote
such number by $\kappa_{k}$. Now, note that $\left|R\hat{\Psi}^{-1}\hat{\Phi}-R\Psi^{-1}\Phi\right|_{{\rm op}}\leq\left|R\right|_{{\rm op}}\left|\hat{\Psi}^{-1}\hat{\Phi}-\Psi^{-1}\Phi\right|_{{\rm op}}$.
Then, $\left|R\right|_{{\rm op}}^{2}$ is the maximum eigenvalue of
$RR'$ and the latter matrix is diagonal with $\left(k,k\right)$
entry equal to $\kappa_{k}$. It is easy to see that $RR'$ is diagonal
because the positions for two different parents cannot overlap, i.e.
$R_{k}^{\left(i\right)}\left(R_{l}^{\left(i\right)}\right)'=0$ when
$k\neq l$. Then, $\left|R\right|_{{\rm op}}=\kappa^{1/2}$, where
$\kappa:=\max_{k}\kappa_{k}$, as defined in the theorem. Hence, it
remains to bound $\left|\hat{\Psi}^{-1}\hat{\Phi}-\Psi^{-1}\Phi\right|_{{\rm op}}$;
note that the singular values of a matrix are invariant of transposition.
Adding and subtracting $\Psi^{-1}\hat{\Phi}$ , using the triangle
inequality, and a basic norm inequality, 
\begin{equation}
\left|\hat{\Psi}^{-1}\hat{\Phi}-\Psi^{-1}\Phi\right|_{{\rm op}}\leq\left|\hat{\Psi}^{-1}-\Psi^{-1}\right|_{{\rm op}}\left|\hat{\Phi}\right|_{{\rm op}}+\left|\Psi^{-1}\right|_{{\rm op}}\left|\hat{\Phi}-\Phi\right|_{{\rm op}}.\label{EQ_impulseResponseBasicBound}
\end{equation}
By Lemma \ref{Lemma_inverseMatrixConvergence}, $\left|\hat{\Psi}^{-1}-\Psi^{-1}\right|_{{\rm op}}\leq\left|\Psi^{-1}\right|_{{\rm op}}^{2}\left|\hat{\Psi}-\Psi\right|_{{\rm op}}$.
The maximum singular value of a block diagonal matrix is the maximum
of the singular values of each of the blocks. By Cauchy's interlacing
theorem, $\left|\hat{\Psi}-\Psi\right|_{{\rm op}}\leq\left|\hat{\Sigma}_{\varepsilon}-\Sigma_{\varepsilon}\right|_{{\rm op}}$
and the latter is $O_{P}\left(s\sqrt{\frac{\ln K}{n}}\right)$ by
Theorem \ref{Theorem_InnovationsCovAutoregressiveMatrix}. Using again
Cauchy's interlacing theorem, we deduce that the largest singular
value of $\Psi^{-1}$ is bounded above by the largest singular value
of $\Theta$, which is finite. Moreover, $\left|\hat{\Phi}\right|_{{\rm op}}\leq\left|\Phi\right|_{{\rm op}}+\left|\hat{\Phi}-\Phi\right|_{{\rm op}}$.
The maximum singular value of $\Phi$ is just the maximum of $\Sigma_{\varepsilon,\mathcal{\hat{V}}\left(i\right),i}'\Sigma_{\varepsilon,\mathcal{\hat{V}}\left(i\right),i}$
w.r.t. $i\in\left[K\right]$. It is increasing in the cardinality
of $\mathcal{\hat{V}}\left(i\right)$. Hence, \textbf{$\Sigma_{\varepsilon,\mathcal{\hat{V}}\left(i\right),i}'\Sigma_{\varepsilon,\mathcal{\hat{V}}\left(i\right),i}\leq\Sigma_{\varepsilon,\cdot,i}'\Sigma_{\varepsilon,\cdot,i}$},
recalling the notation at the start of Section \ref{Section_algos}.\textbf{
}The latter is bounded above by\textbf{ $\max_{x'x\leq1}x'\Sigma_{\varepsilon}'\Sigma_{\varepsilon}x=\left|\Sigma_{\varepsilon}\right|_{{\rm op}}^{2}$},
which is bounded, by the Assumptions. By the same argument as before,
the maximum singular value of $\hat{\Phi}-\Phi$ is the square root
of the largest, w.r.t. $i\in\left[K\right]$, of the maximum eigenvalue
of 
\[
\left(\hat{\Sigma}_{\varepsilon,\mathcal{\hat{V}}\left(i\right),i}-\Sigma_{\varepsilon,\mathcal{V}\left(i\right),i}\right)'\left(\hat{\Sigma}_{\varepsilon,\mathcal{\hat{V}}\left(i\right),i}-\Sigma_{\varepsilon,\mathcal{V}\left(i\right),i}\right)
\]
where on $E_{G}$, $\mathcal{\hat{V}}\left(i\right)=\mathcal{V}\left(i\right)$.
This quantity is increasing in the cardinality of $\mathcal{V}\left(i\right)$
so that the square root of the above display is bounded above by $\left|\hat{\Sigma}_{\varepsilon}-\Sigma_{\varepsilon}\right|_{{\rm op}}$,
which is $O_{P}\left(s\sqrt{\frac{\ln K}{n}}\right)$ by Theorem \ref{Theorem_InnovationsCovAutoregressiveMatrix}.
Using the derived upper bounds, it is easy to deduce that (\ref{EQ_impulseResponseBasicBound})
is $O_{P}\left(s\sqrt{\frac{\kappa\ln K}{n}}\right)$. 

From Lemma \ref{Lemma_SVAR_identification}, deduce that $\Pi\varepsilon_{t}=D\Pi\varepsilon_{t}+\xi_{t}$.
This can be rewritten as $\varepsilon_{t}=\Pi^{-1}D\Pi\varepsilon_{t}+\Pi^{-1}\xi_{t}$.
Hence, $\varepsilon_{t}=\Delta\varepsilon_{t}+\Pi^{-1}\xi_{t}$, where
$\Delta=\Pi^{-1}D\Pi$. Now, note that on the event $E_{G}$, as defined
at the start of the proof, any permutation matrix $\hat{\Pi}$ that
makes $\hat{\Pi}\hat{\Delta}\hat{\Pi}^{-1}$ lower triangular, with
diagonal entries equal to zero, also satisfies (\ref{EQ_SVAR}) when
we replace $\Pi$ with it. According to Algorithm \ref{Algo_ImpulseResponse}
we choose the one that requires the least number of row permutations
of the identity, which is unique. Then, on $E_{G}$, $\hat{\Pi}=\Pi$
because also $\Pi$ is unique. Therefore, on $E_{G}$, $\hat{D}:=\hat{\Pi}\Delta\hat{\Pi}^{-1}$
converges to $D:=\Pi\Delta\Pi^{-1}$. This shows the first statement
of the theorem. The convergence rate of $\hat{H}-H$ to zero can be
deduce from the first statement of the theorem together with Lemma
\ref{Lemma_inverseMatrixConvergence}, and Cauchy's interlacing theorem
and the definition $\Sigma_{\varepsilon}=H\left(\mathbb{E}\xi_{t}\xi_{t}'\right)H'$
in order to bound the singular values of $H^{-1}:=\left(I-D\right)$.

\subsection{Proof of Results in the Appendix}

\subsubsection{Proof of Lemma \ref{Lemma_transformsProperties}}

We prove each point separately. 

\paragraph{Points 1-2. }

It follows from R\"{u}schendorf and de Valk (1993, Proposition 1)
and the fact that $\Phi^{-1}$ is the quantile function of a standard
normal random variable. 

\paragraph{Point 3. }

Recall that $V_{t,1},V_{t,2}$ are independent of $X_{t,1},X_{t,2}$
and uniformly distributed in $\left[0,1\right]$. It is clear that
the population Spearman's rho obtained using the transformation (\ref{EQ_uniformTransformFunction})
depends on $\pi_{V}=\mathbb{E}V_{t,1}V_{t,2}$. When, $\mathbb{E}V_{t,1}\mathbb{E}V_{t,2}=1/2$,
we can deduce the result by computing expectation w.r.t. to $V_{t,1}$
and $V_{t,2}$ and then using simple algebra and the fact that $\tilde{F}_{1}\left(X_{t,1},V_{t,1}\right),\tilde{F}_{2}\left(X_{t,1},V_{t,2}\right)$
are uniformly distributed. 

\paragraph{Point 4. }

Note that $\rho$ is the definition of the population Spearman's rho
(Joe, 1997, p.32) and $Z_{t,1},Z_{t,2}$ are standard normal. Then,
their correlation is the stated function of Spearman's rho (Liu et
al., 2012).

\paragraph{Point 5. }

Let $X_{t,1}'$ and $X_{t,2}'$ be two independent copies of $X_{t,1}$
and $X_{t,2}$, independent of each other. Note that $F_{i}\left(x\right)=\mathbb{E}1_{\left\{ X_{t,i}'\leq x\right\} }$,
$i=1,2$. By these remarks and Fubini's Theorem, 
\[
\mathbb{E}F_{1}\left(X_{t,1}\right)F_{2}\left(X_{t,2}\right)=\mathbb{E}^{X_{t,1}'}\mathbb{E}^{X_{t,2}'}\Pr\left(X_{t,1}\geq X_{t,1}',X_{t,2}\geq X_{t,2}'\right)
\]
where $\mathbb{E}^{X_{t,k}'}$ is expectation w.r.t. the marginal
law of $X_{t,k}'$, $k=1,2$. By the fact that $X_{t,k}$ has same
distribution as $X_{t,k}'$, $k=1,2$, the r.h.s. of the above display
is equal to $\mathbb{E}^{X_{t,1}}\mathbb{E}^{X_{t,2}}\bar{C}\left(F_{1}\left(X_{t,1}\right),F_{2}\left(X_{t,2}\right)\right)$,
where $\bar{C}$ is a survival copula. This will not be unique everywhere,
unless the marginals are continuous. However, by assumption we can
choose $\bar{C}$ as the survival Gaussian copula, among possibly
other copulae. Recall the definition of the bivariate Gaussian copula
with scaling matrix $\Sigma$ with $\left(1,2\right)$ entry $\Sigma_{1,2}=r_{V}$:
$C\left(u_{1},u_{2}\right):=\Phi\left(\Phi^{-1}\left(u_{1}\right),\Phi^{-1}\left(u_{2}\right);r_{V}\right)$.
By symmetry of $C$, we have that 
\[
\bar{C}\left(F_{1}\left(X_{t,1}\right),F_{2}\left(X_{t,2}\right)\right)=C\left(1-F_{1}\left(X_{t,1}-\right),1-F_{2}\left(X_{t,2}-\right)\right).
\]
Taking marginal expectations $\mathbb{E}^{X_{t,1}}\mathbb{E}^{X_{t,2}}$,
the r.h.s. of the above display is exactly $h\left(r_{V}\right)$.
The strict monotonicity of $h\left(r\right)$ w.r.t. $r$ is a property
of the normal distribution and follows from Fan et al. (2017, Lemma
2).

\paragraph{Point 6. }

This follows by repeated use of the triangle inequality and the fact
that $\frac{1}{n}\sum_{t=1}^{n}\left(1-\mathbb{E}\right)F_{1}\left(X_{t,1}\right)F_{2}\left(X_{t,2}\right)$
converges to zero in probability by ergodicity.

\subsubsection{Proof of Lemma \ref{Lemma_impulseResponse}}

By the assumption of the model, $X_{t,k}:=f_{k}^{-1}\left(Z_{t,k}\right)$.
From (\ref{EQ_SVAR}) we deduce that $Z_{t}=AZ_{t-1}+\Pi^{-1}H\xi_{t}$
and in consequence that $Z_{t+s}=A^{s+1}Z_{t-1}+\sum_{r=0}^{s}A^{r}\Pi^{-1}H\xi_{t}$.
Then, (\ref{EQ_impulseResponseConditionalExplicit}) follows by taking
conditional expectation. The second result in the lemma follows by
the chain rule.

\section{Choice of Tuning Parameters\label{Section_ChoiceTuning}}

Algorithms \ref{Algo_LassoThresh} and \ref{Algo_climeThresh} require
to choose the penalty parameter $\lambda$ and the threshold $\tau$.
As shown in Theorems \ref{Theorem_lassoSignConsistency} and \ref{Theorem_ClimeSignConvergence}
we need $\tau>\lambda$. The exact values can be chosen by cross-validation
(CV). CV may not be suitable for time series problems. However, it
has been shown to work for prediction problems in the case of autoregressive
process of finite order (Burmann and Nolan, 1992). To this end, we
divide the sample data into $n_{{\rm CV}}$ nonoverlapping blocks
of equal size each. Each block is a test sample. Given the $i^{th}$
test sample, we use the remaining data as $i^{th}$ estimation sample.
Compute $\hat{\Theta}$ on the $i^{th}$ estimation sample and denote
this by $\hat{\Theta}_{{\rm est}}\left(\lambda,\tau,i\right)$ to
make the dependence on the parameters and block explicit. Compute
the scaling matrix $\hat{\Sigma}$ on the $i^{th}$ test sample using
Algorithm \ref{Algo_scalingMatrix} and denote it by $\Sigma_{{\rm test}}\left(i\right)$
to make the dependence explicit. We minimize the negative loglikelihood:
\[
\frac{1}{n_{{\rm CV}}}\sum_{i=1}^{n_{{\rm CV}}}\left[{\rm Trace}\left(\hat{\Sigma}_{{\rm test}}\left(i\right)\hat{\Theta}_{{\rm est}}\left(\lambda,\tau,i\right)\right)-\ln\det\left(\hat{\Theta}_{{\rm est}}\left(\lambda,\tau,i\right)\right)\right]
\]
w.r.t. $\left(\text{\ensuremath{\lambda},\ensuremath{\tau}}\right)\in\mathcal{T}$
where $\mathcal{T}\subset\left(0,\infty\right)^{2}$. Here, for any
matrix $A$, ${\rm diag}\left(A\right)$ a diagonal matrix with same
diagonal entries as $A$. 

In the simulations the parameter $\tau$ is fixed to $2\lambda$,
and we select $\lambda$ employing CV with $n_{CV}=5$. Starting with
a penalization equal to $\lambda=0.10$, we first search (by dividing
iteratively by two) a value for the minimum $\lambda$ such that all
off-diagonal elements of $\hat{\Theta}_{11}$ are zero (precisely
smaller than 1e-6). We denote this value as $\lambda_{0}$. Then we
search for the optimal $\lambda$ in $\left\{ \lambda_{0}/2,\lambda_{0}/(2^{2}),\ldots,\lambda_{0}/(2^{5})\right\} $. 

Computing both optimal parameters and a causal graph from the PC algorithm
can be time consuming over many simulations. Hence, in our simulations,
we employ an additional simplification. Rather than carrying out CV
for each simulation, we use two separate simulation samples to compute
two values of $\lambda$ according to the aforementioned procedure.
We then use the average of these two values as tuning parameter $\lambda$
in all simulations with the same design. 

\subsection{Choice of VAR Order Using AIC\label{Section_AICLagLength}}

To choose a number of lags greater than one, as in Section \ref{Section_empiricalOilPriceShock},
we can use Akaike's information criterion (AIC). The likelihood of
the latent Gaussian VAR (\ref{EQ_gaussianVAR}) of order greater than
one is proportional to $-\ln\det\left(\bar{\Sigma}_{\varepsilon}\right)$
where $\bar{\Sigma}_{\varepsilon}$ is the estimator computed from
Algorithm \ref{Algo_sparseCopulaParametersEstimation} modifying $\hat{\Omega}$
so that $\hat{\Omega}_{i,j}=1$ for $i,j\in\left[K\right]$. This
means that no zero restriction is imposed on the submatrix $\Theta_{11}=\Sigma_{\varepsilon}^{-1}$.
We can use the number of nonzero elements in $\hat{\Theta}_{12}$,
as number of parameters for the penalty in AIC. 

\section{Finite Sample Analysis via Simulations\label{Section_summarySimulations}}

We assess the finite sample performance of the different estimators
and evaluate their asymptotic properties for various degrees of time
series persistence and cross-sectional dimension. We compare our results
to naive methods that either do not account for sparsity in $\Theta$
or ignore the time series structure of the data. 

\subsection{The True Model}

To generate the time series of equation (\ref{EQ_gaussianVAR}) the
$K$ variables are divided into $\widetilde{K}$ independent clusters.
Each cluster is composed by $N$ variables and shares the same causal
structure as well as the autoregressive matrix. We denote with $\widetilde{A}$
and $\widetilde{H}$ the related coefficients of equation (\ref{EQ_gaussianVAR})
for each cluster. The matrix $\widetilde{H}$ is the matrix which
relates $\varepsilon_{t}$ with the associated structural shocks $\xi_{t}$
of a selected cluster. For the sake of simplicity, for each cluster,
the variables' order coincides with the topological order so that
the matrix $\Pi$ in Lemma \ref{Lemma_SVAR_identification} can be
set equal to the identity. 

We consider $N=3$ and $N=4$. When $N=3$ the three basic causal
structures are selected for each cluster, i.e., the causal chain,
common cause and v-structure. Given three variables $X$, $Y$ and
$Z$, if $X\rightarrow Y\rightarrow Z$, the causal structure is called
causal chain while if $X\leftarrow Y\rightarrow Z$ it is termed common
cause. The causal relation is named v-structure or immorality if $X\rightarrow Y\leftarrow Z$.
We also consider two additional structures when $N=4$: diamond 1
and diamond 2. These are defined as $X\rightarrow Y\leftarrow Z,X\rightarrow U\leftarrow Z$,
and $X\rightarrow Y\leftarrow Z,Y\rightarrow U$, respectively. 

The PC algorithm cannot distinguish between causal chain and common
cause, since these structures are in the same Markov equivalence class.
Then, the PC algorithm will provide the same graph with undirected
edges: $X-Y-Z$. Conversely, the v-structure, diamond 1 and diamond
2 can be identified by the PC algorithm. In this case, the PC algorithm
will return the causal graph with edges correctly oriented. 

To monitor the persistence of the time series, for each cluster, the
autoregressive matrix $\widetilde{A}$ is equal to a lower triangular
matrix with all elements (including the diagonal) equal to a constant
$a$, which describes the persistence of the series. The matrix $\widetilde{H}$
is a function of the selected causal structure. For the v-structure
\[
\widetilde{H}=\begin{bmatrix}1 & 0 & 0\\
0 & 1 & 0\\
1 & 1 & 1
\end{bmatrix}
\]
which is related to the causal structure $\varepsilon_{t,1}\rightarrow\varepsilon_{t,3}\leftarrow\varepsilon_{t,2}$.
Each variable causes itself, but may also affect other variables.
Finally, for simplicity, we suppose that the data have Gaussian marginals.
In this case, simulation of (\ref{EQ_gaussianVAR}) reduces to simulation
of a VAR(1) together with some linear transformations to ensure that
all the covariates have variance equal to one. The details are given
in Algorithm \ref{Algo_recoverASigma}.

\begin{algorithm}
\caption{Simulation of the Gaussian Copula VAR in (\ref{EQ_gaussianVAR}) when
the Marginals are Gaussian.}
\label{Algo_recoverASigma}

Set $N\times N$ matrices $\widetilde{A}$ and $\widetilde{H}$ s.t.
$\widetilde{H}$ is full rank.

For $k=1,2,...\widetilde{K}$:

Simulate i.i.d. $N\times1$ dimensional Gaussian vectors $\left(e_{t}^{(k)}\right)_{t\in\left[n\right]}$
with mean zero and identity covariance matrix. 

Compute $X_{t}^{(k)}=\widetilde{A}X_{t-1}^{(k)}+\widetilde{H}e_{t}^{(k)}$,
$t\in\left[n\right]$.

End of For.

Define the $K$-dimensional VAR(1) $X_{t}=A_{{\rm block}}X_{t-1}+H_{{\rm block}}e_{t}$,
where $X_{t}=\left(\left(X_{t}^{(1)}\right)',\left(X_{t}^{(2)}\right)',\ldots,\left(X_{t}^{(\widetilde{K})}\right)'\right)'$
and similarly for $e_{t}$, $t\in\left[n\right]$; a fortiori, $A_{{\rm block}}$
and $H_{{\rm block}}$are block diagonal matrices, where each block
equals $\widetilde{A}$ and $\widetilde{H}$, respectively.

Define $S=\left[{\rm diag}\left(Var\left(X_{t}\right)\right)\right]^{-1}$
where ${\rm diag}\left(\cdot\right)$ is the diagonal matrix with
diagonal equals to its argument. 

Set $A=SA_{{\rm block}}S^{-1}$, $\Sigma_{\varepsilon}=SH_{{\rm block}}H_{{\rm block}}'S'$.

Define the latent $K\times1$ vector $Z_{t}=SX_{t}$, $t\in\left[n\right]$.
\end{algorithm}

\subsection{Simulation Results}

\begin{figure}[!t]
\centering \includegraphics[scale=0.65]{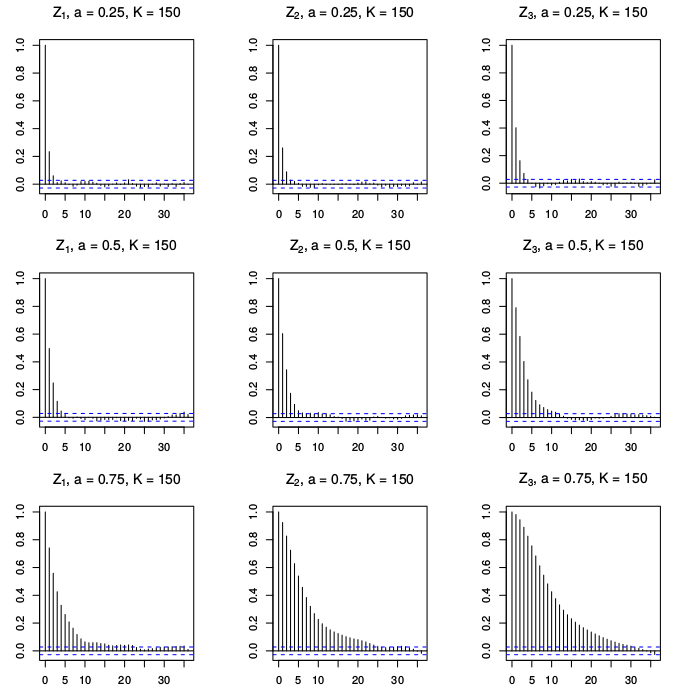}
\caption{Autocorrelation functions of the variable $Z_{t}$ of a cluster where
the contemporaneous causal relations are generated by a v-structure.}
\label{fig:ACF}
\end{figure}

To study the effect of time series persistent, three values of such
parameter $a$ are considered: $0.25$, $0.5$ and $0.75$. These
values of $a$ produce a wide range of time series dependence. For
example, Figure \ref{fig:ACF} shows the autocorrelation function
of a cluster for a v-structure. To analyze the relevance of sparsity
in our approaches, we select $\widetilde{K}=3,30,50$ clusters. We
investigate the finite sample properties of our estimator by considering
a sample size $n=1000,5000$. 

We use Algorithms \ref{Algo_LassoThresh} and \ref{Algo_climeThresh}
find the moral graph. Recall that the moral graph is defined from
the nonzero entries in $\hat{\Theta}$ as in Algorithm \ref{Algo_sparseCopulaParametersEstimation}.
We then follow Algorithms \ref{Algo_PCAlgo} and \ref{Algo_ImpulseResponse}
to estimate any remaining parameters. The tuning parameters for Algorithms
\ref{Algo_LassoThresh} and \ref{Algo_climeThresh} are chosen by
CV as described in Section \ref{Section_ChoiceTuning}. This means
only choosing $\lambda$. We denote the estimated parameter by $\lambda_{CV}$.
We use 250 simulations to compute the performance of our methodology. 

We also test the performance of the PC algorithm when we impose the
restrictions provided by Lasso and CLIME. The elements of $\hat{\Theta}_{11}$
which are equal to zero represent those edges which we exclude from
the skeleton. These restrictions can be embedded in the PC algorithm
using the appropriate ``fixedGaps'' command, which guarantees that
will be no edge between nodes \emph{j} and \emph{i} if the element
of $\hat{\Theta}_{11}$ in position $(i,j)$ is equal to zero. We
obtain improved compute time performance of the PC algorithm in this
case. This is particularly relevant in the high dimensional case.
Imposing the restriction has however nontrivial implications for the
PC algorithm, as an edge is deleted without a test so that no variable
is included in the separation set. We refer to Algorithm 1 and 2 in
Kalisch and B\"{u}hlmann (2007) for the details. In general, imposing
the restrictions might ensure that we obtain a DAG rather than a CPDAG.
It may also be advisable to use a tuning parameter $\lambda$ smaller
than the one suggested by CV. This is because the PC algorithm can
only delete edges, but not add them back. To verify if this is the
case, we also report results for $\lambda_{CV}/2$ and $\lambda_{CV}/4$.
We find no general evidence in favour of this claim. 

We compare our results with two benchmarks. One does not account for
sparsity and is essentially equivalent to choosing $\lambda=0$ in
the estimation. The second does not account for time series dependence,
and carries out the PC algorithm directly on the observed data. We
shall refer to these benchmarks as $\lambda=0$ and $A=0$, respectively.
The case $\lambda=0$ should produce sensible results in the low-dimensional
case. On the other hand, given that the simulated data are Gaussian,
the case $A=0$ should be appropriate when the time series dependence
is low. 

All approaches are compared on their performance to estimate the contemporaneous
causal structure. To achieve this, we report the average structural
Hamming distance (SHD) of the estimated causal graph to the true (Acid
and de Campos, 2003, Tsamardinos et al., 2006). The SHD between two
partially directed acyclic graphs counts how many edge types do not
coincide. For instance, estimating a non-edge instead of a directed
edge contributes an error of one to the overall distance. We remark
that the PC algorithm estimates the Markov equivalence class of a
given graph, i.e., the related CPDAG, and some causal structure, as
common cause and causal chain, shares the same class, i.e., the same
CPDAG, (e.g., for the v-structure the Markov class coincides with
the related DAG). Therefore, as the true causal structure in SHD analysis
we consider the (block) equivalence class attained by the PC algorithm,
with a very high significance level, $1-10^{-13}$, to obtain a deterministic
estimate performed on the theoretical correlation matrix of each cluster.

Tables \ref{tab:SHD_HighDim} and \ref{tab:SHD_LowDim} display the
average SHD and standard errors computed over 250 simulations for
all approaches. For the sake of conciseness we only report results
for the v-structure for the persistency parameter $a\in\left\{ 0.25,0.75\right\} $
and the number of clusters $\widetilde{K}\in\left\{ 3,50\right\} $\footnote{The complete results are available upon request.}.
Our approach produces estimators with superior finite sample performance,
relatively to the benchmarks, regardless of the considered causal
structures. While not reported here, we note that for both the causal
chain and common cause, the performance of the PC algorithm deteriorates
when we impose the a priori restrictions from the zeros of $\hat{\Theta}_{1,1}$
even if we undersmooth. 

The discrepancy among the contemporaneous causal structure is also
investigated by computing the number of nonzero elements of $\Theta_{11}$.
Indeed, we recall that nonzero elements of $\Theta_{11}$ correspond
to possible edges between variables of the corresponding row and column.
We also compute the number of false positive and negative between
the estimated and true $\Theta_{11}$ of nonzero elements\footnote{We say that an element of $\Theta_{11}$ is a false positive, if it
is estimated as nonzero element while it is zero. Vice versa, it is
a false negative, if it is estimated as zero element while it is different
from zero.}. Tables \ref{tab:FalsePositivesNegative_HighDim} and \ref{tab:FalsePositivesNegative_LowDim}
summarize the results for the high and low dimensional case, respectively.
We only report the results for the v-structure, as we can draw similar
conclusions for the other causal structures. 

\begin{table}[H]
\centering

\caption{Structural Hamming Distance for a Causal V-Structure. Expected value
approximated using 250 Monte Carlo simulations (standard errors in
parenthesis) for the SHD between the Lasso and CLIME estimators in
Algorithms \ref{Algo_LassoThresh} and \ref{Algo_climeThresh}, and
the true one. The contemporaneous causal structure is a v-structure
with $K=150$ variables with $\widetilde{K}=50$ clusters. Results
are reported for different values of $\lambda$ , where $\lambda_{CV}$
is the value obtained using cross-validation and denoted by $\lambda_{CV}$.
The columns labelled NR reports the SHD obtained when we do not impose
the restrictions provided by either Lasso or CLIME in the initialization
step of the PC algorithm. The columns $\lambda=0$ and $A=0$ refer
to the benchmarks that do not account for sparsity and time series
dependence, respectively. \label{tab:SHD_HighDim}}

\begin{tabular}{cccccccccc}
\hline 
 &  & \multicolumn{6}{c}{Lasso} &  & \tabularnewline
 &  & \multicolumn{2}{c}{$\lambda_{CV}$} & \multicolumn{2}{c}{$\lambda_{CV}/2$} & \multicolumn{2}{c}{$\lambda_{CV}/4$} & $\lambda=0$ & $A=0$\tabularnewline
$n$ & $a$ &  & NR &  & NR &  & NR &  & \tabularnewline
\hline 
 &  &  &  &  &  &  &  &  & \tabularnewline
1000 & 0.25 & 9.208 & 9.212 & 58.032 & 58.160 & 66.080 & 71.616 & 40.424 & 45.628\tabularnewline
 &  & (0.28) & (0.28) & (0.48) & (0.48) & (0.49) & (0.52) & (0.39) & (0.57)\tabularnewline
 & 0.75 & 1.960 & 95.888 & 4.464 & 4.488 & 29.596 & 29.556 & 131.060 & 225.212\tabularnewline
 &  & (0.14) & (0.2) & (0.2) & (0.2) & (0.35) & (0.34) & (0.93) & (0.48)\tabularnewline
5000 & 0.25 & 3.124 & 3.124 & 44.700 & 44.776 & 31.092 & 32.504 & 22.144 & 144.944\tabularnewline
 &  & (0.16) & (0.16) & (0.43) & (0.43) & (0.37) & (0.38) & (0.29) & (0.29)\tabularnewline
 & 0.75 & 0 & 99.904 & 2.496 & 2.496 & 2.780 & 2.780 & 51.696 & 230.704\tabularnewline
 &  & (0) & (0.03) & (0.17) & (0.17) & (0.17) & (0.17) & (0.47) & (0.62)\tabularnewline
\hline 
 &  & \multicolumn{6}{c}{CLIME} &  & \tabularnewline
 &  & \multicolumn{2}{c}{$\lambda_{CV}$} & \multicolumn{2}{c}{$\lambda_{CV}/2$} & \multicolumn{2}{c}{$\lambda_{CV}/4$} &  & \tabularnewline
$n$ & $a$ &  & NR &  & NR &  & NR &  & \tabularnewline
\hline 
 &  &  &  &  &  &  &  &  & \tabularnewline
1000 & 0.25 & 27.700 & 27.740 & 53.496 & 53.604 & 78.984 & 83.340 & - & \tabularnewline
 &  & (0.51) & (0.51) & (0.47) & (0.47) & (0.51) & (0.51) & - & \tabularnewline
 & 0.75 & 100.012 & 100.012 & 56.776 & 105.104 & 12.880 & 96.220 & - & \tabularnewline
 &  & (0.01) & (0.01) & (0.53) & (0.19) & (0.33) & (0.32) & - & \tabularnewline
5000 & 0.25 & 2.488 & 2.488 & 41.744 & 41.892 & 39.896 & 41.104 & - & \tabularnewline
 &  & (0.15) & (0.15) & (0.45) & (0.45) & (0.37) & (0.38) & - & \tabularnewline
 & 0.75 & 119.440 & 138.064 & 3.192 & 4.392 & 6.348 & 6.348 & - & \tabularnewline
 &  & (0.5) & (0.18) & (0.19) & (0.21) & (0.23) & (0.23) & - & \tabularnewline
\hline 
\end{tabular}
\end{table}

\begin{table}[H]
\centering

\caption{Structural Hamming Distance for a Causal V-Structure. Expected value
approximated using 250 Monte Carlo simulations (standard errors in
parenthesis) for the SHD between the Lasso and CLIME estimators in
Algorithms \ref{Algo_LassoThresh} and \ref{Algo_climeThresh}, and
the true one. The contemporaneous causal structure is a v-structure
with $K=9$ variables with $\widetilde{K}=3$ clusters. Results are
reported for different values of $\lambda$ , where $\lambda_{CV}$
is the value obtained using cross-validation and denoted by $\lambda_{CV}$.
The columns labelled NR reports the SHD obtained when we do not impose
the restrictions provided by either Lasso or CLIME in the initialization
step of the PC algorithm. The columns $\lambda=0$ and $A=0$ refer
to the benchmarks that do not account for sparsity and time series
dependence, respectively.\label{tab:SHD_LowDim}}

\begin{tabular}{cccccccccc}
\hline 
 &  & \multicolumn{6}{c}{Lasso} &  & \tabularnewline
 &  & \multicolumn{2}{c}{$\lambda_{CV}$} & \multicolumn{2}{c}{$\lambda_{CV}/2$} & \multicolumn{2}{c}{$\lambda_{CV}/4$} & $\lambda=0$ & $A=0$\tabularnewline
$n$ & $a$ &  & NR &  & NR &  & NR &  & \tabularnewline
\hline 
 &  &  &  &  &  &  &  &  & \tabularnewline
1000 & 0.25 & 0.184 & 0.184 & 0.16 & 0.172 & 0.156 & 0.16 & 0.16 & 2.756\tabularnewline
 &  & (0.04) & (0.04) & (0.04) & (0.04) & (0.04) & (0.04) & (0.04) & (0.16)\tabularnewline
 & 0.75 & 0.144 & 5.64 & 0.184 & 0.184 & 0.244 & 0.244 & 0.372 & 9.156\tabularnewline
 &  & (0.04) & (0.06) & (0.05) & (0.05) & (0.05) & (0.05) & (0.06) & (0.04)\tabularnewline
5000 & 0.25 & 0.18 & 0.18 & 0.272 & 0.272 & 0.244 & 0.256 & 0.224 & 8.632\tabularnewline
 &  & (0.05) & (0.05) & (0.05) & (0.05) & (0.05) & (0.05) & (0.05) & (0.07)\tabularnewline
 & 0.75 & 0 & 6 & 0.144 & 0.144 & 0.136 & 0.136 & 0.332 & 8.712\tabularnewline
 &  & (0) & (0) & (0.04) & (0.04) & (0.04) & (0.04) & (0.05) & (0.05)\tabularnewline
\hline 
 &  & \multicolumn{6}{c}{CLIME} &  & \tabularnewline
 &  & \multicolumn{2}{c}{$\lambda_{CV}$} & \multicolumn{2}{c}{$\lambda_{CV}/2$} & \multicolumn{2}{c}{$\lambda_{CV}/4$} &  & \tabularnewline
$n$ & $a$ &  & NR &  & NR &  & NR &  & \tabularnewline
\hline 
 &  &  &  &  &  &  &  &  & \tabularnewline
1000 & 0.25 & 0.256 & 0.256 & 0.184 & 0.188 & 0.16 & 0.152 & - & \tabularnewline
 &  & (0.05) & (0.05) & (0.04) & (0.04) & (0.04) & (0.04) & - & \tabularnewline
 & 0.75 & 6.016 & 6.016 & 4.776 & 6.592 & 0.196 & 5.244 & - & \tabularnewline
 &  & (0.01) & (0.01) & (0.12) & (0.05) & (0.05) & (0.08) & - & \tabularnewline
5000 & 0.25 & 0.18 & 0.18 & 0.264 & 0.264 & 0.256 & 0.26 & - & \tabularnewline
 &  & (0.05) & (0.05) & (0.05) & (0.05) & (0.05) & (0.05) & - & \tabularnewline
 & 0.75 & 0.084 & 5.028 & 0.184 & 0.192 & 0.264 & 0.28 & - & \tabularnewline
 &  & (0.03) & (0.09) & (0.05) & (0.05) & (0.05) & (0.05) & - & \tabularnewline
\hline 
\end{tabular}
\end{table}

\begin{sidewaystable}[H]
\centering

\caption{False Positives and Negatives for a Causal V-Structure. Expected number
of true plus false positives (TP+FP), false positives (FP) and false
negatives (FN) for the off-diagonal terms of $\Theta_{11}$ approximated
using 250 Monte Carlo simulations (standard errors in parenthesis).
The contemporaneous causal structure is a v-structure with $K=150$
variables with $\widetilde{K}=50$ clusters. The number of nonzero
off diagonal elements is 300, where the total number of the off-diagonal
elements is 22350. Results are reported for different values of $\lambda$
, where $\lambda_{CV}$ is the value obtained using cross-validation
and denoted by $\lambda_{CV}$. The column $\lambda=0$ refers to
the benchmark that does not account for sparsity.\label{tab:FalsePositivesNegative_HighDim}}

\begin{tabular}{cccccccccccccc}
\hline 
 &  & \multicolumn{9}{c}{Lasso} &  &  & \tabularnewline
 &  & \multicolumn{3}{c}{$\lambda_{CV}$} & \multicolumn{3}{c}{$\lambda_{CV}/2$} & \multicolumn{3}{c}{$\lambda_{CV}/4$} & \multicolumn{3}{c}{$\lambda=0$}\tabularnewline
$n$ & $a$ & TP+FP & FP & FN & TP+FP & FP & FN & TP+FP & FP & FN & TP+FP & FP & FN\tabularnewline
\hline 
 &  &  &  &  &  &  &  &  &  &  &  &  & \tabularnewline
1000 & 0.25 & 313.44 & 13.44 & 0 & 2197.832 & 1897.8 & 0 & 9711.5 & 9411.5 & 0 & 22350 & 22050 & 0\tabularnewline
 &  & (0.33) & (0.33) & (0) & (4.24) & (4.24) & (0) & (7.83) & (7.83) & (0) & (0) & (0) & (0)\tabularnewline
 & 0.75 & 210.344 & 4.72 & 94.376 & 549.104 & 249.1 & 0.024 & 2109.6 & 1809.6 & 0 & 22350 & 22050 & 0\tabularnewline
 &  & (0.3) & (0.19) & (0.23) & (1.31) & (1.31) & (0.01) & (3.35) & (3.35) & (0) & (0) & (0) & (0)\tabularnewline
5000 & 0.25 & 302.52 & 2.52 & 0 & 1472.928 & 1172.9 & 0 & 8488.2 & 8188.2 & 0 & 22350 & 22050 & 0\tabularnewline
 &  & (0.15) & (0.15) & (0) & (3.14) & (3.14) & (0) & (7.81) & (7.81) & (0) & (0) & (0) & (0)\tabularnewline
 & 0.75 & 200.096 & 0 & 99.904 & 300 & 0 & 0 & 343.08 & 43.08 & 0 & 22350 & 22050 & 0\tabularnewline
 &  & (0.03) & (0) & (0.03) & (0) & (0) & (0) & (0.59) & (0.59) & (0) & (0) & (0) & (0)\tabularnewline
\hline 
 &  & \multicolumn{9}{c}{CLIME} &  &  & \tabularnewline
 &  & \multicolumn{3}{c}{$\lambda_{CV}$} & \multicolumn{3}{c}{$\lambda_{CV}/2$} & \multicolumn{3}{c}{$\lambda_{CV}/4$} &  &  & \tabularnewline
$n$ & $a$ & TP+FP & FP & FN & TP+FP & FP & FN & TP+FP & FP & FN &  &  & \tabularnewline
\hline 
 &  &  &  &  &  &  &  &  &  &  &  &  & \tabularnewline
1000 & 0.25 & 300.928 & 0.928 & 0 & 1189.848 & 889.8 & 0 & 6638.7 & 6338.7 & 0 & - & - & -\tabularnewline
 &  & (0.09) & (0.09) & (0) & (3.18) & (3.18) & (0) & (6.96) & (6.96) & (0) & - & - & -\tabularnewline
 & 0.75 & 106.144 & 0 & 193.856 & 187.472 & 13.248 & 125.7 & 1024 & 807.4 & 83.424 & - & - & -\tabularnewline
 &  & (0.21) & (0) & (0.21) & (0.63) & (0.36) & (0.5) & (2.75) & (2.71) & (0.33) & - & - & -\tabularnewline
5000 & 0.25 & 300.56 & 0.56 & 0 & 760.752 & 460.7 & 0 & 4570.88 & 4270.8 & 0 & - & - & -\tabularnewline
 &  & (0.06) & (0.06) & (0) & (2.45) & (2.45) & (0) & (6.59) & (6.59) & (0) & - & - & -\tabularnewline
 & 0.75 & 235.48 & 0.032 & 64.552 & 318.344 & 19.544 & 1.2 & 764.216 & 464.2 & 0 & - & - & -\tabularnewline
 &  & (0.5) & (0.02) & (0.5) & (0.4) & (0.39) & (0.1) & (2) & (2) & (0) & - & - & -\tabularnewline
\end{tabular}
\end{sidewaystable}

\begin{sidewaystable}[H]
\centering

\caption{False Positives and Negatives for a Causal V-Structure. Expected number
of true plus false positives (TP+FP), false positives (FP) and false
negatives (FN) for the off-diagonal terms of $\Theta_{11}$ approximated
using 250 Monte Carlo simulations (standard errors in parenthesis).
The contemporaneous causal structure is a v-structure with $K=9$
variables with $\widetilde{K}=3$ clusters. The number of nonzero
off diagonal elements is 18, where the total number of the off-diagonal
elements is 72. Results are reported for different values of $\lambda$
, where $\lambda_{CV}$ is the value obtained using cross-validation
and denoted by $\lambda_{CV}$. The column $\lambda=0$ refers to
the benchmark that does not account for sparsity.\label{tab:FalsePositivesNegative_LowDim}}

\begin{tabular}{cccccccccccccc}
\hline 
 &  & \multicolumn{9}{c}{Lasso} &  &  & \tabularnewline
 &  & \multicolumn{3}{c}{$\lambda_{CV}$} & \multicolumn{3}{c}{$\lambda_{CV}/2$} & \multicolumn{3}{c}{$\lambda_{CV}/4$} & \multicolumn{3}{c}{$\lambda=0$}\tabularnewline
$n$ & $a$ & TP+FP & FP & FN & TP+FP & FP & FN & TP+FP & FP & FN & TP+FP & FP & FN\tabularnewline
\hline 
 &  &  &  &  &  &  &  &  &  &  &  &  & \tabularnewline
1000 & 0.25 & 22.968 & 4.968 & 0 & 41.584 & 23.584 & 0 & 57.48 & 39.48 & 0 & 72 & 54 & 0\tabularnewline
 &  & (0.24) & (0.24) & (0) & (0.41) & (0.41) & (0) & (0.34) & (0.34) & (0) & (0) & (0) & (0)\tabularnewline
 & 0.75 & 12.504 & 0.016 & 5.512 & 18.456 & 0.456 & 0 & 22.168 & 4.168 & 0 & 72 & 54 & 0\tabularnewline
 &  & (0.06) & (0.01) & (0.06) & (0.06) & (0.06) & (0) & (0.16) & (0.16) & (0) & (0) & (0) & (0)\tabularnewline
5000 & 0.25 & 18 & 0 & 0 & 20.808 & 2.808 & 0 & 38.04 & 20.04 & 0 & 72 & 54 & 0\tabularnewline
 &  & (0) & (0) & (0) & (0.16) & (0.16) & (0) & (0.35) & (0.35) & (0) & (0) & (0) & (0)\tabularnewline
 & 0.75 & 12 & 0 & 6 & 18 & 0 & 0 & 18.12 & 0.12 & 0 & 72 & 54 & 0\tabularnewline
 &  & (0) & (0) & (0) & (0) & (0) & (0) & (0.03) & (0.03) & (0) & (0) & (0) & (0)\tabularnewline
\hline 
 &  & \multicolumn{9}{c}{CLIME} &  &  & \tabularnewline
 &  & \multicolumn{3}{c}{$\lambda_{CV}$} & \multicolumn{3}{c}{$\lambda_{CV}/2$} & \multicolumn{3}{c}{$\lambda_{CV}/4$} &  &  & \tabularnewline
$n$ & $a$ & TP+FP & FP & FN & TP+FP & FP & FN & TP+FP & FP & FN &  &  & \tabularnewline
\hline 
 &  &  &  &  &  &  &  &  &  &  &  &  & \tabularnewline
1000 & 0.25 & 19.256 & 1.256 & 0 & 30.792 & 12.792 & 0 & 46.072 & 28.072 & 0 & - & - & -\tabularnewline
 &  & (0.11) & (0.11) & (0) & (0.34) & (0.34) & (0) & (0.37) & (0.37) & (0) & - & - & -\tabularnewline
 & 0.75 & 6.16 & 0 & 11.84 & 9.456 & 0.016 & 8.56 & 14.104 & 1.152 & 5.048 & - & - & -\tabularnewline
 &  & (0.05) & (0) & (0.05) & (0.12) & (0.01) & (0.11) & (0.12) & (0.1) & (0.08) & - & - & -\tabularnewline
5000 & 0.25 & 18 & 0 & 0 & 19.056 & 1.056 & 0 & 28.872 & 10.872 & 0 & - & - & -\tabularnewline
 &  & (0) & (0) & (0) & (0.1) & (0.1) & (0) & (0.29) & (0.29) & (0) & - & - & -\tabularnewline
 & 0.75 & 13.096 & 0.04 & 4.944 & 19.088 & 1.096 & 0.008 & 22.408 & 4.424 & 0.016 & - & - & -\tabularnewline
 &  & (0.09) & (0.02) & (0.09) & (0.09) & (0.09) & (0.01) & (0.17) & (0.17) & (0.01) & - & - & -\tabularnewline
\end{tabular}
\end{sidewaystable}

Finally, in Tables \ref{tab:distance_HighDim} and \ref{tab:distance_LowDim},
we assess the finite sample performance of the estimators of $A$
and $\Sigma_{\varepsilon}$ and analyse their asymptotic properties
stated in Theorem \ref{Theorem_InnovationsCovAutoregressiveMatrix}.
We compute the average distance from the true matrices, where the
distance is measured in terms of the operator's norm: the largest
singular value. These statistics are compared only to the case $\lambda=0$.

\begin{table}[H]
\centering

\caption{Average distance between $A$ and $\hat{A}$, $\Sigma_{\varepsilon}$
and $\hat{\Sigma}_{\varepsilon}$, respectively, computed over 250
simulations (standard errors in round brackets) when the contemporaneous
causal structure is a v-structure for $K=150$ variables with $\widetilde{K}=50$
clusters. For each method we report the results obtained also when
undersmoohting is performed, i.e., columns $\lambda_{CV}/2$ and $\lambda_{CV}/4$.
The column $\lambda=0$ refers to the benchmark that does not account
for sparsity. \label{tab:distance_HighDim}}

\begin{tabular}{ccccccccc}
\hline 
\multicolumn{2}{c}{} & \multicolumn{6}{c}{$\left|A-\hat{A}\right|_{{\rm op}}$} & \tabularnewline
 &  & \multicolumn{3}{c}{Lasso} & \multicolumn{3}{c}{CLIME} & \tabularnewline
$n$ & $a$ & $\lambda_{CV}$ & $\lambda_{CV}/2$ & $\lambda_{CV}/4$ & $\lambda_{CV}$ & $\lambda_{CV}/2$ & $\lambda_{CV}/4$ & $\lambda=0$\tabularnewline
\hline 
 &  &  &  &  &  &  &  & \tabularnewline
1000 & 0.25 & 0.567 & 1.25 & 2.354 & 0.684 & 1.082 & 2.639 & 314.10\tabularnewline
 &  & (0.003) & (0.003) & (0.006) & (0.006) & (0.003) & (0.012) & (3.106)\tabularnewline
 & 0.75 & 4.265 & 1.093 & 1.095 & 0.798 & 3.290 & 3.812 & >1000\tabularnewline
 &  & (0.048) & (0.006) & (0.003) & (0.016) & (0.083) & (0.054) & (-)\tabularnewline
5000 & 0.25 & 0.131 & 0.369 & 0.722 & 0.297 & 0.307 & 0.644 & 24.610\tabularnewline
 &  & (0.001) & (0.001) & (0.001) & (0.005) & (0.001) & (0.001) & (0.061)\tabularnewline
 & 0.75 & 3.604 & 0.925 & 0.135 & 3.555 & 1.404 & 0.250 & >1000\tabularnewline
 &  & (0.016) & (0.001) & (0.001) & (0.101) & (0.048) & (0.001) & (-)\tabularnewline
\hline 
\multicolumn{2}{c}{} & \multicolumn{6}{c}{$\left|\Sigma_{\varepsilon}-\hat{\Sigma}_{\varepsilon}\right|_{{\rm op}}$} & \tabularnewline
 &  & \multicolumn{3}{c}{Lasso} & \multicolumn{3}{c}{CLIME} & \tabularnewline
$n$ & $a$ & $\lambda_{CV}$ & $\lambda_{CV}/2$ & $\lambda_{CV}/4$ & $\lambda_{CV}$ & $\lambda_{CV}/2$ & $\lambda_{CV}/4$ & $\lambda=0$\tabularnewline
\hline 
 &  &  &  &  &  &  &  & \tabularnewline
1000 & 0.25 & 0.258 & 1.394 & 1.803 & 0.292 & 1.143 & 2.531 & 0.916\tabularnewline
 &  & (0.002) & (0.007) & (0.009) & (0.002) & (0.007) & (0.019) & (0.001)\tabularnewline
 & 0.75 & 0.430 & 0.119 & 0.228 & 0.118 & 0.335 & 0.616 & 0.331\tabularnewline
 &  & (0.006) & (0.001) & (0.0014) & (0.003) & (0.007) & (0.006) & (0.003)\tabularnewline
5000 & 0.25 & 0.081 & 0.367 & 0.511 & 0.076 & 0.316 & 0.582 & 0.395\tabularnewline
 &  & (0.001) & (0.001) & (0.002) & (0.002) & (0.001) & (0.002) & (0.001)\tabularnewline
 & 0.75 & 0.314 & 0.039 & 0.043 & 0.390 & 0.184 & 0.052 & 0.109\tabularnewline
 &  & (0.001) & (0.001) & (0.001) & (0.002) & (0.010) & (0.001) & (0.001)\tabularnewline
\hline 
\end{tabular}
\end{table}

\begin{table}[H]
\centering

\caption{Average distance between $A$ and $\hat{A}$, $\Sigma_{\varepsilon}$
and $\hat{\Sigma}_{\varepsilon}$, respectively, computed over 250
simulations (standard errors in round brackets) when the contemporaneous
causal structure is a v-structure for $K=9$ variables with $\widetilde{K}=3$
clusters. For each method we report the results obtained also when
undersmoohting is performed, i.e., columns $\lambda_{CV}/2$ and $\lambda_{CV}/4$.
The column $\lambda=0$ refers to the benchmark that does not account
for sparsity. \label{tab:distance_LowDim}}

\begin{tabular}{ccccccccc}
\hline 
\multicolumn{2}{c}{} & \multicolumn{6}{c}{$\left|A-\hat{A}\right|_{{\rm op}}$} & \tabularnewline
 &  & \multicolumn{3}{c}{Lasso} & \multicolumn{3}{c}{CLIME} & \tabularnewline
$n$ & $a$ & $\lambda_{CV}$ & $\lambda_{CV}/2$ & $\lambda_{CV}/4$ & $\lambda_{CV}$ & $\lambda_{CV}/2$ & $\lambda_{CV}/4$ & $\lambda=0$\tabularnewline
\hline 
 &  &  &  &  &  &  &  & \tabularnewline
1000 & 0.25 & 0.243 & 0.331 & 0.345 & 0.307 & 0.311 & 0.343 & 12.716\tabularnewline
 &  & (0.004) & (0.004) & (0.004) & (0.007) & (0.004) & (0.004) & (0.122)\tabularnewline
 & 0.75 & 2.835 & 0.845 & 0.182 & 0.673 & 1.774 & 2.186 & >1000\tabularnewline
 &  & (0.039) & (0.005) & (0.004) & (0.012) & (0.079) & (0.055) & (-)\tabularnewline
5000 & 0.25 & 0.079 & 0.097 & 0.140 & 0.078 & 0.091 & 0.129 & 10.495\tabularnewline
 &  & (0.001) & (0.002) & (0.002) & (0.002) & (0.002) & (0.002) & (0.049)\tabularnewline
 & 0.75 & 2.963 & 0.853 & 0.050 & 3.143 & 0.840 & 0.100 & >1000\tabularnewline
 &  & (0.017) & (0.002) & (0.001) & (0.027) & (0.010) & (0.009) & (-)\tabularnewline
\hline 
\multicolumn{2}{c}{} & \multicolumn{6}{c}{$\left|\Sigma_{\varepsilon}-\hat{\Sigma}_{\varepsilon}\right|_{{\rm op}}$} & \tabularnewline
 &  & \multicolumn{3}{c}{Lasso} & \multicolumn{3}{c}{CLIME} & \tabularnewline
$n$ & $a$ & $\lambda_{CV}$ & $\lambda_{CV}/2$ & $\lambda_{CV}/4$ & $\lambda_{CV}$ & $\lambda_{CV}/2$ & $\lambda_{CV}/4$ & $\lambda=0$\tabularnewline
\hline 
 &  &  &  &  &  &  &  & \tabularnewline
1000 & 0.25 & 0.149 & 0.176 & 0.169 & 0.125 & 0.185 & 0.179 & 0.158\tabularnewline
 &  & (0.004) & (0.003) & (0.002) & (0.004) & (0.003) & (0.002) & (0.002)\tabularnewline
 & 0.75 & 0.262 & 0.051 & 0.057 & 0.086 & 0.180 & 0.397 & 0.062\tabularnewline
 &  & (0.005) & (0.001) & (0.001) & (0.002) & (0.007) & (0.008) & (0.001)\tabularnewline
5000 & 0.25 & 0.035 & 0.060 & 0.082 & 0.035 & 0.050 & 0.084 & 0.072\tabularnewline
 &  & (0.001) & (0.001) & (0.001) & (0.001) & (0.001) & (0.001) & (0.001)\tabularnewline
 & 0.75 & 0.257 & 0.022 & 0.022 & 0.415 & 0.024 & 0.028 & 0.027\tabularnewline
 &  & (0.002) & (0.001) & (0.001) & (0.004) & (0.001) & (0.002) & (0.001)\tabularnewline
\hline 
\end{tabular}
\end{table}

\newpage{}


\begin{thebibliography}{10}
\bibitem{key-3} Acid, S. and L.M. de Campos (2003) Searching for
Bayesian Network Structures in the Space of Restricted Acyclic Partially
Directed Graphs. Journal of Artificial Intelligence Research 18, 445\textendash 490.

\bibitem{key-41}Bernanke, B. (1986) Alternative Explanations of the
Money-Income Correlation. In Carnegie-Rochester Conference Series
on Public Policy 25, 49-99. North- Holland.

\bibitem{key-43}Bernanke, B.S., J. Boivin and P. Eliasz (2005) Measuring
the Effects of Monetary Policy: A Factor-Augmented Vector Autoregressive
(FAVAR) Approach. The Quarterly Journal of Economics 120, 387-422.

\bibitem{key-44}Blanchard, O. and D. Quah (1989) The Dynamic Effects
of Aggregate Demand and Supply Disturbances. American Economic Review
79, 655-673.

\bibitem{key-15}B\"{u}hlmann, P., J. Peters and J. Ernest (2014)
CAM: Causal Additive Models, High-Dimensional Order Search and Penalized
Regression. The Annals of Statistics 42, 2526-2556.

\bibitem{key-1}Cai, T., W. Liu and X. Luo (2011) A Constrained $\ell_{1}$
Minimization Approach to Sparse Precision Matrix Estimation. Journal
of the American Statistical Association 106, 594-607.

\bibitem{key-45}Chari, V., P.J. Kehoe and E.R. McGrattan (2008) Are
Structural VARs with Long-Run Restrictions Useful in Developing Business
Cycle Theory?. Journal of Monetary Economics 55, 1337-1352.

\bibitem{key-16}Christiano, L.J., M. Eichenbaum and C. L. Evans (1999)
Monetary Policy Shocks: What Have We Learned and to What End?. Handbook
of Macroeconomics 1, 65\textendash 148.

\bibitem{key-2}Clarke, P.K. (1973) A Subordinated Stochastic Process
Model with Finite Variance for Speculative Prices. Econometrica 41,
135-155. 

\bibitem{key-5} Comon, P. (1994) Independent Component Analysis a
New Concept?. Signal Processing 36, 287\textendash 314. 

\bibitem{key-2}Cont, R., A. Kukanov and S. Stoikov (2014) The Price
Impact of Order Book Events. Journal of Financial Econometrics 12,
47-88.

\bibitem{key-3}Darsow, W.F., B. Nguyen and E.T. Olsen (1992) Copulas
and Markov processes. Illinois Journal of Mathematics 36, 600-642.

\bibitem{key-46}Demiralp, S. and K.D. Hoover (2003) Searching for
the Causal Structure of a Vector Autoregression. Oxford Bulletin of
Economics and Statistics 65, 745-767.

\bibitem{key-4}Donnelly, R. (2022) Optimal Execution: A Review. Applied
Mathematical Finance 29, 181-212.

\bibitem{key-2}Doukhan, P. (1995) Mixing. New York: Springer.

\bibitem{key-5}Fan, J., H. Liu, Y. Ning and H. Zou (2017) High Dimensional
Semiparametric Latent Graphical Model for Mixed Data. Journal of the
Royal Statistical Society B 79, 405-421.

\bibitem{key-1}Fan Y., F. Han and H. Park (2022) Estimation and Inference
in a High-dimensional Semiparametric Gaussian Copula Vector Autoregressive
Model. Preprint.

\bibitem{key-47}Faust, J. and E.M. Leeper (1997) When Do Long-Run
Identifying Restrictions Give Reliable Results?. Journal of Business
\& Economic Statistics 15, 345-353.

\bibitem{key-48}Forni, M., D. Giannone, M. Lippi and L. Reichlin
(2009) Opening the Black Box: Structural Factor Models with Large
Cross-Sections. Econometric Theory 25, 1319-1347.

\bibitem{key-11}Forni, M., M. Hallin, M. Lippi and L. Reichlin (2000)
The Generalized Dynamic-Factor Model: Identification and Estimation.
Review of Economics and Statistics 82, 540-554.

\bibitem{key-49}Gouri\'{e}roux, C., A. Monfort and J.-P. Renne (2017)
Statistical Inference for Independent Component Analysis: Application
to Structural VAR Models. Journal of Econometrics 196, 111-126.

\bibitem{key-9}Han, F. and W.B. Wu (2019) Probability Inequalities
for High Dimensional Time Series Under a Triangular Array Framework.
https://arxiv.org/abs/1907.06577v1.

\bibitem{key-17}Hanson, M. S. (2004) The \textquotedblleft Price
Puzzle\textquotedblright{} Reconsidered. Journal of Monetary Economics
51, 1385\textendash 1413.

\bibitem{key-6}Harris, N. and M. Drton (2013) PC Algorithm for Nonparanormal
Graphical Models. Journal of Machine Learning Research 14, 3365-3383.

\bibitem{key-6}Hyv\"{a}rinen, A., J. Karhunen and E. Oja (2001)
Independent Component Analysis. Wiley, New York.

\bibitem{key-7}Hyv\"{a}rinen, A. and E. Oja (2000) Independent Component
Analysis: Algorithms and Applications. Neural Networks 13, 411\textendash 430.

\bibitem{key-1} Huang, R. and T. Polak (2011) LOBSTER: The Limit
Order Book Reconstructor. School of Business and Economics, Humboldt
Universit\"{a}t zu Berlin, Techenical Report.

\bibitem{key-29}Joe, H. (1997) Multivariate Models and Dependence
Models. London: Chapman \& Hall.

\bibitem{key-27}Kalisch, M. and P. B\"{u}hlmann (2007) Estimating
High-Dimensional Directed Acyclic Graphs with the PC-Algorithm. Journal
of Machine Learning Research 8, 613-636.

\bibitem{key-2}K\"{a}nzig, D. (2021) The Macroeconomic Effects of
Oil Supply News: Evidence from OPEC Announcements. American Economic
Review 111, 1092-1125.

\bibitem{key-3-2}Kercheval, A.N., Y. Zhang (2015) Modelling High-Frequency
Limit Order Book Dynamics with Support Vector Machines. Quantitative
Finance 15, 1-15.

\bibitem{key-1}Kilian, L. and H. L\"{u}tkepohl (2017) Structural
Vector Autoregressive Analysis. Cambridge University Press.

\bibitem{key-1}Koop, G., M.H. Pesaran and S.M. Potter (1996) Impulse
Response Analysis in Non-Linear Multivariate Models. Journal of Econometrics
74, 119\textendash 147.

\bibitem{key-50}Lanne, M., M. Meitz and P. Saikkonen (2017) Identification
and Estimation of NonGaussian Structural Vector Autoregressions. Journal
of Econometrics 196, 288-304.

\bibitem{key-7}Lauritzen, S. L. (1996) Graphical Models. Oxford:
Oxford University Press.

\bibitem{key-5}Leeb, H. and B. M. P\"{o}tscher (2005) Model Selection
and Inference: Facts and Fiction. Econometric Theory 21, 21-59.

\bibitem{key-32}Liu, H., F. Han, M. Yuan, J. Lafferty and L. Wasserman
(2012) High Dimensional Semiparametric Gaussian Copula Graphical Models.
The Annals of Statistics 40, 2293-2326.

\bibitem{key-4}Liu, H., J. Lafferty and L. Wasserman (2009) The Nonparanormal:
Semiparametric Estimation of High Dimensional Undirected Graphs. Journal
of Machine Learning Research 10, 2295-2328.

\bibitem{key-51}L\"{u}tkepohl, H. and A. Net\v{s}unajev (2017)
Structural Vector Autoregressions with Heteroskedasticity: A Review
of Different Volatility Models. Econometrics and Statistics 1, 2-18.

\bibitem{key-1-2}MacKenzie, D. (2017) A Material Political Economy:
Automated Trading Desk and Price Prediction in High - Frequency Trading.
Social Studies of Science 47, 172-194 .

\bibitem{key-1}Mandelbrot, B. (1963) The Variation of Certain Speculative
Prices. Journal of Business 36, 394-419.

\bibitem{key-28}Meinshausen, N. and P. B\"{u}hlmann (2006) High-Dimensional
Graphs and Variable Selection with the Lasso. The Annals of Statistics
34, 1436-1462.

\bibitem{key-2}Mertens, K. and M. O. Ravn (2013) The Dynamic Effects
of Personal and Corporate Income Tax Changes in the United States.
American Economic Review 103, 1212-47.

\bibitem{key-1}Plagborg-M\o{}ller, M. and C.K. Wolf (2021) Local
Projections and VARs Estimate the Same Impulse Responses. Econometrica
89, 955-980. 

\bibitem{key-52}Moneta, A. (2008) Graphical Causal Models and VARs:
An Empirical Assessment of the Real Business Cycles Hypothesis. Empirical
Economics 35, 275-300.

\bibitem{key-52-1}Moneta, A., D. Entner, P. O. Hoyer and A. Coad
(2013) Causal Inference by Independent Component Analysis: Theory
and Applications. Oxford Bulletin of Economics and Statistics 75,
705-730.

\bibitem{key-6}Mucciante, L. and A. Sancetta (2022a) Estimation of
a High Dimensional Counting Process Without Penalty for High Frequency
Events. Econometric Theory: https://doi.org/10.1017/S0266466622000238.

\bibitem{key-7}Mucciante, L. and A. Sancetta (2022b) Estimation of
an Order Book Dependent Hawkes Process for Large Datasets. Preprint.

\bibitem{key-8}Pearl, J. (2000) Causality: Models, Reasoning, and
Inference. Cambridge, UK: Cambridge University Press.

\bibitem{key-10}Peters, J., J. M. Mooij, D. Janzing and B. Sch\"{o}lkopf
(2014) Causal Discovery with Continuous Additive Noise Models. Journal
of Machine Learning Research 15, 2009\textendash 2053.

\bibitem{key-53}Rigobon, R. (2003) Identification through Heteroskedasticity.
The Review of Economics and Statistics 85, 777-792.

\bibitem{key-4}Sancetta, A. (2018) Estimation for the Prediction
of Point Processes with Many Covariates. Econometric Theory 34, 598-627.89-107.

\bibitem{key-10}Sentana, E. and G. Fiorentini (2001) Identification,
Estimation and Testing of Conditionally heteroskedastic Factor Models.
Journal of Econometrics 102, 143-164.

\bibitem{key-9}Shimizu, S., P. O. Hoyer, A. Hyv\"{a}rinen and A.
Kerminen (2006) A Linear Non-Gaussian Acyclic Model for Causal Discovery.
Journal of Machine Learning Research 7, 2003\textendash 2030.

\bibitem{key-54}Sims, C. A. (1980) Macroeconomics and Reality. Econometrica
48, 1-48.

\bibitem{key-18}Sims, C. A. (1992) Interpreting the Macroeconomic
Time Series Facts: The effects of Monetary Policy. European Economic
Review 36, 975\textendash 1000.

\bibitem{key-42}Spirtes, P., C. Glymour and R. Scheines (2000) Causation,
Prediction, and Search. Boston: The MIT Press. 

\bibitem{key-1}Stock, J. H. and M. W. Watson (2018) Identification
and Estimation of Dynamic Causal Effects in Macroeconomics Using External
Instruments. The Economic Journal 128, 917-948.

\bibitem{key-55}Swanson, N. R. and C.W. Granger (1997) Impulse Response
Functions Based on a Causal Approach to Residual Orthogonalization
in Vector Autoregressions. Journal of the American Statistical Association
92, 357-367.

\bibitem{key-5}Tsamardinos, I., L. E. Brown and C. F. Aliferis (2006)
The max-min hill-climbing Bayesian network structure learning algorithm.
Machine Learning 65, 31\textendash 78.

\bibitem{key-6}Uhlig, H. (2005) What are the Effects of Monetary
Policy on Output? Result from an Agnostic Identification procedure.
Journal of Monetary Economics 52, 381-419.

\bibitem{key-13}Zhou, S., P. R\"{u}timann, M. Xu and P. B\"{u}hlmann
(2011) High-dimensional Covariance Estimation Based On Gaussian Graphical
Models. Journal of Machine Learning Research 12, 2975-3026.
\end{thebibliography}

\begin{thebibliography}{10}
\bibitem{key-1-1-1}Bhatia, R. (1996) Matrix Analysis. New York: Springer.

\bibitem{key-7-1}Burman, P. and D. Nolan (1992) Data Dependent Estimation
of Prediction Functions. Journal of Time Series Analysis 13, 189-207.

\bibitem{key-9-1}Cai, T., W. Liu and X. Luo (2011) A Constrained
$\ell_{1}$ Minimization Approach to Sparse Precision Matrix Estimation.
Journal of the American Statistical Association 106, 594-607.

\bibitem{key-2}Han, F. and W.B. Wu (2019) Probability Inequalities
for High Dimensional Time Series Under a Triangular Array Framework.
https://arxiv.org/abs/1907.06577v1.

\bibitem{key-29-1}Joe, H. (1997) Multivariate Models and Dependence
Models. London: Chapman \& Hall.

\bibitem{key-27-1}Kalisch, M. and P. B\"{u}hlmann (2007) Estimating
High-Dimensional Directed Acyclic Graphs with the PC-Algorithm. Journal
of Machine Learning Research 8, 613-636.

\bibitem{key-7-1}Lauritzen, S. L. (1996) Graphical Models. Oxford:
Oxford University Press.

\bibitem{key-32-1}Le, T.-M. and P.-S. Zhong (2021) High-Dimensional
Precision Matrix Estimation with a Known Graphical Structure. Stat
11, e424.

\bibitem{key-3}Liu, H., F. Han, M. Yuan, J. Lafferty and L. Wasserman
(2012) High Dimensional Semiparametric Gaussian Copula Graphical Models.
The Annals of Statistics 40, 2293-2326.

\bibitem{key-36-1}Loh, P.-L. and M. J. Wainwright (2012) High-Dimensional
Regression With Noisy and Missing Data: Provable Guarantees With Nonconvexity.
The Annals of Statistics 40, 1637-1664.

\bibitem{key-28-1}Meinshausen, N. and P. B\"{u}hlmann (2006) High-Dimensional
Graphs and Variable Selection with the Lasso. The Annals of Statistics
34, 1436-1462.

\bibitem{key-24-1}R\"{u}schendorf, L. and V. de Valk (1993) On Regression
Representation of Stochastic Processes. Stochastic Processes and their
Applications 46, 183-198.

\bibitem{key-5-1}van de Geer, S. A. and P. B\"{u}hlmann (2009) On
the conditions used to prove oracle results for the lasso. Electronic
Journal of Statistics 3, 1360-1392.

\bibitem{key-17-1}van der Vaart, A. and J.A. Wellner (2000) Weak
Convergence and Empirical Process Theory. New York: Springer. 
\end{thebibliography}
\end{document}